\documentclass[reprint, aps, groupedaddress, nofootinbib]{revtex4-1}
\usepackage{graphicx}
\usepackage{setspace,transparent}
\usepackage{color}
\usepackage{placeins}
\usepackage{fontenc,subcaption,ragged2e,wrapfig}
\usepackage{mathtools,amssymb,amsmath,nicefrac,mathrsfs}
\usepackage[font=footnotesize]{caption}
\usepackage{soul}
\usepackage{hyperref}
\usepackage[dvipsnames]{xcolor}
\usepackage[utf8]{inputenc}
\usepackage{tikz, tikz-cd}
\usetikzlibrary{arrows.meta, bending}
\usetikzlibrary{decorations, calligraphy}
\usepackage{algorithmicx}
\usepackage{algorithm}
\usepackage{listings,lineno}
\usepackage{algcompatible}
\usepackage[noend]{algpseudocode}

\tikzset{C/.style={
		circle, minimum size=8mm,
                 node contents={},
                 append after command={\pgfextra{%
        \draw [-stealth, semithick] 
        (-0.62, 0.2) arc [start angle=150, end angle=-150, x radius=0.7cm, y radius=0.4cm];
} }}}

\DeclareCaptionJustification{justified}{\justifying}
\DeclareMathAlphabet{\mathcalligra}{T1}{calligra}{m}{n}
\DeclareMathAlphabet{\mathpzc}{OT1}{pzc}{m}{it}
\DeclareMathOperator{\dd}{\mathrm{d}}

\begin{document}

\title{Realizing interdependent couplings as thermal or higher-order interactions}
%\title{Interdependent couplings map to both thermal and higher-order interactions}

\author{\textsc{Ivan Bonamassa}}
\email{Corresponding author: ivan.bms.2011@gmail.com\\ 
I.B.\ \& B.G.\ contributed equally to this work.}
\affiliation{Department of Physics, Bar-Ilan University, 52900 Ramat-Gan, Israel}
\author{\textsc{Bnaya Gross}}
\affiliation{Department of Physics, Bar-Ilan University, 52900 Ramat-Gan, Israel}
\author{\textsc{Shlomo Havlin}}
\affiliation{Department of Physics, Bar-Ilan University, 52900 Ramat-Gan, Israel}

\date{February 13, 2022}
\begin{abstract}
\noindent 
Interdependence is a fundamental ingredient to analyze the stability of many real-world complex systems featuring functional constraints. 
Yet, physical realizations of interdependent couplings are still unknown, lacking a theoretical benchmark to study their foundations. 
Here we present an interdependent magnetization framework and show that dependency links between $K-1$ pairwise networks of Ising spins can be physically realized as directed $K$-spin interactions or as adaptive thermal couplings. 
Supported by analytical results and extensive simulations, we find that interdependent interactions act like entropic forces that fully randomize greedy algorithms by amplifying site-to-site fluctuations, yielding unusual forms of vulnerability and unrecoverability. 
We discover a mapping between percolation in randomly interdependent networks and the zero-temperature ground state of random ferromagnetic multi-spin models, thanks to which we access the free-energy landscape of the former. 
By further exploring this connection, we highlight a surprising isomorphism between the one step replica-symmetry breaking solution of random formulas and the mutual giant component in interdependent networks. 
Besides raising perspectives of cross-fertilization, our results provide unfamiliar methods with practical implications in the study of constraint satisfaction and to control the functional metastability of interdependent systems. 
\end{abstract}
\maketitle

% This brings interdependent systems into a new dimension of interests and perspectives. 

\underline{\emph{Introduction}}. 
Interdependent couplings are network representations~\cite{buldyrev-nature2010} of the functional feedback mechanisms featured by many natural~\cite{pocock2012robustness, pagani2013power, bashan2012network, fornito2015connectomics, klosik2017interdependent, rocha2018cascading} and man-made multilayer ecosystems~\cite{rinaldi-ieee2001, little2002controlling, rosato-criticalinf2008, haldane2011systemic} that make these systems prone to systemic risks and catastrophic collapses~{\cite{may1977thresholds, scheffer2009critical, hokstad2012risk, helbing2013globally, guillen2015architecture}}. Despite the importance of this topic and the vast and interdisciplinary literature focusing on it~\cite{bianconi2018multilayer}, the {\em physics} of interdependent interaction has remained unexplored, lacking a theoretical framework to study its foundations. 

%Paraphrasing Anderson~{\cite{anderson-science1972}}, it could be said today that ``more systems is different'', seeing that cross-systems interactions often yield collective phenomena that differ qualitatively from those of their isolated macro-components~{\cite{andrei2021marvels}}. Interdependent systems~{\cite{rinaldi-ieee2001}} are a paradigmatic example of this thesis. Here, cross-layers couplings set mutual feedbacks between the functioning of their elements, making these adaptive systems prone to systemic risks and sudden collapses~{\cite{may1977thresholds, scheffer2009critical, buldyrev-nature2010, hokstad2012risk, helbing2013globally, guillen2015architecture}}. Despite the vast literature focusing on interdependence~\cite{bianconi2018multilayer} and on its importance for catastrophic shifts in many natural~\cite{pocock2012robustness, fornito2015connectomics, klosik2017interdependent} and man-made ecosystems~\cite{haldane2011systemic, pagani2013power, rocha2018cascading}, the {\em physics} underlying this interaction stands largely unexplored, lacking a theoretical benchmark to study its foundations. 

\begin{figure*}[t]
\centering
    \includegraphics[width=0.97\linewidth]{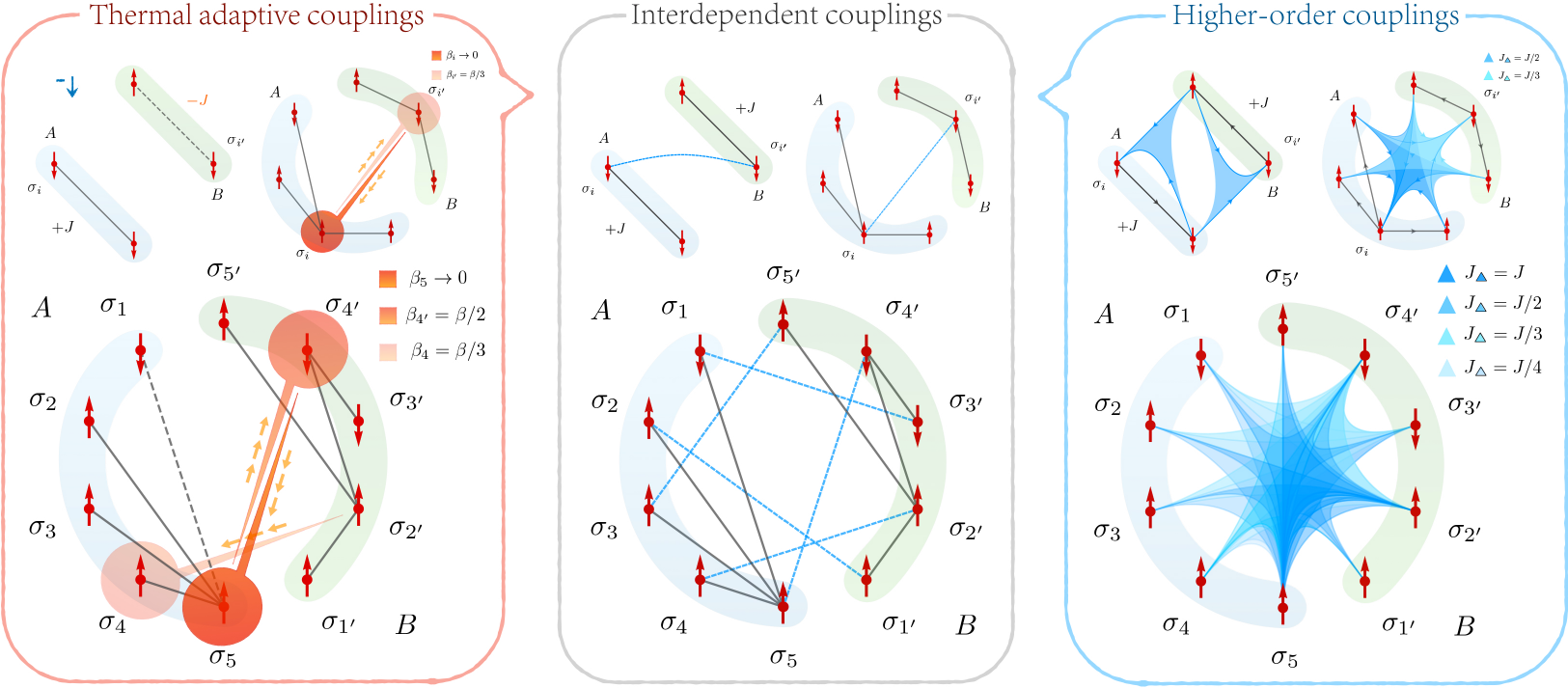}
            \caption{\small \textbf{Frameworks of interdependent spin networks.} (Color online) Illustration of the physical mapping of interdependence to thermal, higher-order interactions. (Center Box) Depending on the time scales governing the system's functioning, dependency links (dashed blue lines) between $M=2$ networks act (see Eq.~\eqref{eq:1}) either as (Left Box) {\em adaptive thermal couplings}, when interdependence is slower then the layers' thermalization, or as (Right Box) $3$-{\em spin couplings} on a directed hypergraph (notice the open hyperlinks and arrows) with Hamiltonian given by Eq.~\eqref{eq:2}, when the layers' thermalization is faster. In the thermal portrait, interdependent links generate an adaptive distribution of local temperatures (see Eq.~\eqref{eq:4}) {\em rarely} accompanied by an antiferromagnetic coupling (dashed gray bond, Left Box) due to local fields ordered in the $-1$ direction (blue arrow).}
        \label{fig:1}\vspace*{-0.2cm}
\end{figure*}

Here, we fill this fundamental gap by integrating interdependent interactions in the study of spin models. 
We show that dependency links matching the nodes of $K-1$ pairwise networks of Ising spins (Fig.\,\ref{fig:1}, center) map to $K$-spin interactions on directed hypergraphs~\cite{battiston2020networks, battiston2021physics} (Fig.\,\ref{fig:1}, right) or to adaptive thermal couplings (Fig.\,\ref{fig:1}, left).
%In so doing, we introduce two limiting frameworks: a {\em thermal} one (Fig.\,\ref{fig:1}, left), where interdependence is slower then the layers' thermalization, and a {\em higher}-{\em order} one (Fig.\,\ref{fig:1}, right), where the opposite occurs. 
We analyze the magnetic stability of randomly coupled Erd\H{o}s-R\'enyi (ER) networks for arbitrary fractions of dependency links and number of layers. 
The model undergoes spontaneous first-order transitions~\cite{binder1987theory, kirkpatrick1987p, biroli2012random} from a ``functioning'' (ferromagnetic/crystal) phase to a ``malfunctioning'' (paramagnetic/liquid) one from where it often fails to recover~{\cite{majdandzic-naturephysics2014, sanhedrai2022reviving}}. 
In fact, we find that collapsed networks remain trapped in a ``supercooled'' state from where they escape only if $K=3$ and the fraction $q$ of interdependent spins is below a certain threshold.

We corroborate these results theoretically within the thermal portrait, where the replica-symmetric (RS) solutions~{\cite{mezard1987spin}} of the isolated networks enable to solve the higher-order Hamiltonian via interdependent population dynamics.  
As a result, we find that interdependent interactions amplify the entropy production by propagating local thermal fluctuations over the available scales, boosting the rate at which greedy algorithms explore the system's configurational space. 
This is manifested in an unusual form of vulnerability~{\cite{bashan-naturephysics2013}} signaled by the loss of purely marginally stable states at low average connectivities and followed by a crossover towards the theoretical (bulk) melting spinodals~\cite{krzakala2011melting1} for increasing edge densities (concretely, graphs with mean degree $\langle k\rangle\gtrsim10)$. \\
\indent
Moreover, we discover a rigorous mapping between the zero temperature ground state (GS) of diluted ferromagnetic multi-spin models~{\cite{derrida1980random, ricci2001simplest, franz2001ferromagnet}}---governing the complexity of random \textsc{xor}-\textsc{sat}~\cite{mezard2009information}, a paradigm in constraint satisfaction---and percolation on randomly interdependent networks. 
We detail this correspondence and adopt the \textsc{xor-sat} entropy~\cite{franz2001exact, leone2001phase, mezard2003two} as {\em free}-{\em energy} to locate the coexistence threshold delimiting the structural metastability of interdependent networks. 
%Surprisingly, we find that their metastability width grows unbounded with the number of layers. 
In this hitherto unknown regime---whose width, most notably, diverges with the number of layers---these systems can almost surely be dismantled. 
However, we demonstrate with theory and simulations that percolation cascades~\cite{schneider2013algorithm, hwang2015efficient} in interdependent networks remain systematically trapped in the local minima (functional states) of their free-energy landscape. 
Besides revealing the physical nature of the interdependent percolation transition and its connection with the onset of complexity in random formulas ensemble~{\cite{krzakala2007gibbs, zdeborova2007phase, semerjian2008freezing,montanari2008clusters}}, %and offering algorithms to study the critical features of the clustering transition in \textsc{xorsat}/\textsc{col} problems, MOREOVER
this mapping raises perspectives of cross-fertilization, e.g.\ to design optimal algorithms~{\cite{morone2015influence, braunstein2016network, marino2016backtracking, osat2017optimal}} for dismantling/recovering interdependent systems as well as to study networks' energy surfaces~{\cite{ros2019complex, liu2021isotopy}} or ergodicity breaking~\cite{foini2012relation} in real-world adaptive systems. 

\emph{\underline{Interdependent spin networks}}. 
Let us consider $M$ networks of $N$ Ising spins $\sigma=\pm1$ embedded in a common heat bath at temperature $T$. 
Each layer's spin configuration $\boldsymbol{\sigma}=\{\sigma_i\}_{i=1,\dots,N}$ has energy $-\mathcal{H}[\boldsymbol{\sigma}, A]=J\sum_{i<j}A_{ij}\sigma_i\sigma_j$,\vspace*{-0.075cm} 
\noindent 
where $A=(A_{ij})_{i,j}$ is the adjacency matrix\vspace*{-0.05cm} of spin-spin interactions and $J\in{\rm I\!R}^+$ is their ferromagnetic strength. 
We model the magnetic evolution of the isolated layers by means of a Glauber single spin-flip process~{\cite{krapivsky2010kinetic}}: a randomly chosen spin, $\sigma_i$, changes its sign with probability $w(\sigma_i)=(1+e^{2\beta J k_i\sigma_i\Sigma_i})^{-1}$, where $\beta\equiv1/T$ (in units of $k_B^{-1}$) and $\Sigma_i\equiv \frac{1}{k_i}\sum_{j} A_{ij}\sigma_j$ is the effective field\vspace*{-0.05cm}
\noindent 
created at node $i$ by its $k_i=\sum_j A_{ij}$ neighbours. 
For easier notation, let $\mathcal{J}_{i\leftarrow j}\equiv\beta J k_i$ be the effective coupling strength of spin $i$.
To set interdependent interactions between layers, we adopt the $\Sigma_i$'s as proxies for the nodes' magnetic state~\cite{danziger2019dynamic} and intertwine their functioning via the spin-flip probabilities. 
Specifically, assuming e.g.\ that vertex $i$ in network $A$ depends on vertex $i'$ in network $B$ (Fig.~\ref{fig:1}, center box), we transform 
\begin{equation}\label{eq:1}
\mathcal{J}_{i\leftarrow j}^A \xmapsto{i \in A\, \leftarrow\, i'\in B} \mathcal{J}_{i\leftarrow j}^A \Sigma_{i'}^B. 
\end{equation}
%In what follows, we will always consider one-to-one mutual feedbacks, i.e.\ a relation similar to \eqref{eq:1} will also hold for $i'$ in $B$. 
In this way, $\sigma_{i}$ gains an {\em adaptive} coupling whose strength depends on the local order around $\sigma_{i'}$ and it flips with probability $w_{i}^{A\leftarrow B}=(1+\mathrm{exp}\{2\mathcal{J}_{i\leftarrow j}^A \sigma_{i}\Sigma_{i}^A\Sigma_{i'}^B\})^{-1}$. 
The role of the interdependent link is transparent: if $\sigma_{i'}$ ``fails'' to be functional---i.e., $\Sigma_{i'}^B\simeq0$---then $w_i^{A\leftarrow B}\simeq1/2$ and $\sigma_i$ fails along, since it flips sign independently of its neighbors' order. 
These ``concurrent malfunctions'' spread across the layers over time scales to be compared with those governing their magnetization evolution. 
In this regard, we introduce two limiting frameworks.

\indent 
$\circ$ \emph{Higher-order scheme} -- Dependency links intertwine the magnetization kinetics of each network at the same rate of single spin-flips, making the processes' time scales indistinguishable. 
An higher-order Hamiltonian $\mathcal{H}_\mathcal{I}$ can then be formed from those of single networks by integrating Eq.~\eqref{eq:1} within the system's structure. 
Let then $\mathcal{A}=A_1\oplus\cdots\oplus A_{M}$ be the direct sum of the networks' matrices and $\mathcal{I}\in{\rm I\!R}^{MN\times MN}$ a block matrix matching dependent nodes. 
For general networks of fully (see Methods for partial) interdependent Ising networks, we write
\begin{equation}\label{eq:2}
-\mathcal{H_I}[\{\boldsymbol{\sigma}_\mu\}_{\mu=1,\dots,M},\mathcal{A}]=J\sum_{i,j}\mathcal{A}_{ij}\sigma_i\sigma_j\prod_\ell \mathcal{I}_{i\ell}\Sigma_\ell,\vspace*{-0.2cm}
\end{equation}
\noindent 
where $i,j,\ell=1,\dots,MN$. 
If nodes of different layers are one-to-one interdependent, Eq.~\eqref{eq:2} simplifies to $-\mathcal{H_I}=J\sum_{i_1,\dots, i_{K}}\mathpzc{W}_{i_1\dots i_{K}}\sigma_{i_1}\!\cdots\sigma_{i_{K}}$, i.e.\ a $K$-spin model (with $K=M+1$) on a hypergraph with adjacency tensor 		
\begin{equation}\label{eq:3}
\mathpzc{W}_{i_1 i_2\dots i_K}\equiv \mathcal{A}_{i_1i_2}\prod_{\mu=1}^{M-1}
\frac{\mathcal{A}_{\ell_\mu(i_1) i_{\mu+2}}}{k_{\ell_\mu(i_1)}},
\end{equation}
\noindent 
$\forall i_1=1,\dots,MN$, where $\{\ell_\mu(i_1)\}_{\mu=1,\dots, M-1}$ is the set of vertices dependent on each $i_1$ that belong to the remaining layers~\cite{note2}. 
Notice that, while $\mathcal{I}$ and $\mathcal{A}$ are symmetric, $\mathpzc{W}$ is weighted and directed (neighboring spins depend on different nodes), generating then {\em incomplete} higher-order interactions (Fig.~\ref{fig:1}, right box). 
In this integrated portrait, sketched in Fig.~\ref{fig:2}\,\textbf{b} for $M=2$, the whole multilayer system evolves towards a bulk equilibrium.

\indent
$\circ$ \emph{Thermal scheme} -- Single networks learn the equilibrium spin configuration of the layers they depend on (dependency step) and adjust theirs (magnetization step) accordingly. 
In this portrait, %~{\cite{note0}}, 
Eq.~\eqref{eq:1} can be physically interpreted as a thermal coupling (Fig.~\ref{fig:1}, left box) that generates a recursive sequence of local temperatures' configurations~\cite{note1}.
This, e.g.\ for $2$ layers, is given by
\begin{equation}\label{eq:4}
\beta\xmapsto{A\to B} 
\boldsymbol{\beta}_{B}^{(1)}\xmapsto{B\to A} 
\boldsymbol{\beta}_{A}^{(1)}\xmapsto{A\to B}
\boldsymbol{\beta}_{B}^{(2)}\cdots,\vspace*{-0.1cm}
\end{equation}
where $\boldsymbol{\beta}\equiv\{\beta_{i}\}_{i\leq N}$ is a configuration of local tempera-\vspace*{-0.1cm}
\noindent 
tures, $\beta_{i,A}^{(n)}\equiv \beta\langle\Sigma_{i'}^B\rangle_{\boldsymbol{\beta}_B^{(n)}}$ and $\langle(\,\cdot\,)\rangle_{\boldsymbol{\beta}}$ is a thermal average.\vspace*{-0.1cm}
%Notice that, w.h.p.~{\cite{note1}}, $\beta_{\mu}^{(n)}\geq \beta_{\mu}^{(n-1)}$ for $\mu=A,\,B$ and $n\geq1$.
\noindent
This process (Fig.~\ref{fig:2}\,\textbf{a}) repeats for a number of iterations (NOIs) $n$, until all the layers mutually equilibrate.\vspace*{+0.2cm}\\
\begin{figure}[t]
\vspace*{-0.35cm}
$$
\hspace*{-4.0cm}\small{\mathbf{a})}\hspace*{+3.8cm} \small{\mathbf{b})}\vspace*{-0.65cm}
$$
\centering
\begin{tikzcd}[row sep=0.75cm, column sep=1.75cm, inner sep=1ex, arrow style=tikz, arrows=semithick]
\boldsymbol{\sigma}_A \arrow{r}[name=U]{\text{thermalize}}
& \boldsymbol{\sigma}_{A}^{eq} \arrow{d}{\boldsymbol{\beta}_B^{(n)}} \\
\boldsymbol{\sigma}_{B}^{eq} \arrow {u}{\boldsymbol{\beta}_A^{(n)}}
& \boldsymbol{\sigma}_{B} \arrow{l}[name=D]{\text{thermalize}} 
\arrow[to path={(U) node[pos=.5, C] (D)}, swap]{}
\arrow[to path={(U) node[midway, scale=3] {\scalebox{0.2}{$\text{NOIs}=n$}}  (D)}]{} 
\end{tikzcd}\,\,\hspace*{+0.1cm}\,\,
\begin{tikzcd}[row sep=0.16cm, arrow style=tikz, arrows=semithick]
  \boldsymbol{\sigma}_A \arrow[drr, bend left = 7.5] \arrow[swap]{dd}{\mathpzc{W}} & & \\
     & \hspace{-0.9cm}\scalebox{0.75}{\text{thermalize}} \hspace{-0.5cm} & \boldsymbol{\sigma}_{A\oplus B}^{eq}\\
  \boldsymbol{\sigma}_B \arrow[urr, bend right = 7.5] \arrow{uu} & &  
\end{tikzcd}
\caption{\small \textbf{Magnetization evolution.} 
Diagrammatic representation of the interdependent magnetization kinetics in \textbf{a}) {\em slow} (thermal) and $\textbf{b})$ {\em fast} (higher-order) schemes, for $M=2$ layers. 
While in the thermal framework the system equilibrates with a sequence of layer-by-layer thermalization stages, in the higher-order one the 2 networks concurrently equilibrate. } \label{fig:2}\vspace*{-0.5cm}
\end{figure}
\indent 
\underline{\emph{Metastable functioning}}. 
Monte Carlo simulations on ER networks (Fig.~\ref{fig:3}) highly support that thermal and higher-order frameworks yield identical equilibria. 
Based on this equivalence, we adopt the thermal scheme to develop an analytical approach for computing thermodynamic observables of the Hamiltonian $\mathcal{H_I}$, Eq.~\eqref{eq:2}.\\
\indent
Slow interdependence, in fact, endows each network with a distribution $\boldsymbol{\beta}$ of local temperatures whose disorder is adaptively quenched by the (relative) equilibrium configurations reached by its interdependent layers. 
We can then apply the replica method~{\cite{mezard1987spin}} to find the distribution of local fields $\langle \Sigma_i\rangle_{\boldsymbol{\beta}}$ of each isolated network before a dependency step occurs. 
Based on this, we develop (see Methods for details) an {\em interdependent population dynamics} (iPD) algorithm to solve iteratively the sequence of local temperatures in Eq.~\eqref{eq:4}.
For $n\to\infty$,\vspace*{-0.05cm} the distributions $\{\boldsymbol{\beta}_{\mu}^{(n)}\}_{\mu=1,\dots,M}$ converge (w.h.p.)\ to a set of steady state configurations at which the networks' average magnetizations $\{\mathcal{M}_\mu^{\infty}\}_{\mu=1,\dots,M}$ become co-stable and a mutual giant magnetic component $\mathcal{M}$ is found.\\
%Fig.~\ref{fig:3} portrays the results of interdependent population dynamics against Monte Carlo simulations. 
%We have examined the spinodal singularities~\cite{binder1987theory} limiting the metastable areas of $M=2$ (Fig.~\ref{fig:3}\textbf{a}) and $M=3$ (Fig.~\ref{fig:3}\textbf{b}) randomly {\em fully}-coupled ER graphs with equal average degree $\langle k\rangle$. 
%Theoretical thresholds coincide (within numerical accuracy) with the RS-ones featured by ferromagnetic $(M+1)$-spin models on random hypergraphs~\cite{franz2001ferromagnet, franz2001exact} with average hyperdegree $\langle k\rangle$ (dotted curves, Fig.~\ref{fig:3}\textbf{a},\textbf{b}). 
%Both models host (bulk) melting transitions above structural thresholds $\langle k\rangle_{sp}$ (dot-dashed lines), 
\indent 
Given the iPD algorithm above, we first examine  the (bulk) melting transitions~\cite{krzakala2011melting1} of $M=2$ (Fig.~\ref{fig:3}\textbf{b}) and $M=3$ (Fig.~\ref{fig:3}\textbf{a}) randomly {\em fully}-coupled ER Ising networks with equal average degree $\langle k\rangle$. 
Theoretical thresholds coincide (within numerical accuracy) with the RS-ones featured by ferromagnetic $(M+1)$-spin models on random hypergraphs~\cite{franz2001ferromagnet} with average hyperdegree $\langle k\rangle$ (dotted curves, Fig.~\ref{fig:3}\textbf{a},\textbf{b}). 
Notably, both models host ferromagnetic spinodals~\cite{binder1987theory} above structural thresholds $\langle k\rangle_{sp}$ (dot-dashed lines, Fig.~\ref{fig:3}\textbf{a},\textbf{b}) matching the tipping points of percolation cascades~\cite{buldyrev-nature2010} in fully-coupled ER networks. 
We explore this surprising connection in the upcoming section. 
Let us focus now on the agreement of the above theory with Monte Carlo simulations.\\
\indent
Full symbols in Fig.~\ref{fig:3}\textbf{a},\textbf{b} show that simulations on ER graphs with $\langle k\rangle\!\gtrsim\!10$ fit nicely the analytical predictions, while sparser structures are extremely more fragile.
The theoretical equivalence with hypergraphs suggests that this stronger low-$\langle k\rangle$ vulnerability stems from a systemic amplification of thermal fluctuations by interdependence. 
Diluted multi-spin models, in fact, undergo two relevant ground state (GS) transitions~\cite{ricci2001simplest, franz2001ferromagnet}: a {\em dynamical} (clustering) one, precisely at $\langle k\rangle_{sp}$, where marginally stable ferromagnetic solutions appear~\cite{franz2001exact, leone2001phase, mezard2003two}, followed by a {\em static} one at $\langle k\rangle_{cx}$ (double-dot-dashed lines, Fig.~\ref{fig:3}\textbf{a},\textbf{b}), where the ferro-para transition takes place. 
Greedy algorithms generally remain trapped in the local energy minima %populating the area 
between these thresholds~{\cite{mezard2009information}}, resulting in (exponentially) long-lived metastable functional states~{\cite{binder1987theory}}. 
Interdependent interactions, on the other hand, randomly spread the thermal fluctuations of the fields $\Sigma_i$, acting as infinite-range entropic forces that fully randomize the kinetics of the Glauber process. 
%induce local algorithms (like Glauber's) to escape marginal energy minima by randomizing their kinetics. 
This surprising phenomenon of noise ``self-injection'' boosts the phase-space exploration and~\cite{motwani1995randomized}, analogously with the enhanced performances recently reported in noise-enriched Ising machines~\cite{pierangeli2020noise, yamamoto2021recent, dutta2021ising}, it enables the spin-flip algorithm to escape the region of purely metastable states between $\langle k\rangle_{sp}$ and $\langle k\rangle_{cx}$, for then crossing over to the expected spinodals as the local fluctuations weaken. 
%To verify this intriguing scenario, we start from noticing that RS solutions--hence, interdependent population dynamics--are thermal averages and, as such, they are blind to site-to-site thermal fluctuations. We exploit this fact in two independent ways. First, we analyze the case of all-to-one interdependence, i.e.\ we replace the effective fields $\Sigma_i^{\mu}$ in Eq.~\eqref{eq:1} with the layers' average magnetizations $\{\mathcal{M}_\mu^{\infty}\}_{\mu=1,\dots,M}$. 
To verify the above scenario, we notice that iPD algorithms root on a thermal averaging, so they are blind to site-to-site thermal fluctuations. 
This motivates two experiments: $i)$~setting global (all-to-one) interdependence by replacing $\Sigma_i^{\mu}$ in Eq.~\eqref{eq:1} with the global magnetizations $\mathcal{M}_\mu^{\infty}$, so that thermal fluctuations are surely negligible; $ii)$~injecting synthetic Gaussian noise as $\Sigma_i^\mu+\xi_i$, with $\xi_i\in\mathcal{N}(0,a/\langle k\rangle)$, whose strength decreases with $\langle k\rangle$. 
\begin{figure*}
	\centering	
	\begin{tikzpicture}[      
	every node/.style={anchor=south west, inner sep=0pt},
	x=1mm, y=1mm,]   
	\node (fig1) at (0,-51.5)
	{\includegraphics[width=0.435\linewidth]{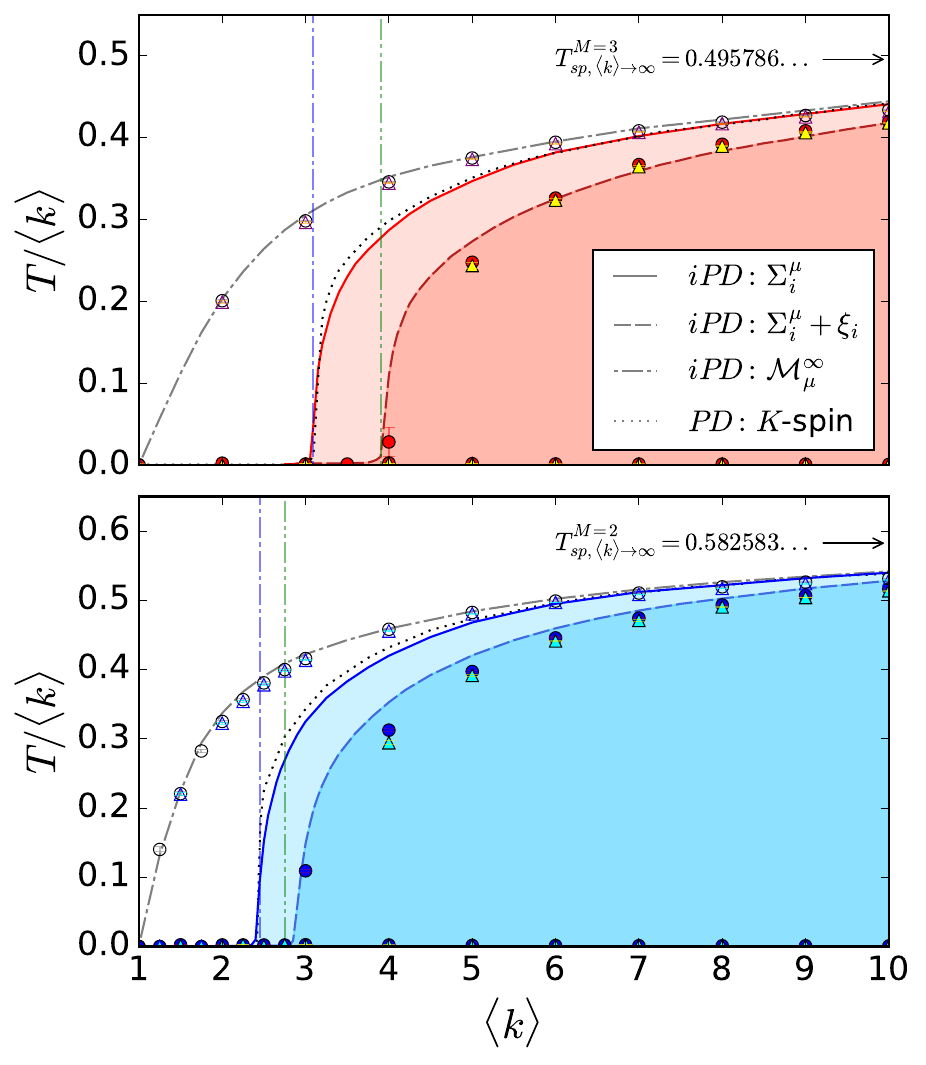}};
	\node at (14, -9) {\small{\textbf{b})}};
	\node at (37, -10) {\scriptsize{\textcolor{OrangeRed}{($\textrm{P}$)}}};
	\node at (35, 2) {\scriptsize{(\textcolor{OliveGreen}{$\textrm{F}$}+\textcolor{OrangeRed}{$\textrm{P}$}})};
	\node at (14, +31) {\small{\textbf{a})}};
	\node (fig2) at (+42,-41)
	{\includegraphics[scale=0.29]{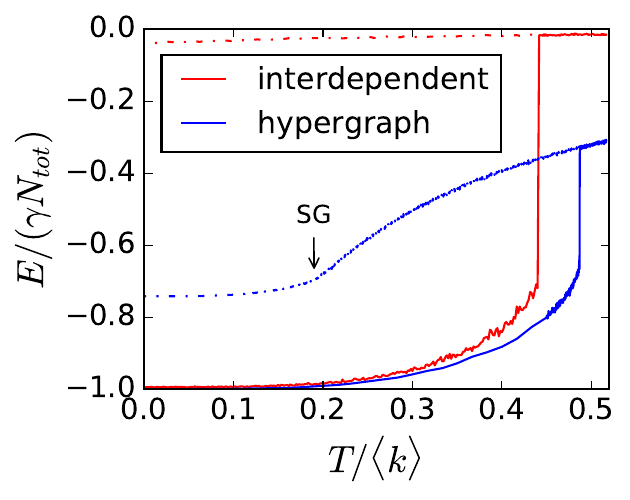}};
	\end{tikzpicture}\hspace*{-0.1cm}
	\begin{tikzpicture}[
	every node/.style={anchor=south west, inner sep=0pt},
	x=1mm, y=1mm,]   
	\node (fig1) at (0,-58.5)
	{\includegraphics[width=0.53\linewidth]{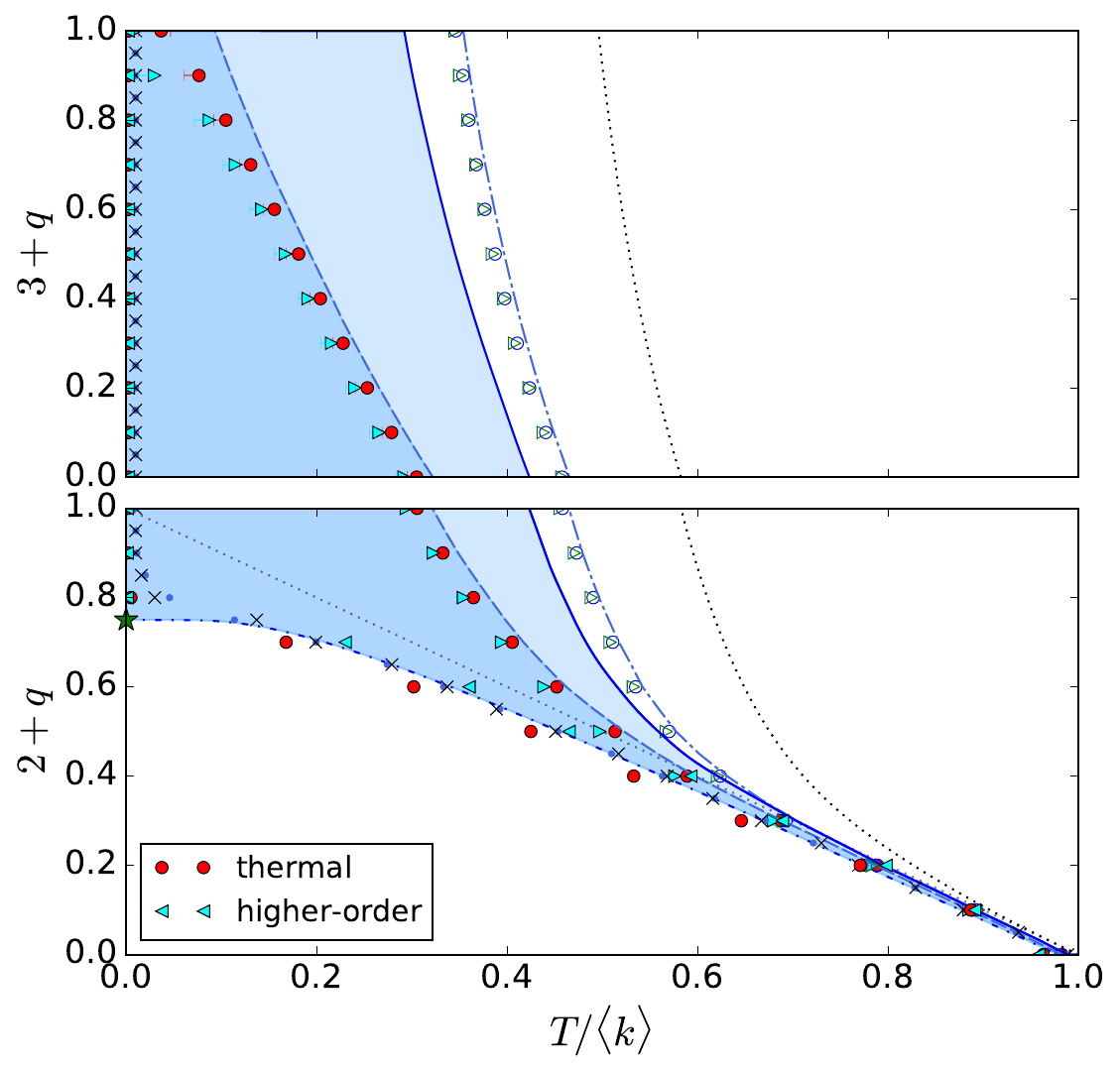}}; 
	\node at (86.5, -6) {\small{\textbf{c})}};
	\node at (86.5, -45) {\small{\textbf{d})}};
	\node at (19, -30) {\scriptsize{\textcolor{OliveGreen}{($\textrm{F}$)}}};
	\node at (17, -5) {\scriptsize{(\textcolor{OliveGreen}{$\textrm{F}$}+\textcolor{OrangeRed}{$\textrm{P}$}})};
	\node at (70, -5) {\scriptsize{\textcolor{OrangeRed}{($\textrm{P}$)}}};
	\node (fig2) at (+42.5,-0.5)
	{\includegraphics[scale=0.34]{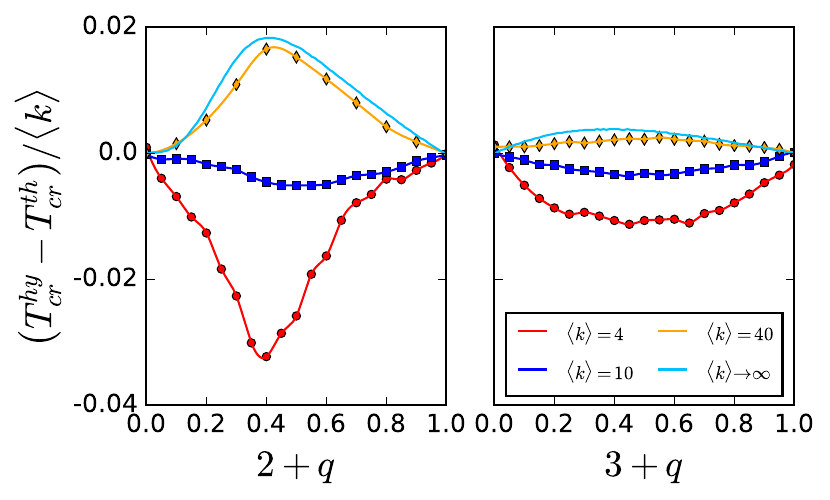}};
	\node (fig3) at (+59.5,-36)
	{\includegraphics[scale=0.30]{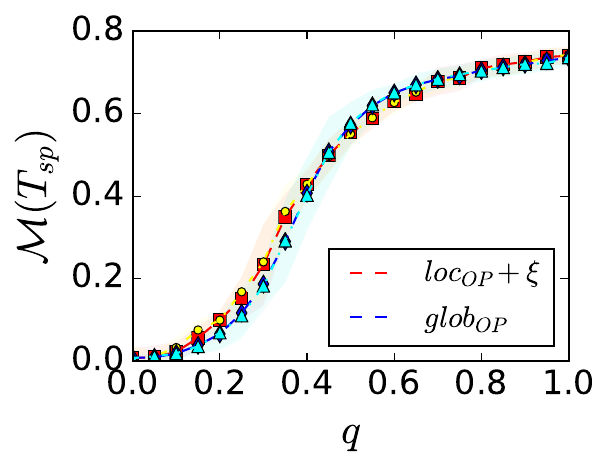}};
	\end{tikzpicture}\vspace*{-0.3cm}
\caption{\small{\textbf{Extreme vulnerability and non-recoverability.} (Color online) Analytical (iPD) and numerical (Monte Carlo) phase diagrams for interdependent magnetization in (left column) fully-coupled and (right column) partially-coupled ER networks of Ising spins with (bottom row) $M=2$ and (top row) $M=3$ layers. Coloured phases correspond to areas of functional (ferro-para) metastability. 
\textbf{(a},\,\textbf{b)}~Melting spinodals obtained via the iPD in Eq.~\eqref{eq:M4} (full curves), noise-injected iPD with variances $a_{M=2}=1.22$ and $a_{M=3}=1.55$ (dashed curves), iPD with global OP couplings (dot-dashed curve), and standard PD for a diluted $(M+1)$-spin model (dotted curve). Thresholds are averaged over $100$ runs with networked populations of size $N=10^4$ and convergence is found after $T=10^3\,\text{--}\,10^4$ iterations. The dot-dashed (double-dot-dashed) line marks the hyperbackbone [hyperloop] percolation threshold $\langle k\rangle_{sp}=2.4554(1)$ [$\langle k\rangle_{cx}=2.7538(0)$] for $M=2$\vspace*{-0.05cm} and $\langle k\rangle_{sp}=3.0891(1)$ [$\langle k\rangle_{cx}=3.9870(8)$] for $M=3$. For $\langle k\rangle\gtrsim 10^2$, the iPD thresholds converge to $\widetilde{T}_{sp,\langle k\rangle\to\infty}(M)$ calculated within the annealed network approximation (see Methods). 
\textbf{b}) (Inset) $3$-spin energy density as in Eq.~\eqref{eq:2} for $M=2$ fully interdependent ER graphs with $\langle k\rangle=6$ and for a random hypergraph with edge density $\gamma=2$. Notice that $\gamma=\langle k\rangle/3$ and $N_{tot}=N$ for hypergraphs while $\gamma=\langle k\rangle^2$ and $N_{tot}=2N$ in interdependent graphs. 
\textbf{(c},\,\textbf{d)}~Results of iPD for $M=2$ (bottom) and $M=3$ (top) ER graphs with $\langle k\rangle=4$ and partial $q\in[0,1]$ couplings; in noise-injected iPD, we adopted the variances $a_{2+q}=1.22 q$ and $a_{3+q}=(1.22+q/3)$. In $2+q$ layers, functional recovery (i.e.\ a para-ferro transitions at finite $T$ occurs only below $q'=3/4$ (star symbol) with tricritical point $q_c\to0^+$, while in $3+q$ layers the loss of functionality is always irreversible. 
\textbf{d}) (Inset) Critical magnetizations $\mathcal{M}(T_{sp})$ versus $q$ for global and noise-injected iPDs. 
\textbf{c}) (Inset) Difference between the melting spinodals of interdependent spin networks and of ferromagnetic $(M+q)$-spin models on random hypergraphs at finite connectivities (see $\S$S1.2, SM). Notice the inversion taking place around $\langle k\rangle=10$. 
\underline{\emph{Simulations}}. Circles display the results (averaged over $500$ runs) of simulations in the thermal scheme (Fig.~\ref{fig:2}\textbf{a}) on ER graphs of size $N=10^5$ after $MCSs=10^5\times N$ Monte Carlo steps and $NOIs=10^2$ dependency iterations, while triangles depict the thresholds in the higher-order scheme (see Fig.~\ref{fig:2}\textbf{b}) with $N_{tot}=2N$ and $MCSs=10^5\times N\times NOIs$.}}
        \label{fig:3}\vspace*{-0.4cm}
\end{figure*} 
The resulting analytical behaviors, depicted in Fig.~\ref{fig:3}\textbf{a},\textbf{b} with $i)$~dot-dashed and $ii)$~dashed curves, show excellent agreement with numerical simulations (open symbols) and corroborate in two independent ways the thermal fluctuations' scenario.

\underline{\emph{Functional revival}}.
Another aspect of practical relevance emerges when attempting the revival of failed states~{\cite{sanhedrai2022reviving}}. 
Large-scale collapses turn out to be {\em irreversible} in fully-interdependent networks, with systems staying unmagnetized even when quenched to near-zero temperatures. 
The hypergraph equivalence comes again in hand, associating the lack of spontaneous recovery to a {\em supercooling} process, during which the system's evolution becomes insensitive to the functional GS. 
However, if hypergraphs undergo a glass transition~\cite{ricci2001simplest, franz2001ferromagnet} at temperatures accessible by simulated annealing (Fig.~\ref{fig:3}\textbf{b}, inset), interdependent spin networks do not, due to their high hyperedge density, $\gamma=\langle k\rangle^{M}$ (see Methods).\\
\indent
The case of partial interdependence is summarized in Fig.~\ref{fig:3}\textbf{c},\textbf{d} for $\langle k\rangle=4$. 
We find that the melting spinodals (full curve) follow here only qualitatively the RS-ones of diluted $(2+q)$- and $(3+q)$-spin models for $q\in(0,1)$~{\cite{leone2001phase}}, with small deviations (Fig.~\ref{fig:3}\textbf{c}, inset) whose sign reverses as $\langle k\rangle$ increases. 
Alike the $q=1$ case, local thermal fluctuations erode the functional metastability---as we verify in Fig.~\ref{fig:3}\textbf{c},\textbf{d} via iPD endowed with synthetic noise (dashed curve) and with global interdependence (dot-dashed curve)---and ``supercooled'' failed phases set in for any $M>2$ and any fraction $q$ of cross-couplings. 
Spontaneous recovery takes place only in $M=2$ weakly coupled graphs. 
Numerical thresholds reported in Fig.~\ref{fig:3}\textbf{c},\textbf{d} by crosses (noise-injected iPD) and points (global iPD), reveal that para-ferro transitions occur at finite $T$ only below $q'=1-1/\langle k\rangle$ (Fig.~\ref{fig:3}\textbf{d}, green star). 
In fact, the whole line $(q_{tc}, T_{tc}/\langle k\rangle)$ of para-ferro transitions can be characterized analytically as $\langle k\rangle(1-q_{tc})\mathrm{tanh}(\beta_{tc})=1$, identifying the transcritical bifurcations governed by the fraction $(1-q)$ of independent/pairwise-coupled spins in both the interdependent and the $(2+q)$-spin models (see $\S$S1.2, SM). 
This equivalence further enables ($\S$S2, SM) to locate the tricritical point, $q_{c}$, in the $(q,T/\langle k\rangle)$ phase plane. 
We find that $q_c\to0^+$, i.e.\ interdependent spin networks undergo first-order transitions for {\em any} finite fraction of cross-layers couplings, as verified via iPD (Fig.~\ref{fig:3}\textbf{d}, inset). 
\begin{figure*}
	\resizebox{.93\linewidth}{!}{
	\centering
	\begin{tikzpicture}[      
	every node/.style={anchor=south west,inner sep=0pt},
	x=1mm, y=1mm,]   
	%\node (fig0) at (17.5,-5.5)
	%{\includegraphics[width=0.10\linewidth]{Figures/factor_graph.pdf}};
	\node (fig0) at (0, -5.5)
	{\includegraphics[width=0.25\linewidth]{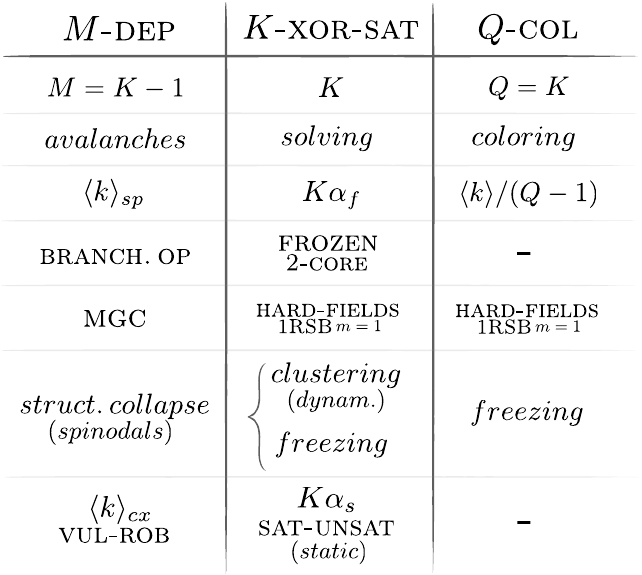}};
	\node (fig1) at (0,-51.5)
	{\includegraphics[width=0.35\linewidth]{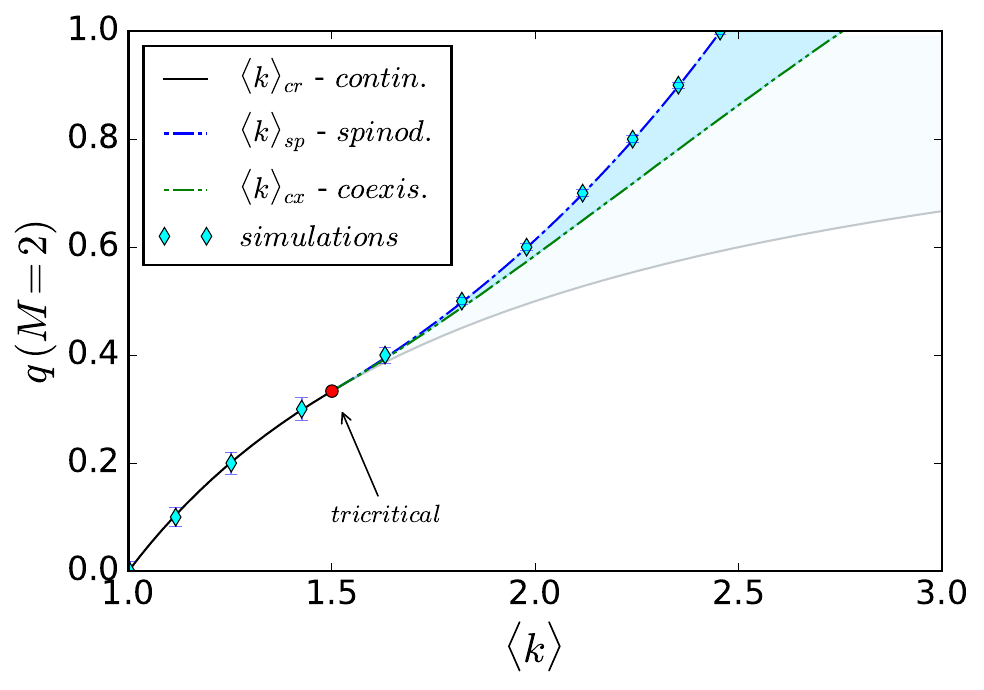}};
	\node (fig2) at (31.5,-43.5)
	{\includegraphics[scale=0.24]{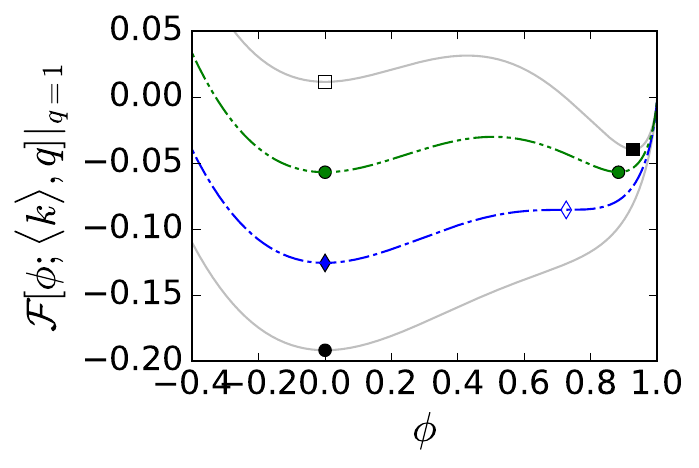}};
	\node (fig3) at (45,-9.5)
	{\includegraphics[width=0.35\linewidth]{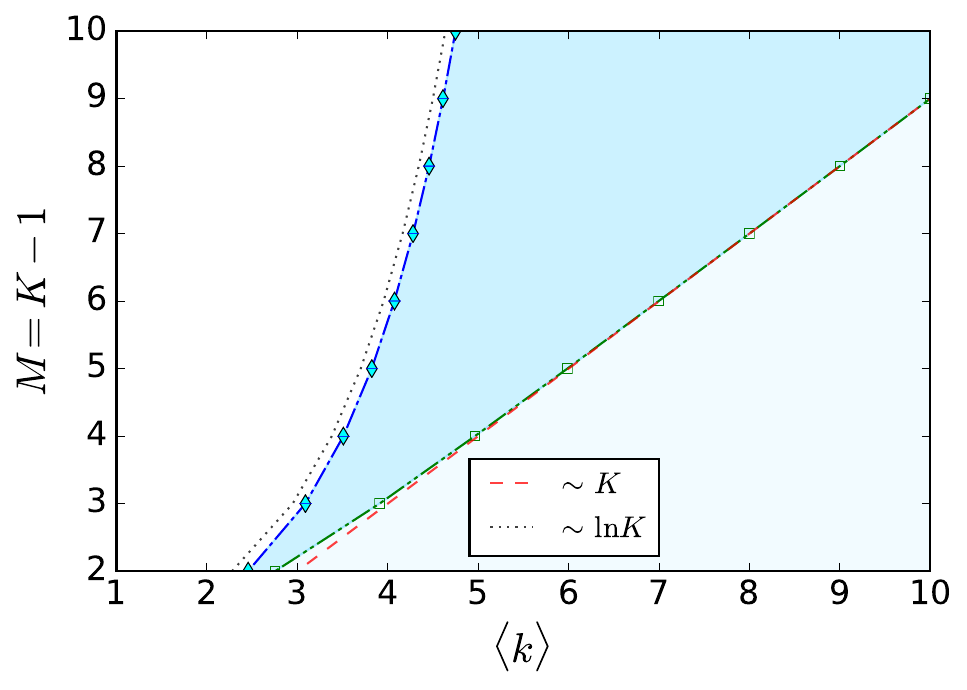}};
	\node (fig4) at (54,+17)
	{\includegraphics[width=0.082\linewidth]{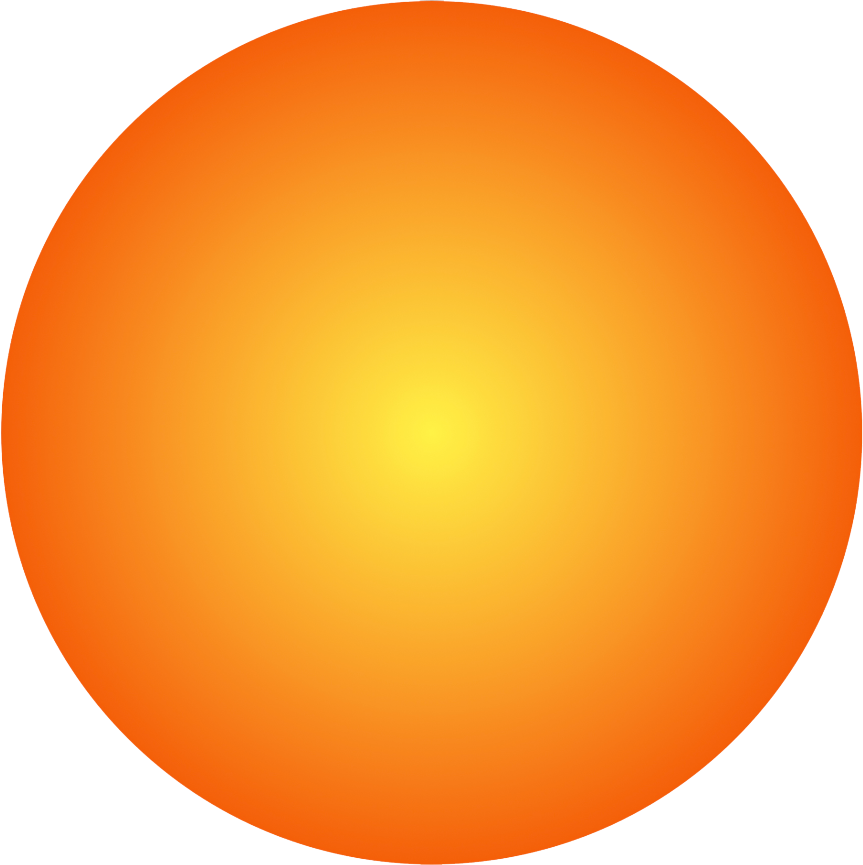}};
	\node (fig5) at (76, 17)
	{\includegraphics[width=0.082\linewidth]{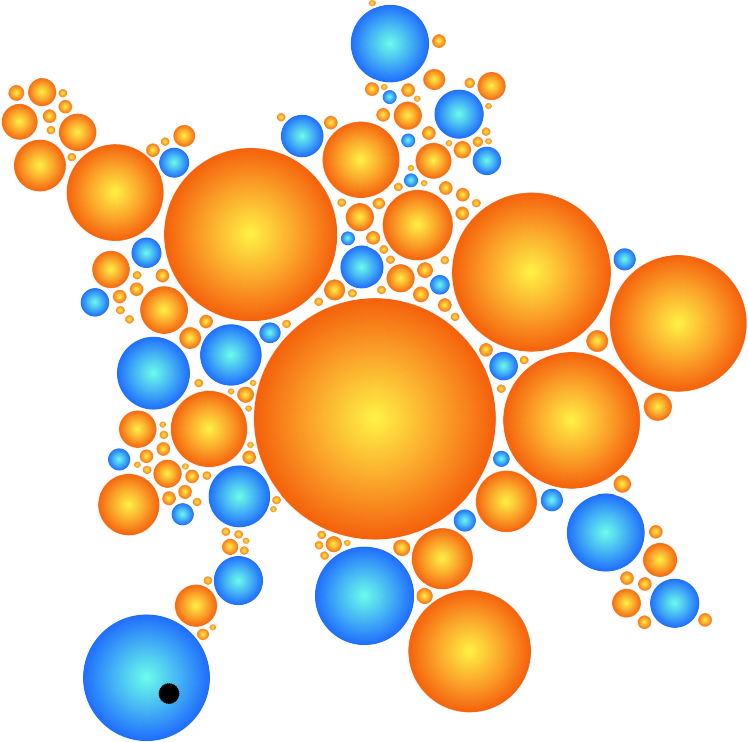}};
	\node (fig6) at (89, -1.5)
	{\includegraphics[width=0.082\linewidth]{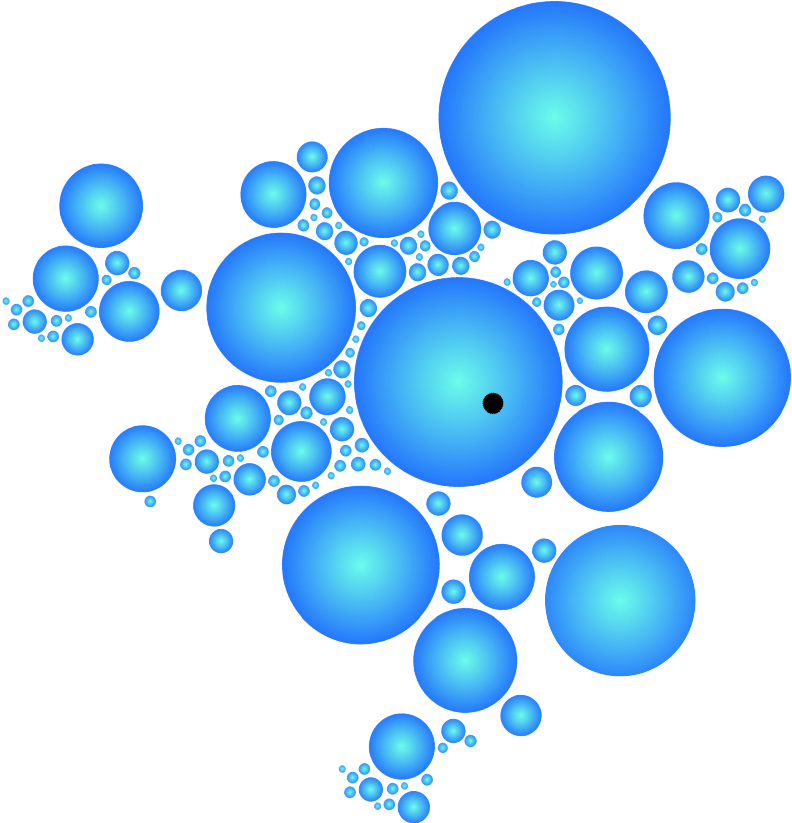}};
	\node (fig7) at (90, 14)
	{\includegraphics[width=0.08\linewidth]{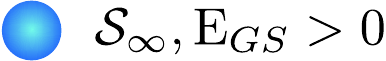}};
	\node (fig8) at (54, 14)
	{\includegraphics[width=0.08\linewidth]{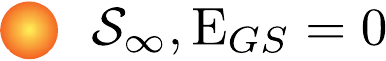}};
	\node (fig9) at (110, -50)
	{\includegraphics[width=0.223\linewidth]{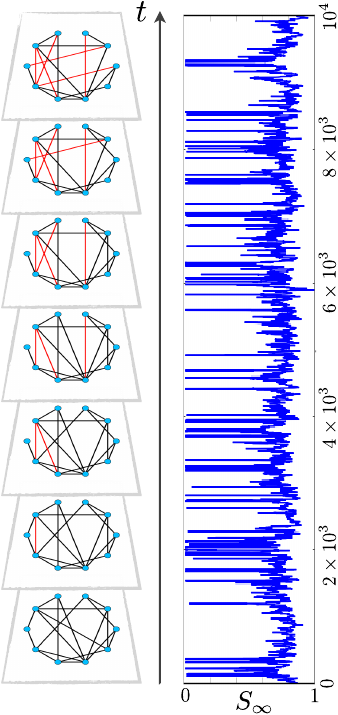}};
	\node (fig10) at (62.5, -49.5)
	{\includegraphics[width=0.25\linewidth]{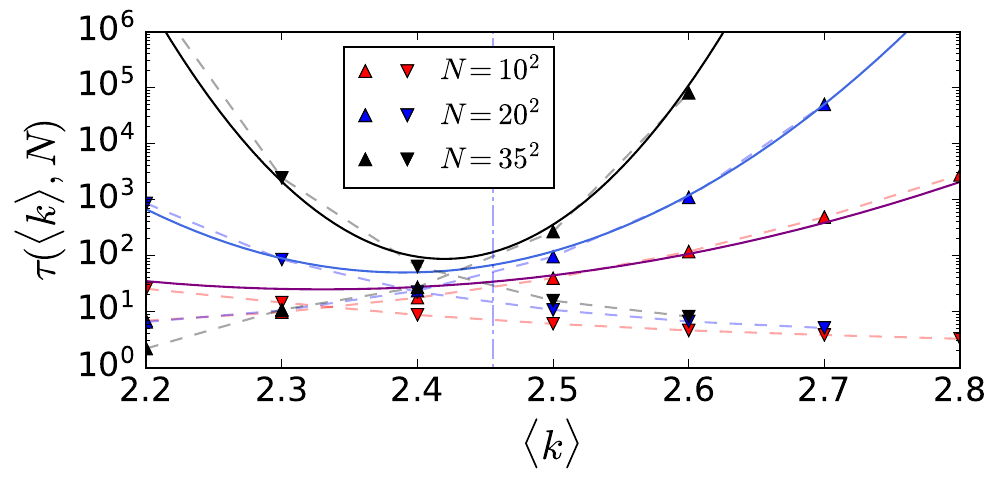}};
	\node (fig11) at (62.5, -29.5)
	{\includegraphics[width=0.255\linewidth]{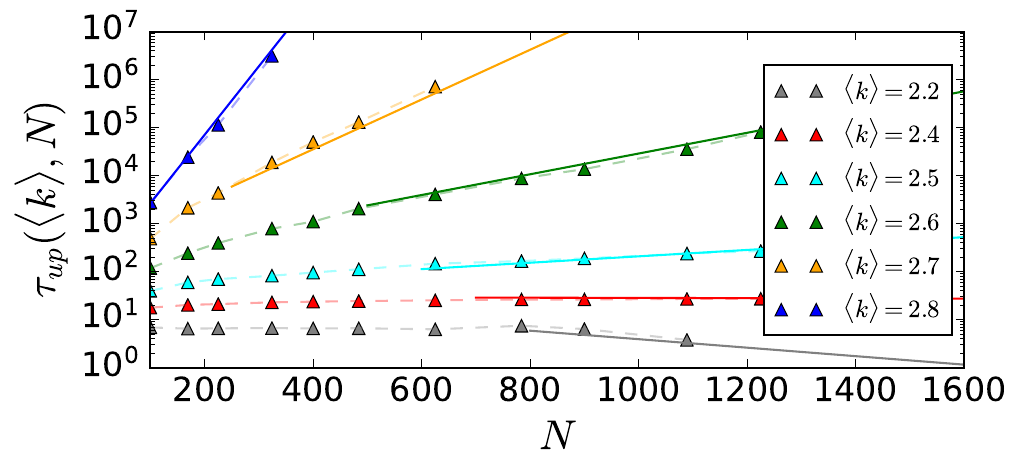}};
	\node at (56, -14) {\footnotesize{\textbf{a})}};
	\node at (101, 29) {\footnotesize{\textbf{b})}};
	\node at (133, 29) {\footnotesize{\textbf{c})}};
	\node at (92.5, -14) {\footnotesize{\textbf{d})}};
	\node at (102, -33) {\footnotesize{\textbf{e})}};
	\end{tikzpicture}}\vspace*{-0.3cm}
\caption{\small{\textbf{Structural metastability of interdependent networks.} (Color online) Colored phases portrait areas of metastability. \textbf{a})~Phase diagram of $M=2$ partially-coupled ER networks displaying the analytical spinodals (dot-dashed curve) and the \textsc{vul-rob} coexistence (double-dot-dashed curve). Notice the finite tricritical point $(q_c,\langle k\rangle_c)=(3/2,1/3)$ and the formal line of transcritical bifurcations (full grey curve). (Inset) RS-free energy density, Eq.~\eqref{eq:6}, for $q=1$ and $\langle k\rangle=2.17,2.46,2.75,3.05$ (from bottom to top), indicating stable (full symbols) and marginally stable (empty symbols) states. 
\textbf{b})~Extension of the above to $M\geq2$ fully-coupled ER graphs [values of $\langle k\rangle_{cx}(M)$ are obtained from Eq.~\eqref{eq:M7} in Methods]; notice the linear growth of the \textsc{vul-rob} thresholds (dashed curve) versus the logarithmic one of the spinodals (dotted curve). (Inset) Sketch of the configurational space of cascading failures (see text for its description);
notice the planted \textsc{unsat} configuration (black dot). 
\textbf{c)}~Evolution of the \textsc{mgc}, $\mathcal{S}_\infty$, for $M=2$ small ($N=10^2$) fully-coupled temporal ER graphs at $\langle k\rangle=2.6$ undergoing a random rewiring of single links at each time step; single runs are obtained by regenerating the cascading process. 
\textbf{d)}~Finite-size scaling of the \textsc{mgc}'s average lifetime $\tau$ in the alive (connected) phase for varying $\langle k\rangle$. Critical slowing down is marked by the sub-exponential scaling at $\langle k\rangle=2.4\,\text{--}\,2.5$; full lines correspond to the first-passage time estimate in the text. 
\textbf{e)} Scaling of $\tau$ versus $\langle k\rangle$ for given $N$; as predicted by the first-passage time estimate, an exponential inversion between the alive and collapsed phases occurs precisely at the structural spinodal $\langle k\rangle_{sp}=2.455(1)$ (dot-dashed line). 
\underline{\emph{Table}}.~Connections among interdependent percolation on $M$ randomly coupled ER networks ($M$-\textsc{dep}), random \textsc{K}-\textsc{xor-sat}~\cite{semerjian2008freezing} and random $Q$-\textsc{col}oring~\cite{zdeborova2007phase}. 
\underline{\emph{Simulations}}.~Symbols in $\textbf{(a,b)}$ show the thresholds (averaged over $10^3$ runs) of simulated cascading failures on ER graphs with $N=10^5\,\text{--}\,10^6$.}}\vspace*{-0.3cm}
        \label{fig:4}
\end{figure*} 
%Similarly to interdependent percolation in randomly coupled $2D$-lattices~\cite{bashan-naturephysics2013}, this acute fragility can be explained in light of the Ising mean-field universality class in ER graphs, characterized by an exponent $\beta=1/2 < 1$.\\
\indent 
\underline{\emph{Interdependence meets} \textsc{xor-sat}}. 
Figs.~\ref{fig:3}\textbf{a},\textbf{b} show that the RS-GS of fully interdependent spin networks at $T=0$ coincides numerically with the critical connectivity, $\langle k\rangle_{sp}$, of cascading failures in their percolation analogue~\cite{buldyrev-nature2010}. 
At the very same threshold, diluted $K$-spin models with $K\equiv M+1$ (dotted curves, Fig.~\ref{fig:3}\textbf{a},\textbf{b})---and, more generally, $(M+q)$-spin models~\cite{leone2001phase}---undergo a dynamic transition marking the percolation of the $2$-core in the underlying hypergraph~\cite{mezard2003two}. 
We find that this surprising correspondence stems, in fact, from an analytical mapping between the models. 
At $T=0$, the RS-magnetization $m$ (i.e., the hyperbackbone) of the $K$-spin model equals the largest root of $m=1-\mathrm{exp}\{-\langle k\rangle m^{K-1}\}$, which is precisely the branching order parameter (OP)---called $1-f$ in Ref.~\cite{buldyrev-nature2010}---of interdependent percolation in $M$ fully-coupled ER graphs. 
An analogous mapping is found also in the case of partial couplings (see Methods). 
Thanks to this connection, the RS-GS entropy of multi-spin models becomes a proxy for the {\em free energy}, $\mathcal{F}$, of interdependent networks, enabling to delimit their structural metastability. 
E.g., for two partially coupled ER graphs 
\begin{equation}\label{eq:6}
-\mathcal{F}[\phi;\langle k\rangle, q]/\ln2=\phi-\tfrac{1}{2}\langle k\rangle(1+q)\phi^2+\tfrac{2}{3}\langle k\rangle q\phi^3,
\end{equation}
\noindent 
s.t.\ $\phi=\mathrm{exp}\{-\langle k\rangle(1-\phi)(1-q\phi)\}$~\cite{leone2001phase}; a general formula for $M$ layers is given in the Methods. 
The landscape of $\mathcal{F}$ (Fig.~\ref{fig:4}\textbf{a}, inset) unfolds the coexistence threshold, $\langle k\rangle_{cx}$ (Figs.~\ref{fig:4}\textbf{a},\textbf{b}), where hyperloops (frustration) percolate in hypergraphs~{\cite{mezard2003two}} and interdependent networks transition from vulnerable (\textsc{vul}) to robust (\textsc{rob}) structures. \\
\indent
A relevant implication of the above is that, in randomly interdependent ER graphs, the metastability width grows {\em linearly} with the number of layers. 
One finds~\cite{{krzakala2007gibbs, zdeborova2007phase, semerjian2008freezing}} that $\langle k\rangle_{sp}=\ln K + \mathcal{O}(\ln\ln K)$ (Fig.~\ref{fig:4}\textbf{b}, dotted curve) while $\langle k\rangle_{cx}\sim K$ (Fig.~\ref{fig:4}\textbf{b}, dashed curve) for $K\gg1$, unveiling that typical interdependent ER graphs can almost surely be dismantled over much broader regimes of connectivity than those found in previous studies~\cite{bianconi2018multilayer}! \\
\indent
The comparison against simulated cascading failures raises, nevertheless, some intriguing puzzles. 
Typical runs on large ER graphs (Fig.~\ref{fig:4}\textbf{a},\textbf{b}, filled symbols) appear completely insensitive to the paramagnetic GS, remaining mutually connected even when quenched much below the \textsc{vul-rob} threshold. 
Hallmarks of criticality appear only in the vicinity of the spinodals $\langle k\rangle_{sp}$, where local cascades yield extensive avalanches. 
This is nicely confirmed by simulations on two fully-coupled temporal ER networks of increasing size, $N$, whose structure evolves by randomly rewiring single links at each time step (Fig.~\ref{fig:4}\textbf{c}). 
In this case, the average lifetime $\tau_{\mathrm{up}}$ of the system in the alive ($\mathcal{S}_{\infty}>0$) phase---defined by the relative size, $\mathcal{S}_\infty$, of the mutual giant component (\textsc{mgc})---marks the onset of criticality at $\langle k\rangle_{sp}=2.4554(1)$, displaying critical slowing down therein (Fig.~\ref{fig:4}\textbf{d}) and an exponential growth (decay) with $N$ above (below) it. 
Similar results arise from the scaling of $\tau$ with $\langle k\rangle$ at fixed $N$ (Fig.~\ref{fig:4}\textbf{e}), further supported by the first-passage\vspace*{-0.05cm} time estimate $\tau \sim \sqrt{N}\text{exp}\{\kappa N \big(\langle k\rangle-\langle k\rangle_{sp}(N)\big)^2\}$. 
Why then interdependent networks do not fragment below $\langle k\rangle_{cx}$? 
Most curiously, why cascading failures reach the spinodals, seeing that these singularities are notoriously inaccessible due to the instability of systems deep quenched in their metastable regime~\cite{binder1976clusters, herrmann1982spinodals, klein2007structure}?\\
\indent 
Both puzzles can be resolved by an isomorphism between interdependent percolation and \textsc{xor-sat}, a prominent constraint satisfaction problem (CSP)~\cite{mezard2009information}. 
%Owing to a notable mapping~\cite{kirkpatrick1983optimization, monasson1992relation, monasson1999determining, mezard2002analytic}, 
Solutions to random $K$-\textsc{xor-sat} (see Methods) are in one-to-one correspondence with the $T=0$ GSs of ferro-diluted $K$-spin models and, as such, they exist for any density, $\alpha$, of combinatorial constraints (clauses) smaller then the \textsc{sat-unsat} threshold $\alpha_s\equiv\langle k\rangle_{\textrm{cx}}/K$, where hyperloops percolate~\cite{franz2001exact, leone2001phase, mezard2003two}. 
The computational cost of finding solutions, however, undergoes an abrupt easy-hard transition at the so-called {\em freezing} threshold $\alpha_f<\alpha_s$~\cite{krzakala2007gibbs, zdeborova2007phase, semerjian2008freezing}, marking the onset of complexity in local algorithms. 
The OP, $\psi$, of the $K$-\textsc{xor-sat} freezing transition---called $1-t$ in Ref.~\cite{mezard2003two}---measures the fraction of hard cavity fields in the 1RSB solution with Parisi parameter $m=1$~\cite{semerjian2008freezing} and satisfies the self-consistent equation 
\begin{equation}\label{eq:7}
\psi=\big(1-e^{-K\alpha\psi}\big)^{K-1},
\end{equation}
\noindent 
which also defines the \textsc{mgc} of interdependent percolation in $K-1$ fully-coupled ER graphs with $\langle k\rangle\equiv K\alpha$~\cite{gao-naturephysics2012}. 
We stress that Eq.~\eqref{eq:7} governs also the freezing transition in random $Q$-\textsc{col}oring~\cite{krzakala2009hiding} (see Table in Fig.~\ref{fig:4}).\\
\indent 
This bijection makes $\langle k\rangle_{sp}$ the threshold locating the spontaneous (ergodicity) breaking of GSs into an exponential number of clusters (Fig.~\ref{fig:4}\textbf{b}, orange disks), accompanied by the proliferation of metastable states with positive energy (Fig.~\ref{fig:4}\textbf{b}, blue disks). 
%In this reasoning, interdependent percolation could be read as an unconventional whitening process~\cite{parisi2002local, krzakala2009hiding} for random satisfiability problems where $K-1$ planted configurations, sampled from the \textsc{unsat} phase (i.e.\ connected ER graphs) of random $2$-\textsc{xor-sat} (Fig.~\ref{fig:4}\textbf{b}, black dots), are adaptively modified by cascading failures in order to solve a random (directed\footnote{Directions are irrelevant in symmetric interdependent ER graphs: $A\to B$ or $B\to A$ cascades are statistically equivalent.}) $K$-\textsc{xor-sat} problem. In this, interdependent links constrain the boolean variables (functioning states) of different $2$-\textsc{xor-sar} ``layers'' by setting a logical \textsc{and} operation between them.
A connection arises: interdependent networks collapse whenever cascading failures find a \textsc{xor-sat} solution. 
Percolation cascades attempt at solving random %(directed\footnote{Directions are irrelevant in symmetric interdependent ER graphs: $A\to B$ or $B\to A$ cascades are statistically equivalent.}) 
$K$-\textsc{xor-sat} by adaptively switching the boolean variables of $K-1$ planted configurations (Fig.~\ref{fig:4}\textbf{b}, black dot), sampled at random from the \textsc{unsat} phase of a random $2$-\textsc{xor-sat} (i.e.\ connected ER graphs) with the same density of constraints. 
In this portrait, hence, interdependent links constrain the functioning states of different $2$-\textsc{xor-sat} ``layers'' by setting a logical \textsc{and} between them. \\
\indent 
We arrive to an answer for both the above conundra.
Below $\langle k\rangle_{sp}$, planted \textsc{unsat} configurations are {\em quiet}~\cite{krzakala2009hiding}, since metastable states are absent and the failed (liquid) phase remains a stable fixed point of the cascading process.  
Above $\langle k\rangle_{sp}$, instead, planting irreversibly biases the avalanches towards one of the many metastable (functional) states. 
Clusters of \textsc{xor-sat} solutions remain, hence, always hidden from percolation cascades so that giant avalanches never occur below the coexistence connectivity $\langle k\rangle_{cx}$. 
The structural spinodal $\langle k\rangle_{sp}$ can then be reached, emerging as the only critical singularity of the process. 
This is reminiscent of quiet planting in random graph coloring~\cite{krzakala2009hiding} and suggests that interdependent percolation can be adopted as an unconventional whitening protocol~\cite{parisi2002local} for random satisfiability problems.

\underline{\emph{Discussion}}. 
Connections between disciplines are powerful sources of new syntheses.  
Our physical realization of interdependence has unveiled profound connections among multilayer networks, spin glasses and CSPs, offering concrete chances to bridge deep theoretical ideas with areas of both empirical and practical depth. 
For instance, mapping cascading failures to \textsc{xor-sat} solutions bonds the design of efficient protocols for controlling the functioning of interdependent systems---e.g.\ to avoid systemic risks in ecosystems~\cite{levin2019architecture}, minimizing disruption in urban infrastructures~\cite{pregnolato2016assessing} or influencing multi-scale systems like the brain~\cite{fornito2015connectomics} or the cell~\cite{klosik2017interdependent}---with powerful SAT-solvers like backtracking~\cite{marino2016backtracking} or conflict-driven clause learning~\cite{marques2021conflict}. 
%Their applicability, in this respect, would be broad and interdisciplinary, providing lateral viewpoints for e.g.\ controlling systemic risks and collapses in real-world ecosystems~\cite{levin2019architecture}, solving AI or cryptography problems~\cite{sun2020neurogift} and more~\cite{ercsey2011optimization}. 
Along these lines, the gained knowledge about the free energy of interdependent networks (Eq.~\eqref{eq:6} and Eqs.~\eqref{eq:M6},\eqref{eq:M7} in the Methods) enables to exactly bound the structural metastability regime of these systems, disclosing a range of vulnerability dramatically broader (Fig.~\ref{fig:4}\textbf{a},\textbf{b}) then what known from previous studies. 
This framework can be readily exported to models of interdependent ecologies~\cite{levin2019architecture}, urban traffic~\cite{zeng2020multiple} or climate changes~\cite{thornton2014climate}, which lack a thermodynamic formulation, to develop free energy analogues to exactly delimit their metastable functioning, rooting this primary task on more solid grounds compared to current approaches based on early warning indicators~\cite{scheffer2012anticipating} of regime shifts and other {\em ad hoc} techniques~\cite{arani2021exit}.
Let us also stress that the surprising isomorphism of Eq.~\eqref{eq:7} with the \textsc{mgc} of percolation in randomly interdependent networks brings unorthodox testbeds of both theoretical and practical nature where to study and further develop the universal features~\cite{coniglio2017universal} of ergodicity breaking mechanisms at the RSB (e.g.\ hard-field distributions, point-to-set correlations, {\em etc}). 
The above perspectives become even brighter if we consider that the {\em thermal} interpretation of interdependence raises the unprecedented opportunity of its experimental realization in e.g.\ thermally-coupled metallic arrays where local Joule heating can embody the formation of dependency links. 
%In granular ferromagnets~\cite{strelniker2004percolation}, e.g., electrons hopping from grain to grain (i.e.\ Ising spins) experience strong scattering (high resistance) if the neighboring grains are magnetically disordered (i.e.\ $\Sigma_i\simeq0$) producing, as in Eq.~\eqref{eq:1}, a local increase in the temperature of the superposed grain that can promote its probability to become disordered w.r.t.\ its own neighbors. 
In this respect, extending the present analysis to other (e.g.\ spatially-embedded) topologies is certainly desirable and we expect that mappings similar to those we detailed here in random structures will emerge, fostering the potential of more cross-disciplinary connections. 
%Finally, seeing that interdependent links are a network representation of the feedback mechanisms observed in ecosystems and infrastructures, our work indicates that higher-order models~\cite{battiston2020networks, battiston2021physics} are tailored frameworks for studying the cascading shifts~\cite{guillen2015architecture,rocha2018cascading} occurring in many real-world social-ecological systems. 

\FloatBarrier
\bibliographystyle{unsrt}
\bibliography{interSAT.bib}

\begin{thebibliography}{10}
\bibitem{note2}
A {\em feedback loop} condition is understood, i.e.\ if $i$ in {A} is
  interdependent with $j$ in {B}, and $j$ in {B} is interdependent with $k$ in
  {C}, then $k$ and $i$ are also interdependent.
\bibitem{note1}
Negative values of the effective fields induce, additionally, an
  antiferromagnetic coupling (dashed lines in {Fig}.~\ref{fig:1}, left box);
  these configurations are, nevertheless, transient and energetically
  disfavored.
\end{thebibliography}

\section*{Methods}\vspace*{-0.25cm}
\underline{\textbf{Partial interdependence}.} For partially interdependent Ising spin networks, Eq.~\eqref{eq:2} can be generalized to\vspace*{-0.2cm}
\begin{equation}\label{eq:M1}
-\mathcal{H}_\mathcal{I}=\sum_{i,j}\mathcal{A}_{ij}\sigma_i\sigma_j\bigg(\delta_{\sum_\ell\mathcal{I}_{i\ell},0}+\prod_\ell \mathcal{I}_{i\ell}\Sigma_\ell\bigg),\vspace*{-0.2cm}
\end{equation}
\noindent 
with $J\equiv1$ for simplicity. 
In Eq.~\eqref{eq:M1}, the Kronecker delta sets pairwise links incoming in {\em independent} spins--i.e., indices $i=1,\dots,MN$ satisfying the zero row sum condition $\sum_{\ell}\mathcal{I}_{i\ell}=0$--from their nearest neighbors. 

For $M=2$ networks with a fraction $q\in[0,1]$ of one-to-one interdependent spins (e.g.\ the ER graphs studied in Fig.~\ref{fig:3}\textbf{d}), Eq.~\eqref{eq:M1} simplifies to
\begin{equation}\label{eq:M2}
-\mathcal{H}_{q,\mathcal{I}}=\sum_{\mu=\mathrm{A},\mathrm{B}}\sum_{i,j}\,\!\!'A^{\mu}_{ij}\sigma^\mu_{i}\sigma^\mu_{j}+J\sum_{i,j,k}\mathpzc{W}_{ijk}\sigma_i\sigma_j\sigma_k\vspace*{-0.2cm}
\end{equation}
\noindent 
where $\sum'$ runs over the indices $i=1,\dots,N$ of independent sites and the tensor $\mathpzc{W}_{ijk}$ is given by Eq.~\eqref{eq:3}. 
Notice that Eq.~\eqref{eq:M1} is, thus, a mixture of directed $2$-spin and $3$-spin interactions (as illustrated in Fig.~\ref{fig:1}). 

For increasing number of layers, the degrees of freedom of the cross-layers couplings grow. 
E.g., for $M=3$ networks, partial couplings can tune the system from $3$ isolated layers to a mixture, like Eq.~\eqref{eq:M2}, of directed $2$-spin and $4$-spin couplings. 
Another option is a mixture of directed $3$- and $4$-spin interactions. 
In this case (i.e.\ results of Fig.~\ref{fig:3}\textbf{c}), one can consider a fraction $q$ of {\em directed} interdependent links and a fraction $1-q$ of one-to-one interdependent links (left diagram below). \vspace*{-0.25cm} 

\begin{figure}[h]
\begin{tikzcd}[row sep=0.75cm, column sep = 0.3cm, arrow style=tikz, arrows=semithick]
{} & \mathcal{A} \arrow{dr} \arrow[dl, shift left=0.25ex, blue] \\
\mathcal{C} \arrow{ur} \arrow[-]{rr}{\textcolor{blue}{q}} \arrow[rr, shift left=0.25ex, blue]{q} && \mathcal{B} \arrow{ll}{1-q} \arrow[ul, shift left = 0.25ex, blue]
\end{tikzcd}\,\hspace*{+0.2cm}\,
\begin{tikzcd}[row sep=0.75cm, column sep = 0.4cm, arrow style=tikz, arrows=semithick]
\mathcal{A} \arrow[<->]{drr} \arrow{rr}{1-q} \arrow[d, shift left=0.25ex, blue] && \mathcal{B} \arrow{d} \arrow[ll, shift left=0.25ex, blue]  \\
\mathcal{D} \arrow[<->]{urr} \arrow{u} \arrow[rr, shift left=0.25ex, blue] && \mathcal{C} \arrow{ll}{\textcolor{blue}{q}} \arrow[u, shift left=0.25ex, blue]
\end{tikzcd} \vspace*{-0.25cm}
\end{figure}
\noindent
Similar ideas can be applied at larger values of $M$ e.g.\ to create mixtures of directed $4$- and $5$-spin interactions by coupling $M=4$ layers (right diagram above).\vspace*{+0.25cm}

\underline{\textbf{Interdependent population dynamics}.} 
RS calculations (see $\S$S1.1, SM) for isolated Ising ER graphs with average degree $\langle k\rangle$ yield the distributional equation 
\begin{equation}\label{eq:M2}
P(h)=e^{-\langle k\rangle}\sum_{k\in{\rm I\!N}_0}\frac{\langle k\rangle^k}{k!}\int \dd\boldsymbol{\mu}\,\delta\big( h-\mathpzc{T}_k(\beta; \textbf{h})\big),
\end{equation}
\noindent 
with integral mesure $\dd\boldsymbol{\mu}\equiv\prod_{i=1}^k P(h_i)\dd h_i$ and kernel 
\begin{equation}\label{eq:M3}
\mathpzc{T}_k(\beta; \mathbf{h})\equiv\frac{1}{2\beta}\sum_{i=1}^k\ln\left(\frac{f_+(\beta; h_i)}{f_-(\beta; h_i)}\right),
\end{equation}
\noindent 
where $f_{\pm}(\beta; \mathpzc{x})=\mathrm{cosh}(\beta(\mathpzc{x}\pm1))$. 
Eq.~\eqref{eq:M2} can be solved self-consistently for the distribution $P(h)$ of cavity fields $\{h_i\}_{i=1,\dots,N}$ via population dynamics methods~\cite{hartmann2005phase}. 
For large enough populations (i.e.\ $N\sim10^4\,\text{--}\,10^5$) and number of iterations (i.e.\ $T_{eq}\sim10^3\,\text{--}\,10^4$), this yields also the statistics of the (thermally averaged) local effective fields $\langle\Sigma_i\rangle_{\boldsymbol{\beta}}=\sum_j A_{ij}\langle\sigma_j\rangle_{\boldsymbol{\beta}}/k_i$, since the spins' magnetization is related to their cavities by $\langle \sigma_i\rangle_{\boldsymbol{\beta}}=\mathrm{tanh}(\beta h_i)$. \\
\indent
In light of Eq.~\eqref{eq:1}, we can then solve iteratively the sequence in Eq.~\eqref{eq:4} as follows. 
At each equilibrium stage (see Fig.~\ref{fig:2}\,\textbf{a}) we calculate the RS-cavity fields $\boldsymbol{h}^{\mu}$ of the $\mu$-th network~{\cite{mezard2001bethe, parisi2020theory}} w.r.t.\ the $\boldsymbol{\beta}$-distribution generated by its dependent layers, and then we update the $\boldsymbol{\beta}$'s of the latter ones as in Eq.~\eqref{eq:4}. 
Concretely, for an interdependent pair of spins $\sigma_i$ in $A$ and $\sigma_{i'}$ in $B$, the above can be represented as a recursion of local temperatures
\begin{equation}\label{eq:M4}
\begin{aligned}
\beta_{i', B}^{(n)}  &\,= \beta k_i^{-1}\sum_j \mathcal{A}_{ij}\mathrm{tanh}\left(\beta_{j,A}^{(n-1)}h_{j,A}^{(n-1)}\right),\\
\beta_{i, A}^{(n)} &\,= \beta k_{i'}^{-1}\sum_{j'} \mathcal{A}_{i'j'}\mathrm{tanh}\left( \beta_{j',B}^{(n)} h_{j',B}^{(n)}\right), 
\end{aligned}\vspace*{-0.3cm}
\end{equation}
\noindent 
where $n\in{\rm I\!N}$ and $\beta_{i, A}^{(0)}\equiv \beta$ is the initial seed. 
More generally, for an arbitrary network of fully-interdependent Ising networks, the recursion Eq.~\eqref{eq:M4} reads as
\begin{equation}\label{eq:M5}
\beta_{i_\mu,\mu}^{(n)}=\beta\prod_{\ell_\nu(i_\mu)}\mathcal{I}_{i_\mu\ell_\nu}\left\langle\Sigma_{\ell_\mu}\right\rangle_{\boldsymbol{\beta}_\nu^{(n-1)}}\vspace*{-0.2cm}
\end{equation}
As $n\to\infty$, the $\boldsymbol{\beta}$-distributions converge to configurations of local temperatures at which the layers' average magnetization $\{\mathcal{M}^\mu_\infty\}_{\mu=1,\dots,M}$ are mutually stable. \\
\indent 
Given the above procedures, we have developed the interdependent population dynamics (iPD) algorithm summarized in the following pseudocode. \vspace*{-0.1cm}  
\begin{algorithm}[H]
\caption*{\textsc{InterPopDyn($\beta, M, N,\mathcal{A}, \mathcal{I},T_{eq},NOI$)}}
\begin{algorithmic}[1]
\State Initialize the tensor $\mathpzc{W}_{i_1,\dots,i_M}$ from $\mathcal{A},\mathcal{I}$ as in Eq.~\eqref{eq:3};
\State Randomly initialize $M$ populations $\{\mathbf{h}^\mu\}_{\mu=1,\dots,M}$; 
\For{$n=1,\dots,NOI$:}
	\For{$\mu=1,\dots,M$:}
		\State Compute $\{\beta^{(n)}_{i_\mu,\mu}\}_{i_\mu=1,\dots,N}$ as in Eq.~\eqref{eq:M5};
		\For{$t=1,\dots, T_{eq}$:}
    			\For {$i_\mu=1,\dots, N$:}\vspace*{-0.05cm}
        				\State  Compute $h_{i_\mu}=\mathpzc{T}_{k_{i_\mu}}\big(\beta_{i_\mu,\mu}^{(n)},\{h_{j_\mu}\}_{j_\mu\in\partial i_\mu}\big)$;\vspace*{-0.15cm}
    			\EndFor 
		\EndFor
		\State Return the local fields $\{\langle\Sigma_{i_\mu}\rangle_{\boldsymbol{\beta}_\mu^{(n)}}\}_{i_\mu=1,\dots,N}$; \vspace*{-0.2cm}
	\EndFor
\EndFor\\
\Return{array $\{\mathcal{M}^\mu_\infty\}_{\mu=1,\dots,M}$}
\end{algorithmic}
\end{algorithm}
\noindent 
Since iPD builds on the {\em equilibrium} RS-solution of the diluted $2$-spin model, $\mathcal{O}(MN)$ computational time can be saved by replacing the dependency steps (line 3) with the populations' equilibration (line 6). 
Partially interdependent systems are solved analogously, with the additional condition that the local temperature of independent spins stays $T$, i.e.\ the temperature of the heat bath.\\ 
\indent 
Noisy iPD runs alike the iPD above, with white noise added to the effective fields (line 9), i.e.\ by computing $\langle\Sigma_{i_\mu}\rangle_{\boldsymbol{\beta}}+\xi_{i_\mu}$ with $\xi_{i_\mu}\!\!\in\mathcal{N}(0, a/\langle k\rangle_\mu)$. 
Finally, iPD with global OP interactions evolves analogously to the regular local iPD above, with Eq.~\eqref{eq:M5} being replaced by
$$
\beta^{(n)}_{i_\mu,\mu}=\beta\!\prod_{\ell_\nu(i_\mu)}\!\mathcal{I}_{i_\mu\ell_\nu(i_\mu)}\mathcal{M}_\nu^{(n)}\!,\,\,\,
\mathcal{M}_\nu^{(n)}=\frac{1}{N}\sum_{i_\nu=1}^N\langle \Sigma_{i_\nu}\rangle_{\boldsymbol{\beta}_\nu^{(n-1)}}.
$$

\underline{\textbf{Random} \textsc{xor-sat} \textbf{in a nutshell}}. Random $K$-\textsc{xor-sat}~\cite{mezard2009information} searches for assignments to a Boolean vector $\mathbf{x}\in\{0,1\}^N$ satisfying $\mathcal{C}\mathbf{x}=\mathbf{b}(\!\!\!\!\mod2)$, where $\mathbf{b}\in\{0,1\}^L$ is randomly given and $\mathcal{C}$ is an $L\times N$ matrix underlying the structure of the $L=\alpha N$ randomly chosen $K$-tuple logical constraints. 
For $K=3$, e.g., this corresponds to finding solutions to $F=\bigwedge_{\{i,j,k\}\in \mathcal{E}_3}x_i\oplus x_j\oplus x_k$, where\vspace*{-0.05cm} $\oplus$ is the \textsc{xor} operation and the \textsc{and} sum, $\wedge$, runs over all the triples $\{i,j,k\}$ in the edge list $\mathcal{E}_3$ of a random hypergraph. 
Through the mapping $\sigma_i=(-1)^{x_i}$~\cite{ricci2001simplest}, $K$-\textsc{xor-sat} can be studied as the $T=0$ limit of the diluted $K$-spin model $\mathcal{H}=\sum_{\{i_1,\dots,i_K\}\in\mathcal{E}_K}\big(1-\sigma_{i_1}\cdots\sigma_{i_K}\big)$.\vspace*{-0.05cm} 
Similarly, random $(M+q)$-\textsc{xor-sat} with $M=K-1$, i.e.\ a mixture of $qL$ $K$-tuples and $(1-q)L$ $M$-tuples, can be solved by studying the $T=0$ limit of a diluted $(M+q)$-spin model. 

Within the RS-ansatz (see SM for details), the complexity of random $(M+q)$-\textsc{xor-sat} can be fully characterized via the zero-temperature free-energy density of a diluted $(M+q)$-spin Hamiltonian, whose expression is 
\begin{equation}\label{eq:M6}
\begin{aligned}
-\mathcal{F}/\ln2&\,=\phi(1-\ln\phi)+\\ 
&\,-\tfrac{\langle k\rangle}{M+1}q\big[1-(1-\phi)^{M+1}\big]+\\
&\,-\tfrac{\langle k\rangle}{M}(1-q)\big[1-(1-\phi)^M\big]
\end{aligned}
\end{equation}
\noindent 
where $\langle k\rangle\equiv K\alpha$ is the mean degree of the underlying hypergraph. 
%Notice that Eq.~\eqref{eq:6} is given by Eq.~\eqref{eq:M6} for $M=2$.
Eq.~\eqref{eq:M6} generalizes Eq.~\eqref{eq:6} (and other formulas discussed therein) to an arbitrary number $M$ of ER graphs with the same average degree $\langle k\rangle$.\vspace*{+0.25cm}

\underline{\textbf{Asymptotics of} $M$-\textsc{dep} \textbf{at large $M$}}. For $q=1$, the branching OP for $M$ interdependent ER graphs with average degree $\langle k\rangle$ minimizes Eq.~\eqref{eq:M6} yielding the equation $m=\mathpzc{G}(\langle k\rangle,m)$ with $\mathpzc{G}(x,y)=1-\mathrm{exp}\{-x y^M\}$. 
At the spinodal $\partial_m\mathpzc{G}|_{(\langle k\rangle_{sp},m_{sp})}=1$, leading to the closed formula $1/M=-(1-m_{sp})\ln(1-m_{sp})/m_{sp}$. 
The latter yields an expansion for $m_{sp}(M)$ that, once inserted into $m_{sp}=\mathpzc{G}(\langle k\rangle_{sp},m_{sp})$ gives~\cite{semerjian2008freezing} $\langle k\rangle_{sp}=\ln K+\ln\ln K+1+\mathcal{O}(\ln\ln K /\ln K)$ with $K=M-1$. \\
\indent
The asymptotics of the \textsc{vul-rob} threshold, $\langle k\rangle_{cx}(M)$, can be found instead by studying the zeroes of the configurational entropy (i.e.\ the $K$-\textsc{xor-sat} complexity) 
\begin{equation}\label{eq:M7}
\begin{aligned}
\Sigma/\ln2=&\,m\left[1+\langle k\rangle(1-m)^M\right]+\\
-&\,\tfrac{\langle k\rangle}{M+1}\left[1-(1-m)^{M+1}\right],
\end{aligned}
\end{equation}
\noindent 
with $m=\mathpzc{G}(\langle k\rangle,m)$. 
Since $(1-m)^{M+1}\to0$ and $\phi\to1$ for $M\gg1$, the null condition $\Sigma|_{(\langle k\rangle_{cx}, m_{cx})}=0$ leads to $\langle k\rangle_{cx}\sim M+1$. 
It follows that the metastability width has scaling $\langle k\rangle_{cx}-\langle k\rangle_{sp}\sim M-1$ for $M\gg1$, i.e.\ interdependent networks become less and less structurally stable as the number of interacting layers grows. \vspace*{+0.25cm}

\underline{\textbf{Annealed network approximation}}. 
For dense structures, the mapping revealed by the iPD between $M$ randomly interdependent Ising networks and diluted $K$-spin models (Fig.~\ref{fig:3}\textbf{a},\textbf{b}) can be verified analytically within the annealed network approximation~{\cite{dorogovtsev-revmodphys2008}}. 
Let us recall that, for a typical ER graph $G\in\mathfrak{G}(N,p)$ with adjacency matrix $A=(A_{ij})_{i,j=1,\dots,N}$, the matrix of wiring probabilities has entries $\pi_{ij}:=\langle A_{ij}\rangle_{\mathfrak{G}}\simeq \langle k\rangle/N\equiv p$, where $\langle\,\cdots\rangle_\mathfrak{G}$ is the average over the graph ensemble. 
Similarly, for configurational model networks, one finds $\pi_{ij}\simeq k_i k_j/\langle k\rangle N$. 
The annealed network approximation sums up to averaging over $\mathfrak{G}$\vspace*{-0.05cm} the Hamiltonian, s.t.\ $A_{ij}\xrightarrow{a.a.}\pi_{ij}$. 
To apply the above to Eq.~\eqref{eq:2} we must, therefore, average over all the graph ensembles from where the networks composing the multilayer system are extracted. 
For one-to-one interdependence, this operation is straightforward due to trivial disorder governing the cross-layers pairings. 
In light of Eq.~\eqref{eq:3}, the integration of interdependent interactions within the multilayer's structure yields the wiring tensor
\begin{displaymath}
\begin{aligned}
\pi_{i_1\dots i_K}
=&\,\left\langle\cdots\big\langle\mathpzc{W}_{i_1\dots i_K}\big\rangle_{\mathfrak{G}_K}\cdots\right\rangle_{\mathfrak{G}_{1}}\\
=&\,\big\langle\mathcal{A}_{i_1i_2}\big\rangle_{\mathfrak{G}_1}\prod_{\mu=1}^{M-1}\frac{\big\langle \mathcal{A}_{\ell_\mu(i_1)i_{\mu+2}}\big\rangle_{\mathfrak{G}_\mu}}{k_{\ell_\mu(i_1)}}\\
=&\,\frac{k_{i_1}k_{i_2}}{\langle k\rangle_1 N}\prod_{\mu=1}^{M-1}\frac{k_{i_{}\mu+2}}{\langle k\rangle_\mu N}
\simeq\frac{k_{i_1}\cdots k_{i_{M+1}}}{\langle k\rangle_1\cdots\langle k\rangle_{M}N^{M}},
\end{aligned}
\end{displaymath}
\noindent 
with $K\equiv M+1$, i.e.\ the wiring probability of an hypergraph in $d=M$ dimensions within the generalized configurational model~\cite{ghoshal2009random, courtney2016generalized}. 
For ER graphs with equal average degree $\langle k\rangle$, analogous arguments yield $p\simeq\langle k\rangle/N^M$. 
One can then solve the Hamiltonian, Eq.~\eqref{eq:2}, within the traditional Bragg-Williams approach, leading e.g.\ to the average magnetization $\mathcal{M}=\mathrm{tanh}\big(\beta\langle k\rangle \mathcal{M}^M\big)$ from where the (bulk) melting spinodals shown by arrows in Fig.~\ref{fig:3}\textbf{a},\textbf{b} for $M=2,3$ can be readily calculated. \\
\indent 
Contrarily with the above, partially-coupled networks break down the mapping. 
%With partial cross-layers couplings, the mapping between the melting transitions of interdependent Ising networks and diluted multi-spin models breaks down. 
In the inset to Fig.~\ref{fig:3}\textbf{c} we supported this property via iPD numerics on ER graphs with finite connectivities, revealing an inversion in the difference of the (bulk) melting spinodals of the two models for increasing values of $\langle k\rangle$. 
In the annealed approximation, the differences between the two models become manifest. 
E.g.\ for $M=2$ partially interdependent ER graphs and for a diluted $(2+q)$-spin model, one finds (see SM)
\begin{align}
\mathcal{M}&\,=q\mathrm{tanh}\big(\beta\langle k\rangle \mathcal{M}^2\big)+(1-q)\mathrm{tanh}\big(\beta\langle k\rangle\mathcal{M}\big),\label{eq:M8}\\
\mathcal{M}&\,=\mathrm{tanh}\big(\beta\langle k\rangle\mathcal{M}(1-q(1-\mathcal{M}))\big),\label{eq:M9}
\end{align}
\noindent 
respectively. 
Notice that the mapping holds for any $\langle k\rangle$ and for any $M$ (see Fig.~\ref{fig:3}\textbf{c}, inset) when $q=1$. \\
\indent
The annealed approximation provides only a coarse-grained view of the integrated structure underlying the interdependent Hamiltonian, Eq.~\eqref{eq:2}. 
In fact, it neglects some essential microscopic details, such as the {\em actual} density of the higher-order interactions. 
Because of Eq.~\eqref{eq:1}, the hypergraph integrating $M$ fully-coupled ER graphs with average degrees $\{\langle k\rangle_\mu\}_{\mu=1,\dots,M}$ has, by construction, an average hyperedge density $\gamma_M=\prod_\mu\langle k\rangle_\mu$. 
Besides being directed, these hyperedges are also suitably weighted, s.t.\ a spin $i_\mu$ in the $\mu$-th layer has, on average, $\gamma_M$ incoming $K$-tuples with an average weight $\prod_{\nu\neq\mu}\langle k\rangle_\nu$\vspace*{-0.05cm}. 
Hence $i_\mu$ sees, effectively, an average number $\langle k\rangle_{\mu}$ of incoming $K$-tuples, i.e.\ the same average degree of its own layer. 
While these microscopic details are negligible when heating up the system from its ferromagnetic phase, they become highly relevant when annealing it from its paramagnetic phase at low cooling rates. 
In this case, in fact, interdependent spin networks remain trapped in a supercooled malfunctioning (i.e.\ paramagnetic) phase, without undergoing the spin-glass transition expected in their ferromagnetic multi-spin analogue~\cite{franz2001ferromagnet}. 
This is best shown in the inset of Fig.~\ref{fig:3}\textbf{b} for $M=2$ and $q=1$, where the spin-glass (SG) transition (indeed, its Kauzmann temperature) is marked by an arrow, whereas no similar threshold is observed in the interdependent spin model. \vspace*{+0.25cm}

\underline{\textbf{Code availability}} \\
Source codes can be freely accessed at the GitHub repository: \url{https://github.com/BnayaGross/High_order_interdependent_ising}.

\underline{\textbf{Data availability}} \\
All data supporting our findings are available from the corresponding author upon reasonable request.

\underline{\textbf{Acknowledgments}} \\
\noindent
I.B.\ warmly thanks W.~Klein, H.~Gould, S.~Kirkpatrick, F.~Radicchi and A.~Frydman for fruitful comments. 
S.H.\ acknowledges financial support from the ISF, the China-Israel SF, the ONR, the BIU Center for Research in Applied Cryptography and Cyber Security, the EU project RISE, the NSF-BSF Grant No. 2019740, and the DTRA Grant No. HDTRA-1-19-1-0016. 
%I.B.\ and S.H.\ acknowledge partial financial support from the Italy-Israel grant ``Explics''. 

\underline{\textbf{Author contributions}} \\
\noindent 
I.B.~initiated the work and designed the research, with contributions from B.G.. 
I.B.~developed the algorithms and carried out the numerical and theoretical analyses. 
B.G.~designed the codes and carried out the Monte Carlo simulations. 
S.H.~supervised the research. 
I.B.~wrote the paper with contributions from B.G.\ and S.H.. 
All authors critically reviewed and approved the manuscript.

% --------------------------------------------------------------------------------------------------
% ************* SUPPLEMENTARY MATERIAL STARTS HERE ***************
% --------------------------------------------------------------------------------------------------

\newpage
\,
\newpage
\onecolumngrid

\renewcommand\thesection{S\arabic{section}}
\renewcommand\thesubsection{S\arabic{section}.\arabic{subsection}}
\setcounter{section}{0}
\setcounter{equation}{0}
\setcounter{figure}{0}
\setcounter{page}{1}
\renewcommand{\theequation}{S\arabic{equation}}
\renewcommand{\thefigure}{S\arabic{figure}}
\renewcommand{\thepage}{S\arabic{page}}

\begin{center}
{\large \textsc{\bf SUPPLEMENTARY MATERIAL}}\vspace*{-1.5cm}
\end{center}

\section{Replica-symmetric solutions for random multi-spin models}\label{S1}\vspace*{-0.5cm}
\subsection{Background: ferromagnetic random $2$-spin model}\label{S11}\vspace*{-0.5cm}
Let us start by considering a $2$-spin Ising model on an Erd\H{o}s-R\'enyi (ER) network $G=(V,E)$ randomly drown from the statistical ensemble $\mathfrak{G}(N,\langle k\rangle N/2)$ of random graphs with $N=|V|$ nodes and average degree $\langle k\rangle$. 
The energy of a spin configuration $\boldsymbol{\sigma}\equiv\{\sigma_i\}_{i}\in\mathbb{Z}_2^N$ is given by the Hamiltonian $-\mathcal{H}[\boldsymbol{\sigma},A]=J\sum_{i,j}A_{ij}\sigma_i\sigma_j$, where $J\in\mathbb{R}^+$ is the coupling strength and the structure of pairwise interactions follows the (symmetric) adjacency matrix $A=(A_{ij})_{i,j}$, with entries $A_{ij}=+1$ if $(i,j)\in E$ and $A_{ij}=0$ otherwise. 
The corresponding wiring probability $\mathbb{P}(A_{ij})$ is then
\begin{equation}\label{eq:S1}
\mathbb{P}(\{A_{ij}\})=\prod_{i<j}\pi_{ij},\qquad 
\pi_{ij}=\frac{\langle k\rangle}{N}\delta(A_{ij}-1)+\bigg(1-\frac{\langle k\rangle}{N}\bigg)\delta(A_{ij}).
\end{equation}
\noindent 
The average of a physical observable $\mathcal{O}(\{A_{ij}\})$ over the quenched disorder generated by the underlying graph is
$$
\langle \mathcal{O}\rangle_{\mathfrak{G}}=\int_{A\in\mathfrak{G}}\mathcal{O}(\{A_{ij}\})\mathbb{P}(\{A_{ij}\})\prod_{i<j}\dd A_{ij}. 
$$
Following the replica approach to (self-averaging) disordered systems~\cite{parisi2020theory}, we compute the free-energy density 
\begin{equation}\label{eq:S2}
-\beta f=\lim_{N\to\infty}\frac{1}{N}\Big\langle\ln\mathpzc{Z}[\{A_{ij}\}]\Big\rangle_{\mathfrak{G}},\qquad 
\mathpzc{Z}[\{A_{ij}\}]=\sum_{\{\boldsymbol{\sigma}\}}\mathrm{exp}\bigg\{\beta\sum_{i<j}A_{ij}\sigma_i\sigma_j\bigg\}
\end{equation}
by exploiting the well celebrated replica trick, i.e.\ the identity $\ln\langle\mathpzc{Z}^n\rangle_\mathfrak{G}=1+n\langle \ln\mathpzc{Z}\rangle_\mathfrak{G}+\mathcal{O}(n^2)$ in the $n\to0$ limit. 
To this aim, let us unfold the $n$-th power of the partition function in terms of the replicated spin configurations $\{\boldsymbol{\sigma}^a\}_{a=1,\dots,n}$ under the same realization of the disorder (i.e.\ for the same adjacency matrix $A$), so that 
$$
\mathpzc{Z}^n[A]
=\sum_{\{\{ \boldsymbol{\sigma}^a\}\}}\mathrm{exp}\bigg\{-\beta\sum_{a=1}^n\mathcal{H}[\boldsymbol{\sigma}^a,A]\bigg\}
=\sum_{\{\{ \boldsymbol{\sigma}^a\}\}}\mathrm{exp}\bigg\{\mathcal{J}\sum_{i<j}A_{ij}\vec{\sigma}_i\cdot\vec{\sigma}_j\bigg\}
$$
\noindent
where $\mathcal{J}\equiv\beta J$, $\{\{\boldsymbol{\sigma}^a\}\}$ is the set of all the possible $2^{nN}$ replicated spin configurations and $\vec\sigma_i\cdot\vec\sigma_j$ is a scalar product over the replicas. 
To perform the average over $\mathfrak{G}(M, N\langle k\rangle/2)$, let us first compute the following element 
\begin{equation}\label{eq:S3}
\Big\langle e^{\mathcal{J}A_{ij}\vec{\sigma}_i\cdot\vec{\sigma}_j}\Big\rangle_{\mathfrak{G}}=
\int_{A\in\mathfrak{G}}e^{\mathcal{J}A_{ij}\vec{\sigma}_i\cdot\vec{\sigma}_j}\mathbb{P}(\{A_{ij}\})\prod_{i<j}\dd A_{ij}
=\prod_{i<j}\bigg(1-\frac{\langle k\rangle}{N}+\frac{\langle k\rangle}{N}e^{\mathcal{J}\vec{\sigma}_i\cdot\vec{\sigma}_j}\bigg).
\end{equation}
\noindent 
Inserting the above in the expression for $\mathcal{Z}^n[A]$, we find
\begin{equation}\label{eq:S4}
\Big\langle\mathcal{Z}^n[A]\Big\rangle_{\mathfrak{G}}=\sum_{\{\{\boldsymbol{\sigma}^a\}\}}\prod_{i<j}e^{-\frac{\langle k\rangle}{N}+\frac{\langle k\rangle}{N}e^{\mathcal{J}\vec{\sigma}_i\cdot\vec{\sigma}_j}+\mathcal{O}(N^{-2})}=\sum_{\{\{\boldsymbol{\sigma}^a\}\}}\mathrm{exp}\bigg\{-\frac{1}{2}\langle k\rangle N + \frac{\langle k\rangle}{2N}\sum_{i<j}e^{\mathcal{J}\vec{\sigma}_i\cdot\vec{\sigma}_j}+\mathcal{O}(1)\bigg\},
\end{equation}
\noindent
having noticed that diagonal elements yield only $\mathcal{O}(1)$ corrections.  
Since the sum runs over the interacting replicas, it turns convenient to introduce~\cite{monasson1998optimization} a suitable order parameter (OP) generalizing the average magnetization $\mathcal{M}=\frac{1}{N}\sum_{i\leq N}\langle\sigma_i\rangle_\beta$ and counting the number of sites $i$ with replicated spins $\vec{s}\in\mathbb{Z}_2^n$, i.e.\ $c(\vec{s})=\frac{1}{N}\sum_{i\leq N}\delta\big(\vec{s},\vec{\sigma}_i\big)$, where $\delta(\cdot\,,\,\cdot)$ is the $n$-dimensional Kronecker symbol. 
Indeed, notice that $\sum_{\vec{s}}c(\vec{s})=1$. 
With this definition, let us rewrite the interaction over the replicas by adopting two resolutions of the identity, that is 
\begin{equation}\label{eq:S5}
\sum_{i<j}e^{\mathcal{J}\vec{\sigma}_i\cdot\vec{\sigma}_j}=
\sum_{i<j}\sum_{\vec{s}_1}\delta\big(\vec{s}_1,\vec{\sigma}_i\big)\sum_{\vec{s}_2}\delta\big(\vec{s}_2,\vec{\sigma}_j\big)e^{\mathcal{J}\vec{s}_2\cdot\vec{s}_2}=N^2\sum_{\vec{s}_1,\vec{s}_2}c(\vec{s}_1)c(\vec{s}_2)e^{\mathcal{J}\vec{s}_1\cdot\vec{s}_2}.
\end{equation}
\noindent 
Inserting Eq.~\eqref{eq:S5} into Eq.~\eqref{eq:S4}, we can write the replicated partition function as 
\begin{equation}\label{eq:S6}
\Big\langle\mathcal{Z}^n[A]\Big\rangle_{\mathfrak{G}}=\sum_{\{\boldsymbol{\sigma}^a\}}\sum_{\{c(\,\cdot\,)\}}\prod_{\{\vec{s}\}}\delta\bigg(c(\vec{s}),\frac{1}{N}\sum_i\delta\big(\vec{s},\vec{\sigma}_i\big)\bigg)\mathrm{exp}\bigg\{\frac{\langle k\rangle N}{2}-\frac{\langle k\rangle N}{2}\sum_{\vec{s}_1,\vec{s}_2}c(\vec{s}_1)c(\vec{s}_2)e^{\mathcal{J}\vec{s}_1\cdot\vec{s}_2}\bigg\}
\end{equation}
By exchanging the order of the sums, we can calculate a purely combinatorial term of the partition function characterizing the number of replicated configurations $\{\boldsymbol{\sigma}^a\}$ corresponding to a certain set of OPs $c(\vec{s})$, i.e.\ 
$$
\sum_{\{\boldsymbol{\sigma}^a\}}\prod_{\vec{s}}\delta\bigg(c(\vec{s}),\tfrac{1}{N}\sum_{i\leq N}\delta\big(\vec{s},\vec{\sigma}_i\big)\bigg)=\frac{N!}{\prod_{\vec{s}}\big(c(\vec{s})N\big)!}=e^{-N\sum_{\vec{s}}c(\vec{s})\ln c(\vec{s})+\mathcal{O}(\ln N)},
$$
\noindent 
where we have used the Stirling approximation. 
Inserting the above into Eq.~\eqref{eq:S6}, we arrive at the functional relation
$$
\Big\langle\mathcal{Z}^n[A]\Big\rangle_{\mathfrak{G}}=
\int\prod_{\vec{s}}\dd c(\vec{s})e^{-N\mathcal{J}\mathcal{F}_n[c(\vec{s})]},\qquad
-\mathcal{J}\mathcal{F}_n[c(\vec{s})]\equiv -\frac{\langle k\rangle }{2}+\frac{\langle k\rangle }{2}\sum_{\vec{s}_1,\vec{s}_2}c(\vec{s}_1)c(\vec{s}_2)^{\mathcal{J}\vec{s}_1\cdot\vec{s}_2}-\sum_{\vec{s}}c(\vec{s})\ln c(\vec{s}),
$$
which can be integrated with the saddle-point method. 
The normalization constraint $\sum_{\vec{s}}c(\vec{s})=1$ is ensured via a Lagrange multiplier $\Lambda$, s.t.\ $\delta_{c(\vec{s})}[\mathcal{J}\mathcal{F}_n[c(\vec{s}')+\Lambda(\sum_{\vec{s}'}c(\vec{s}')-1)]]=0$. 
Performing the functional derivative, one finds 
$$
\mathcal{J}\frac{\delta \mathcal{F}_n[c(\vec{s}\,')]}{\delta c(\vec{s})}=
-\frac{\langle k\rangle}{2}\sum_{\vec{s}_1,\vec{s}_2}\bigg(\delta(\vec{s}_1,\vec{s})c(\vec{s}_2)+c(\vec{s}_1)\delta(\vec{s},\vec{s}_2)\bigg)e^{\mathcal{J}\vec{s}_1\cdot\vec{s}_2}+\sum_{\vec{s}\,'}\Big(\delta(\vec{s},\vec{s}\,')\ln c(\vec{s}\,')+\delta(\vec{s},\vec{s}\,')\Big)
$$
and so we can calculate the free-energy density via the saddle-point equations 
\begin{equation}\label{eq:S7}
c(\vec{s})=\mathrm{exp}\bigg\{-1-\Lambda+\langle k\rangle\sum_{\vec{s}\,'}c(\vec{s}\,')e^{\mathcal{J}\vec{s}\cdot\vec{s}\,'}\bigg\}. 
\end{equation}
\indent 
The ferromagnetic phase of the model can then be studied by breaking the $\mathbb{Z}_2$ symmetry of the OP, while preserving its permutation over the replicas.
In such replica-symmetric (RS) ansatz, $c(\vec{s})$ must be a function of the components of $\vec{s}$ and the simplest invariant form is a sum over the components, i.e.\ $c(\vec{s})=f(\sum_a s^a)$. 
The RS-solution can then be described in terms of a set of effective (cavity) fields $\{h_i\}_{i=1,2,\dots,N}$ whose probability distribution $P(h)$ is defined via the generalized Laplace transform 
\begin{equation}\label{eq:S8}
c(\vec{s})=\int_{\mathbb{R}}\frac{e^{\mathcal{J}h\sum_as^a}}{(2\mathrm{cosh}\mathcal{J}h)^n} P(h)\dd h.
\end{equation}
\indent 
Eq.~\eqref{eq:S8} exhaustively describes the model's thermodynamics. 
In particular, the thermally-averaged magnetization density $\mathcal{M}=\frac{1}{N}\sum_{i\leq N}m_i=\lim_{n\to0}\sum_{\vec{s}}s^1c(\vec{s})$ can be written via the magnetic moments $m_i=\langle\sigma_i\rangle_\beta=\mathrm{tanh}(\beta J h_i)$ as $\mathcal{M}=\int_{\mathbb{R}}P(h)\mathrm{tanh}(\beta Jh)\dd h$. 
An analytical computation of $P(h)$ is found by inserting Eq.~\eqref{eq:S8} into the saddle-point equations, Eq.~\eqref{eq:S7}. 
For convenience, let us call $\mu\equiv\mathcal{J}\sum_as^a$ and $\mathcal{C}_n(h)\equiv(2\mathrm{cosh}\mathcal{J}h)^n$, so that 
\begin{displaymath}
\int_\mathbb{R} \frac{\dd h P(h)}{\mathcal{C}_n(h)}e^{h\mu}
= \mathrm{exp}\bigg\{-1-\Lambda+\langle k\rangle\int_\mathbb{R}
\frac{\dd h P(h)}{\mathcal{C}_n(h)}\prod_{a\leq n}\sum_{s_a'=\pm1}e^{\mathcal{J}(s_a+h)s_a'}\bigg\}. 
\end{displaymath}
Since $\sum_{s_a\pm1}e^{\mathcal{J}(s_a+h)s_a'}=2\mathrm{cosh}\mathcal{J}(s_a+h)$, we can rearrange the product over the replicated spins as follows: 
\begin{displaymath}
\prod_{a\leq n}\sum_{s_a'\pm1}e^{\mathcal{J}(s_a+h)s_a'} =2^n\prod_{a\leq n} \mathrm{cosh}\mathcal{J}(s_a+h)=2^n\mathrm{cosh}^{a_+}\mathcal{J}(h+1)\mathrm{cosh}^{a_-}\mathcal{J}(h-1),
\end{displaymath}
\noindent 
where $a_{\pm}=\frac{n}{2}\pm\frac{\mu}{2\mathcal{J}}$. 
Sending $n\to0$ limit, one finds $\int_\mathbb{R}\dd h P(h)e^{h\mu}=\mathrm{exp}\{-1-\Lambda+\langle k\rangle\int_\mathbb{R}\dd h P(h)[\frac{\mathrm{cosh}\mathcal{J}(h+1)}{\mathrm{cosh}\mathcal{J}(h-1)}]^{\mu/2\mathcal{J}}\}$, where the Lagrange multiplier $\Lambda$ can be fixed at $\mu=0$ so to ensure the normalization of $P(h)$, i.e.\ $\int_\mathbb{R}\dd h P(h)=\mathrm{exp}\{-1-\Lambda+\langle k\rangle\int_\mathbb{R}\dd h P(h)\}$ which implies that $\Lambda=\langle k\rangle-1$. 
We finally arrive at the integral equation
\begin{displaymath}
\int_\mathbb{R}e^{h\mu}P(h)\dd h=\mathrm{exp}\bigg\{ -\langle k\rangle +\langle k\rangle\int_\mathbb{R}\bigg(\frac{\mathrm{cosh}\mathcal{J}(h+1)}{\mathrm{cosh}\mathcal{J}(h-1)}\bigg)^{\nicefrac{\mu}{2\mathcal{J}}}P(h)\dd h\bigg\},
\end{displaymath}
\noindent 
which has to be solved for every $\mu$. 
Expanding the exponential on the r.h.s.\ yields 
\begin{displaymath}
\int_\mathbb{R}\dd h P(h)e^{h\mu}=e^{-\langle k\rangle}\sum_{d\in\mathbb{N}_0}\frac{\langle k\rangle^d}{d!}\int_\mathbb{R}\dd h_1P(h_1)\cdots\int_\mathbb{R}\dd h_d P(h_d)e^{\frac{\mu}{2\mathcal{J}}\sum_{\ell\leq d}\ln\frac{\mathrm{cosh}\mathcal{J}(h_\ell+1)}{\mathrm{cosh}\mathcal{J}(h_\ell-1)}},
\end{displaymath}
\noindent
and so, after comparing the coefficients of the exponentials in $\mu$, we finally find
\begin{equation}\label{eq:S9}
P(h)=e^{-\langle k\rangle}\sum_{d\in\mathbb{N}_0}\frac{\langle k\rangle^d}{d!}\int_\mathbb{R}\dd h_1 P(h_1)\cdots\int_\mathbb{R}\dd h_d P(h_d)\delta\bigg(h-\frac{1}{2\mathcal{J}}\sum_{\ell=1}^d\ln\frac{\mathrm{cosh}\mathcal{J}(h_\ell+1)}{\mathrm{cosh}\mathcal{J}(h_\ell-1)}\bigg).
\end{equation}
\noindent 
For $J\equiv1$, Eq.~\eqref{eq:S9} corresponds to the Eqs.~\eqref{eq:M2}, \eqref{eq:M3} in the main text which can be solved numerically via population dynamics methods~\cite{hartmann2005phase}. 
The line of critical thresholds can instead be found exactly by linearizing around the RS-paramagnetic solution, i.e.\ $P(h)=\delta(h)$, which always satisfies Eq.~\eqref{eq:S9}. 
Taking then as initial seed a function $P_0$ allowing small cavity fields with average $\varepsilon_0=\int_\mathbb{R}hP_0(h)\dd h\ll1$, at the next iteration in Eq.~\eqref{eq:S9} we find 
\begin{displaymath}
\begin{aligned}
\varepsilon_1=\int_\mathbb{R}h P_1(h)\dd h
&\,=e^{-\langle k\rangle}\sum_{d=0}^{\infty}\frac{\langle k\rangle^d}{d!}\int h\dd h\int P_0(h_1)\dd h_1\cdots\int P_0(h_d) \dd h_d \delta\Big(h-\mathcal{T}_d(h_1,\dots,h_d)\Big)\\
&\,=e^{-\langle k\rangle}\sum_{d=0}^{+\infty}\frac{\langle k\rangle^d}{d!}\int\dd h_1 P_0(h_1)\cdots\int \dd h_dP_0(h_d) \mathcal{T}_d(h_1,\dots,h_d)\\
&\,=\frac{1}{2\mathcal{J}}e^{-\langle k\rangle}\sum_{d=0}^{+\infty}\frac{\langle k\rangle^d d}{d!}\int\dd \tilde{h}P_0(\tilde{h})\ln\frac{\mathrm{cosh}\mathcal{J}(\tilde{h}+1)}{\mathrm{cosh}\mathcal{J}(\tilde{h}-1)},
\end{aligned}
\end{displaymath}
where $\mathcal{T}_d\equiv\frac{1}{2\mathcal{J}}\sum_{\ell\leq d}\ln\frac{f_+(h_\ell)}{f_-(h_\ell)}$ and $f_{\pm}(x)\equiv\mathrm{cosh}\mathcal{J}(x\pm1)$. 
Expanding around $h=0$ yields $\frac{1}{2\mathcal{J}}\ln\frac{f_+(h)}{f_-(h)}=h\mathrm{th}\mathcal{J}+\mathcal{O}(h^2)$\vspace*{-0.05cm} so that, neglecting $\mathcal{O}(h^2)$ terms, one has $\varepsilon_1=\langle k\rangle \mathrm{tanh}\mathcal{J} \varepsilon_0$. 
The paramagnetic phase loses then stability at 
\begin{equation}\label{eq:S10}
\langle k\rangle\mathrm{tanh}\mathcal{J}_c=1,
\end{equation}
\noindent 
identifying the line of continuous para-ferro transitions in the $(\langle k\rangle,\mathcal{J})$ phase plane. \\
\indent 
Notice that, for pairwise couplings, a ferromagnetic phase spontaneously emerge only above the random graph's percolation threshold~\cite{leone2002ferromagnetic}, i.e.\ for any average connectivity $\langle k\rangle>1$.

\subsection{Ferromagnetic random $(K+q)$-spin models}\label{S12}\vspace*{-0.25cm}
Having recalled the details underlying the RS-solution of the Ising Hamiltonian on random graphs, let us generalize the above to the case of random higher-order interactions, i.e.\ $K$-spin and $(K+q)$-spin Hamiltonians, i.e.
\begin{displaymath}
\begin{aligned}
-\mathcal{H}_K[\{\boldsymbol{\sigma}\},A]&\,=J_K\sum_{i_1,\dots,i_K}A_{i_1\cdots i_K}\sigma_{i_1}\cdots\sigma_{i_K},\\
-\mathcal{H}_{K+q}[\{\boldsymbol{\sigma}\},A]&\,=J_K\sum_{i_1,\dots,i_K\in E_{1-q,K}}A_{i_1\cdots i_K}\sigma_{i_1}\cdots\sigma_{i_K}+J_{K+1}\sum_{i_1,\dots,i_{K+1}\in E_{q,K+1}}A_{i_1\dots i_{K+1}}\sigma_{i_1}\dots\sigma_{i_{K+1}},
\end{aligned}
\end{displaymath}
respectively.  
We will hence refer to a $(K+q)$-spin model as to a mixture ensemble~\cite{leone2001phase} of random (hyper)graphs composed by a fraction $1-q$ of random $K$-spin interactions and a remaining fraction $q$ of random $(K+1)$-spin interactions, respectively extracted from a list of $K$-tuple and $(K+1)$-tuple hyperedges $E_{1-q,K}$ and $E_{q,K+1}$. \\
 
\begin{figure*}[t]
\centering
    \includegraphics[width=0.45\linewidth]{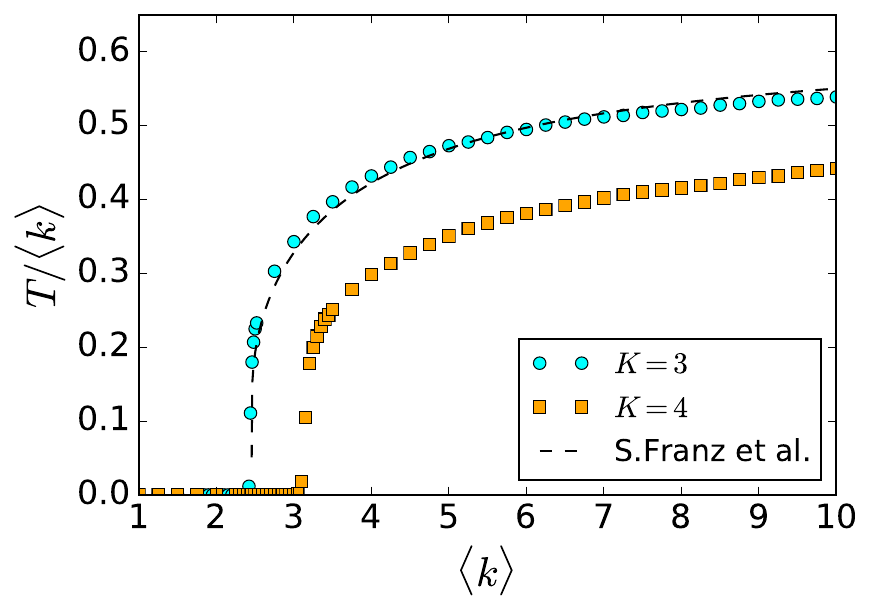}
    \includegraphics[width=0.45\linewidth]{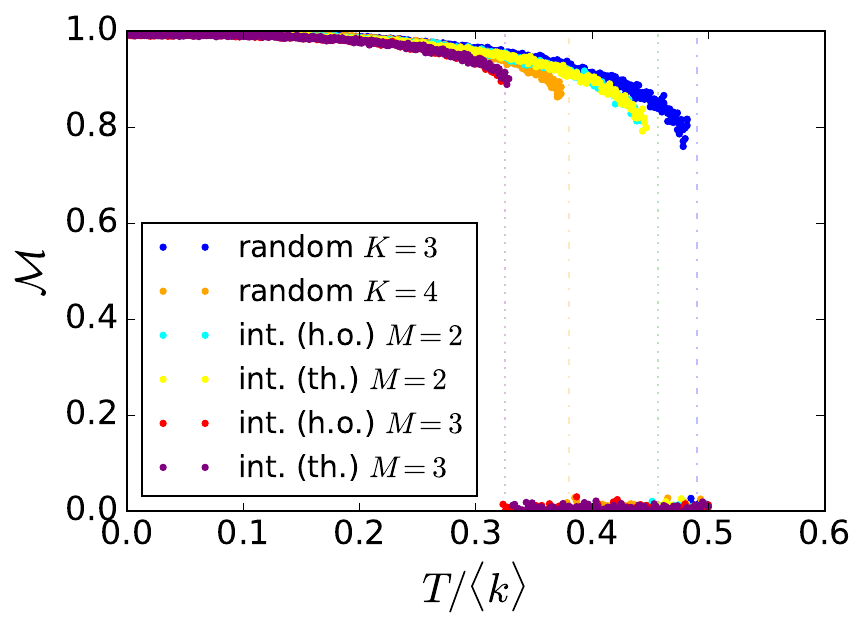}
            \caption{\small \textbf{Bulk-melting spinodal thresholds in random $K$-spin models.} (Color online) \emph{Left panel}) Comparison between the bulk-melting spinodals found via the population dynamics algorithm in the above for (cyan symbols) $K=3$ and (orange symbols) $K=4$, with equilibration time $T_{eq}=10^4$ and a networked population of $N=10^4$ nodes. For checking the quality of our results, we compared our $K=3$ spinodal thresholds with those (dashed line) reported in the literature in Ref.~\cite{franz2001ferromagnet}. \emph{Right panel}) Global average magnetization curves obtained for increasing temperatures via simulated annealing on hypergraphs with average hyperdegree $\langle k\rangle=6$ with $N=10^4$ nodes and for $10^2N$ number of MCSs. The dot-dashed lines represent the bulk-melting thresholds obtained via PDs and shown in the left panel. A comparison has been made with respect to the curves describing randomly fully-coupled interdependent spin models (both thermal and higher-order settings, see Fig.~\ref{fig:2} in the main text) with degree $\langle k\rangle=6$ (here $NOI=10^2$) and their comparison with the noise-injected iPDs thresholds (see Methods). }
        \label{fig:S1}\vspace*{-0.2cm}
\end{figure*}

\indent 
$\bullet$ \emph{\bf $K$-spins}. Let us first consider the wiring probability of a random hypergraph of $K$-tuples:
\begin{equation}\label{eq:S11}
\mathbb{P}(\{A_{i_1\dots i_K}\})
=\prod_{i_1<\dots\,<i_K}\pi_{i_1\dots i_K},\qquad
\pi_{i_1\dots i_K}=\Big(1-\langle k\rangle\gamma_K(N)\Big)\delta\big(A_{i_1\dots i_K}\big)+\langle k\rangle\gamma_K(N)\delta\big(1-A_{i_1\dots i_K}\big),
\end{equation}
\noindent 
where $\gamma_K(N)=M_K/\binom{N}{K}$ is a combinatorial factor given by the probability of hyper-wiring a $K$-tuple of randomly chosen nodes with one of the $M_K=N\langle k\rangle/K$ edges. 
Analogously to the above, we aim at calculating the disorder average of the replicated partition function, that is
\begin{displaymath}
\begin{aligned}
\Big\langle\mathpzc{Z}^n_K[A]\Big\rangle_{\mathfrak{G}_K}
&\,=\sum_{\{\{\boldsymbol{\sigma}^a\}\}} 
\prod_{i_1<\dots\, < i_K}\bigg(1-\langle k\rangle\gamma_K(N)+\langle k\rangle\gamma_K(N)e^{\mathcal{J}\sum_{a\leq n}\sigma_{i_1}^a\sigma_{i_2}^a\cdots\,\sigma_{i_K}^a}\bigg)\\
&\,=\sum_{\{\{\boldsymbol{\sigma}^a\}\}}\mathrm{exp}\bigg\{-\frac{N\langle k\rangle}{K}+\langle k\rangle\frac{\gamma_K(N)}{K!}\sum_{i_1,\dots,i_K}e^{\mathcal{J}\sum_{a=1}^n\sigma_{i_1}^a\cdots\sigma_{i_K}^a}+\mathcal{O}(1)\bigg\},
\end{aligned}
\end{displaymath}
\noindent 
where we have added $\mathcal{O}(1)$ diagonal corrections. 
Introducing the occupation density over the replicas, we have 
$$
\sum_{i_1,\dots,i_K}e^{\mathcal{J}\sum_{a\leq n}\sigma_{i_1}^a\cdots\sigma_{i_K}^a}=N^K\sum_{\vec{s}_1,\dots,\vec{s}_K}c(\vec{s}_1)\cdots c(\vec{s}_K)e^{\mathcal{J}\sum_{a\leq n}s_{1}^a\cdots\, s_{K}^a},
$$
\noindent 
so that, rearranging the sum in a way perfectly analogous to Eq.~\eqref{eq:S6} and following steps, we find 
\begin{equation}\label{eq:S12}
\begin{aligned}
\Big\langle \mathcal{Z}^n_K[A]\Big\rangle_{\mathfrak{G}_K}
&\,=\int \prod_{\vec{s}}\dd c(\vec{s})e^{-N\mathcal{J}\mathcal{F}_{n,K}[c(\vec{s})]},\\
-\mathcal{J}\mathcal{F}_{n,K}[c(\vec{s})]
&\,\equiv
\frac{\langle k\rangle}{K}-\frac{\langle k\rangle}{K}\sum_{\vec{s}_1,\dots,\vec{s}_K}c(\vec{s}_1)\cdots\vec{s}_K e^{\mathcal{J}\sum_{a}s_1^{a}\cdots s_{K}^a}+\sum_{\vec{s}}c(\vec{s})\ln c(\vec{s}).
\end{aligned}
\end{equation}
\noindent 
Eq.~\eqref{eq:S12} can be solved via saddle-point methods, yielding the set of saddle-point equations
\begin{equation}\label{eq:S13}
c(\vec{s})=\mathrm{exp}\bigg\{-\langle k\rangle+\langle k\rangle\sum_{\vec{s}\,'_1,\dots\vec{s}\,'_{K-1}}c(\vec{s}'_1)\cdots c(\vec{s}'_{K-1})e^{\mathcal{J}\sum_{a=1}^n s^a{s'}_1^{a}\cdots\,{s'}_{K-1}^{a}}\bigg\}.
\end{equation}
\noindent 
In the RS-ansatz, Eq.~\eqref{eq:S8}, we can rewrite these equations via the cavity distribution $P(h)$ so that, as $n\to0$, one has
\begin{displaymath}   
\int_\mathbb{R}e^{h\mu}P(h)\dd h
=\mathrm{exp}\bigg\{-\langle k\rangle+\langle k\rangle \int_\mathbb{R}\dd h_1 P(h_1)\cdots\int_\mathbb{R}\dd h_{K-1}P(h_{K-1})\mathcal{G}_K(\mathcal{J};h_1,\dots,h_{K-1})\bigg\},
\end{displaymath}
\noindent 
where $\mu\equiv\mathcal{J}\sum_as_a$ and $\mathcal{G}_K$ is a function of the general form $\mathcal{G}_K(x;\boldsymbol{y})\equiv(\alpha_{K,+}(x;\boldsymbol{y})/\alpha_{K,-}(x;\boldsymbol{y}))^{\nicefrac{\mu}{2\mathcal{J}}}$ and $\alpha_{K,\pm}$ are composed hyperbolic function whose specific expressions depends on the order $K$ of the model. 
\begin{equation}\label{eq:S14}
\begin{aligned}
P(h)=e^{-\langle k\rangle}\sum_{d=0}^{+\infty}\frac{\langle k\rangle^d}{d!}&\,
\prod_{i\leq d}\int\cdots\int \dd{h_{1,i}}\cdots\dd h_{K-1,i} P(h_{1,i})\cdots P(h_{K-1,i})\delta\bigg(h-\mathpzc{T}_K(\mathcal{J};\{h_{1,i}\}_i,\dots,\{h_{K-1,i}\}_{i})\bigg),\\
\mathpzc{T}_K&\,(\mathcal{J};\{h_{1,i}\}_i,\dots,\{h_{K-1,i}\}_{i})\equiv \frac{1}{2\mathcal{J}}\sum_{\ell=1}^d\ln
\frac{\alpha_{K,+}(\mathcal{J};h_{1,\ell},\dots,h_{K-1,\ell})}{\alpha_{K,-}(\mathcal{J};h_{1,\ell},\dots,h_{K-1,\ell})}
\end{aligned}
\end{equation}
\indent
For the sake of simplicity, let us report in what follows only the cases $K=3$ and $K=4$, whose integral kernels $\mathcal{T}_K$ can be written succinctly. 
In particular, after some algebra, we arrive at the following expressions: 
\begin{align}
\alpha_{3,\pm}(\mathcal{J};h_{1,\ell},h_{2,\ell})&\,=\mathrm{cosh}\mathcal{J}\big(h_{1,\ell}\pm h_{2,\ell}\big)+e^{-2\mathcal{J}}\mathrm{cosh}\mathcal{J}\big(h_{1,\ell}\mp h_{2,\ell}\big),\label{eq:S15}\\
\alpha_{4,\pm}(\mathcal{J};h_{1,\ell},h_{2,\ell},h_{3,\ell})&\,=\mathrm{cosh}\mathcal{J}\big(\pm1+h_{3,\ell}\big)\mathrm{cosh}\mathcal{J}(h_{1,\ell}+h_{2,\ell})+\mathrm{cosh}\mathcal{J}\big(\pm1-h_{3,\ell}\big)\mathrm{cosh}\mathcal{J}(h_{1,\ell}-h_{2,\ell}).\label{eq:S16}
\end{align}
\noindent 
A population dynamics algorithm solving the RS-phases of the ferromagnetic $3$-spin and $4$-spin models via the integral kernels in Eq.~\eqref{eq:S15} and Eq.~\eqref{eq:S16} is available on GitHub at \url{https://github.com/BnayaGross/High_order_interdependent_ising}. 
The main steps are summarized in the following pseudocode (notice that, here, $J=1$): 
\begin{algorithm}[H]
\caption*{\textsc{PopDyn($\beta, K, N,\mathcal{A}, T_{eq}$)}}
\begin{algorithmic}[1]
\State Initialize the tensor $\mathpzc{A}_{i_1,\dots,i_K}$ from the random hypergraph ($K$-tuple) with a given average hyperdegree $\langle k\rangle$;
\State Randomly initialize a population of cavity fields $\mathbf{h}=\{h_i\}_{i=1,\dots,N}$; 
\For{$t=1,\dots, T_{eq}$:}
	\For {$i=1,\dots, N$:}
        		\State  Compute $h_{i}=\frac{1}{2\mathcal{J}}\sum_{\ell=1}^{|\partial i|}\ln\frac{\alpha_{K,+}(\beta; h_{1,\ell},\dots,h_{K-1,\ell})}{\alpha_{K,-}(\beta; h_{1,\ell},\dots,h_{K-1,\ell})}$;
    	\EndFor 
\EndFor\\
\Return{Compute the local OPs $m_i=\mathrm{tanh}(\beta h_i)$ and the global OP $\mathcal{M}_\infty=\frac{1}{N}\sum_{i\leq N}m_i$}
\end{algorithmic}
\end{algorithm}
\noindent 
By contrast with pairwise couplings, these random higher-order models always undergo a spontaneous (temperature-driven) first-order transition from ferro- to para-magnetism whose threshold can be found via the RS-solution~\cite{franz2001ferromagnet}. 
In particular, the line of (bulk-melting~\cite{krzakala2011melting1}) spinodal singularities (i.e.\ the limit of ferro-para metastability) can be obtained (see Fig.~\ref{fig:S1}) by analyzing the ferromagnetic stability while cooling the system adiabatically. \\

\begin{figure*}[t]
\centering
    \includegraphics[width=0.45\linewidth]{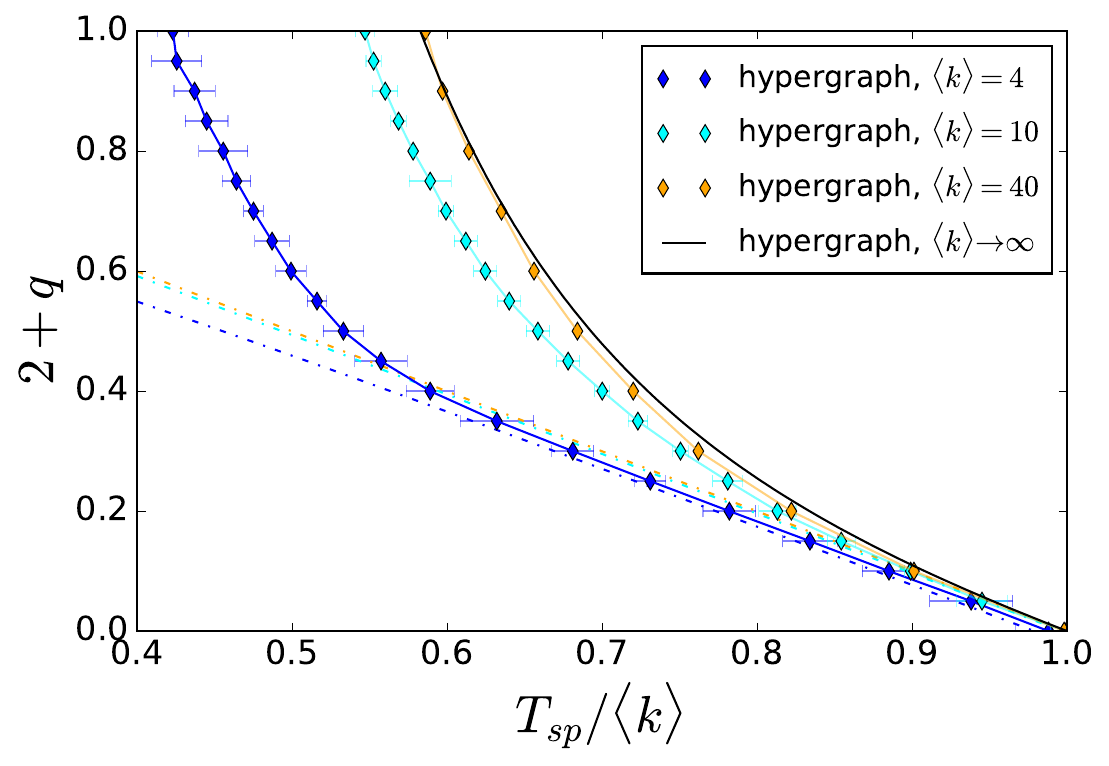}
    \includegraphics[width=0.45\linewidth]{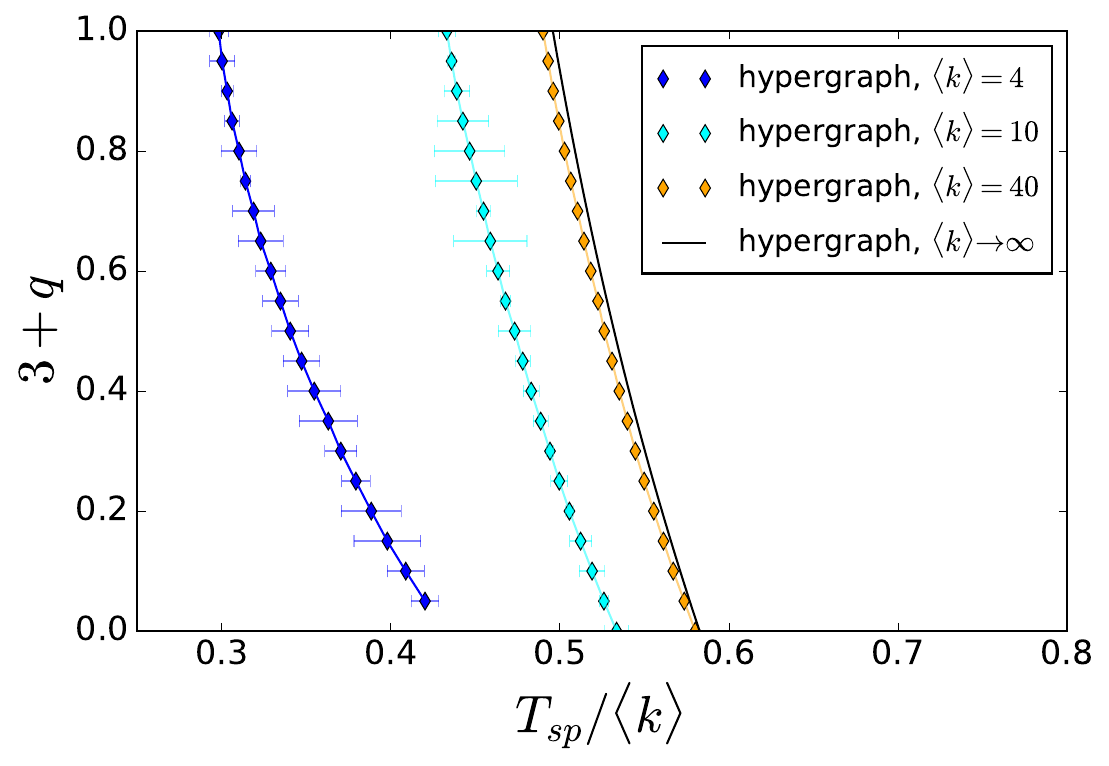}
            \caption{\small \textbf{Bulk-melting thresholds in random $(K+q)$-spin models.} (Color online) Ferromagnetic spinodal thresholds obtained via a population dynamics algorithm analogous to the above for the $(2+q)$-spin case (left panel) and for the $(3+q)$-spin case (right panel), for ER hypergraphs with different average connectivities: $\langle k\rangle=4$ (blue symbols), $\langle k\rangle=10$ (cyan symbols), $\langle k\rangle=40$ (orange symbols), and finally for the fully mean-field case, i.e.\ $\langle k\rangle\to\infty$ (black line). PDs simulations settings: equilibration time $T_{eq}=10^4$, networked population of $N=10^4$ nodes. (Left panel) The dot-dashed curves identify the line of transcritical (para-ferro) bifurcations, i.e.\ Eq.~\eqref{eq:S20} for $\langle{k}\rangle=4$ (blue), $\langle{k}\rangle=10$ (cyan) and $\langle{k}\rangle=40$ (orange).}
        \label{fig:S2}\vspace*{-0.2cm}
\end{figure*}

\indent 
$\bullet$ \emph{\bf $(K+q)$-spins}. Analogously to the $K$-spin case, let us consider the wiring probability of a random hypergraph mixture composed by a fraction $(1-q)\in[0,1]$ of randomly coupled $K$-tuples $\{i_1,\dots,i_K\}\in E_K$ and a remaining fraction $q$ of randomly coupled $(K+1)$-tuples $\{i_1,\dots,i_{K+1}\}\in E_{K+1}$. 
Since hyperedges are drawn independently and at random, their wiring probabilities have independent distributions given by Eq.~\eqref{eq:S11} with elements 
\begin{displaymath}
\begin{aligned}
\pi_{i_1\dots i_K}
&\,=\Big(1-(1-q)\langle k\rangle\gamma_K(N)\Big)\delta\big(A_{i_1\dots i_K}\big)+(1-q)\langle k\rangle\gamma_K(N)\delta\big(1-A_{i_1\dots i_K}\big),\\
\pi_{i_1\dots i_{K+1}}
&\,=\Big(1-q\langle k\rangle\gamma_{K+1}(N)\Big)\delta\big(A_{i_1\dots i_{K+1}}\big)+q\langle k\rangle\gamma_{K+1}(N)\delta\big(1-A_{i_1\dots i_{K+1}}\big).
\end{aligned}
\end{displaymath}
\noindent 
Notice that the hypergraphs belonging to this mixture graph ensemble are generated by fixing the average hyperedge density $\gamma\in[0,1]$ while keeping different average degrees of $K$- and $(K+1)$-tuples, so that 
$$
\gamma N=\frac{1}{K+1}q\bar{k}_{K+1} N+\frac{1}{K}(1-q)\bar{k}_KN,
$$
i.e.\ $\bar{k}_K=\gamma K$. 
In $\S$~\ref{sec:S4} we will return to this point when studying the mapping of these models' ground states (GSs) with the equations governing interdependent percolation on their coupled networks analogues.\\ 
\indent 
To calculate the disorder-averaged replicated partition function, we can factorize its $n$-th power as
$$
\Big\langle\mathpzc{Z}^n_{K,q}[A]\Big\rangle_{\mathfrak{G}_{K+q}}
=\Big\langle\mathpzc{Z}^n_{K}[A]\Big\rangle_{\mathfrak{G}_{K,1-q}}
\Big\langle\mathpzc{Z}^n_{K+1}[A]\Big\rangle_{\mathfrak{G}_{K+1,q}},
$$
so that we can separately follow the main steps that solve the random $K$-spin case. 
One readily finds that
$$
\Big\langle\mathpzc{Z}^n_{K,q}[A]\Big\rangle_{\mathfrak{G}_{K,q}}
=\sum_{\{\{\boldsymbol{\sigma}\}\}}\mathrm{exp}\bigg\{ -\gamma N+\frac{q\bar{k}_{K+1}}{(K+1)N^K}\sum_{i_1,\dots,i_{K+1}}e^{\mathcal{J}_{K+1}\sum_a\sigma_{i_1}^a\cdots\,\sigma_{i_{K+1}}^a}+\frac{(1-q)\bar{k}_K}{KN^{K-1}}\sum_{i_1,\dots,i_K}e^{\mathcal{J}_K\sum_a\sigma_{i_1}^a\cdots\,\sigma_{i_K}^a}\bigg\},
$$
\noindent 
in which case the saddle-point equations for the density of replicas are given by 
\begin{equation}\label{eq:S17}
\begin{aligned}
c(\vec{s})=\mathrm{exp}\bigg\{
-\gamma(K+q)&\,+(1-q)\bar{k}_K\sum_{\vec{s}_1,\dots,\vec{s}_{K-1}}c(\vec{s}_1)\cdots c(\vec{s}_{K-1})e^{\mathcal{J}_K\sum_as^as_1^a\cdots\, s_{K-1}^a}+\\ 
&\,+q\bar{k}_{K+1}\sum_{\vec{s}_1,\dots,\vec{s}_K}c(\vec{s}_1)\cdots c(\vec{s}_{K})e^{\mathcal{J}_{K+1}\sum_as^as_1^a\cdots\, s_{K}^a}
\bigg\}.
\end{aligned}
\end{equation}
\noindent 
In perfect analogy with the previous cases, we can solve Eqs.~\eqref{eq:S17} within the RS-ansatz, so that (for $n\to0$)
\begin{equation}\label{eq:S18}
\begin{aligned}
P(h)=&\,e^{-\gamma(K+q)}\sum_{d\in\mathbb{N}_0}\frac{\bar{k}_K^d}{d!}(1-q)^d\sum_{\ell\in\mathbb{N}_0}\frac{\bar{k}_{K+1}^\ell}{\ell!}q^\ell\int\cdots\int\mathcal{D}\tilde{h}_1\cdots\mathcal{D}\tilde{h}_{K-1}\int\cdots\int\mathcal{D}h'_1\cdots\mathcal{D}h'_{K}\times\\ 
&\quad\times\delta\bigg(h-\frac{1}{2\mathcal{J}}\sum_{i=1}^d\mathpzc{g}_{K}\big(\mathcal{J}_K;\{\tilde{h}_{1,i}\}_i,\dots,\{\tilde{h}_{K-1i}\}_i\big)-\frac{1}{2\mathcal{J}}\sum_{j=1}^\ell\mathpzc{g}_{K+1}\big(\mathcal{J}_{K+1};\{h'_{1,j}\}_j,\dots,\{h'_{K,j}\}_j\big)\bigg),
\end{aligned}
\end{equation}
\noindent 
where $\mathcal{D}\tilde{h}_{\nu}\equiv\prod_{i\leq d}\dd \tilde{h}_{\nu,i}$ with $\nu=1,\dots,K-1$ and $\mathcal{D}h'_{\eta}\equiv\prod_{j\leq \ell}\dd h'_{\eta,j}$ with $\eta=1,\dots,K$, while 
$$
\mathpzc{g}_\mu(\mathcal{J}_\mu;\mathbf{x})\equiv\ln\frac{\alpha_{\mu,+}(\mathcal{J}_\mu;\mathbf{x})}{\alpha_{\mu,-}(\mathcal{J}_\mu;\mathbf{x})},\qquad \mu=1,2,\dots,K+1
$$
\noindent 
where $\alpha_{\mu,\pm}$ are composed transcendental functions whose form depends on the order $\mu=1,\dots,K$ of the spins' interactions. 
Similarly to the $K$-spin case, we focus on the $K=3$ and $K=4$ cases, i.e.\ on random $(2+q)$-spin (Fig.~\ref{fig:S2}, left) and random $(3+q)$-spin (Fig.~\ref{fig:S2}, right) mixtures.
A population dynamics algorithm solving their RS-phases is available on GitHub at \url{https://github.com/BnayaGross/High_order_interdependent_ising}. \\

\indent 
$\bullet$ \emph{\bf Transcritical line in the random $(2+q)$-spin model}. For $K=2$, the above equations describe a mixture of pairwise and $3$-spin couplings characterizing an Hamiltonian crossing over from the classical Ising model on ER graphs and a random $3$-spin (Baxter-Wu) model. 
For the sake of convenience, let us re-write their RS-kernels
\begin{equation}\label{eq:S19}
\mathpzc{g}_2(J; h)=\ln\frac{\mathrm{cosh}\beta(h+1)}{\mathrm{cosh}\beta(h-1)}\qquad
\mathpzc{g}_3(J; h, h')=\ln\frac{\mathrm{cosh}\beta(h'+h)+e^{-2\beta}\mathrm{cosh}\beta(h'-h)}{\mathrm{cosh}\beta(h'-h)+e^{-2\beta}\mathrm{cosh}\beta(h'+h)},
\end{equation}
\noindent 
where we set $J_3= J_2\equiv 1$ to simplify the notation. 
While the bulk-melting spinodals can only be computed numerically via PDs algorithms, the bulk-solidification (para-ferro) thresholds delimiting the loss of paramagnetism can be found analytically by linearizing around the solution $P(h)=\delta(h)$. 
Do this aim, let us follow the approach adopted in the random $2$-spin case and calculate the first moment $\langle h\rangle$ under a perturbation around $h,h'=0$, so that 
\begin{displaymath}
\begin{aligned}
\langle h\rangle=\int hP(h)\dd h=\frac{1}{2\beta}\bigg(e^{-\bar{k}_2(1-q)}&\,\sum_{d\in\mathbb{N}_0}\frac{\bar{k}_2^d}{d!}(1-q)^dd\int\dd\tilde{h}P(\tilde{h})\mathpzc{g}_2(\beta; \tilde{h})+\\
&\,+e^{-q\bar{k}_3}\sum_{\ell\in\mathbb{N}_0}\frac{\bar{k}_3^\ell}{\ell!}q^\ell \ell \int\dd hP(h)\int \dd h' P(h')\mathpzc{g}_3(\beta; h,h')\bigg). 
\end{aligned}
\end{displaymath}
\noindent 
Linearizing around $h,h'=0$, the kernel $\mathpzc{g}_3$ yields $\mathcal{O}(\langle h\rangle^2)$ terms and it can be easily verified that 
$$
\partial_h\mathpzc{g}_3(\beta; h,h')|_{h=h'=0}=0=\partial_{h'}\mathpzc{g}_3(\beta; h,h')|_{h,h'=0},
$$
\noindent 
so that $\langle h\rangle=(\bar{k}_2(1-q)\mathrm{tanh}\beta)\langle h\rangle+\mathcal{O}(\langle h\rangle^2)$. 
The line of transcritical bifurcations in the $(\beta, q)$ phase plane is then
\begin{equation}\label{eq:S20}
\bar{k}_{2}(1-q_{tc})\mathrm{tanh}\beta_{tc}=1,
\end{equation}
\noindent
and their location is shown in the left panel of Fig.~\ref{fig:S2}, for average pairwise connectivities $\bar{k}=4,10,40$. \\
\indent 
Notice that the transcritical line, Eq.~\eqref{eq:S20}, is governed only by the parameters characterizing the $2$-spin Hamiltonian, i.e.\ pairwise interactions determine the recoverability limit of higher-order mixtures towards their functioning phases. 
The non-recoverability threshold can be found via Eq.~\eqref{eq:S20} by calculating its zero-temperature limit, yielding $\bar{k}(1-q')=1$, which has a genuinely geometrical interpretation. 
$q'=1-1/\bar{k}$ is, in fact, nothing but the line of continuous phase transitions of interdependent percolation in partially coupled ER graphs~\cite{parshani-prl2010}. \\

\begin{figure}[h]
\centering
    \includegraphics[width=0.245\linewidth]{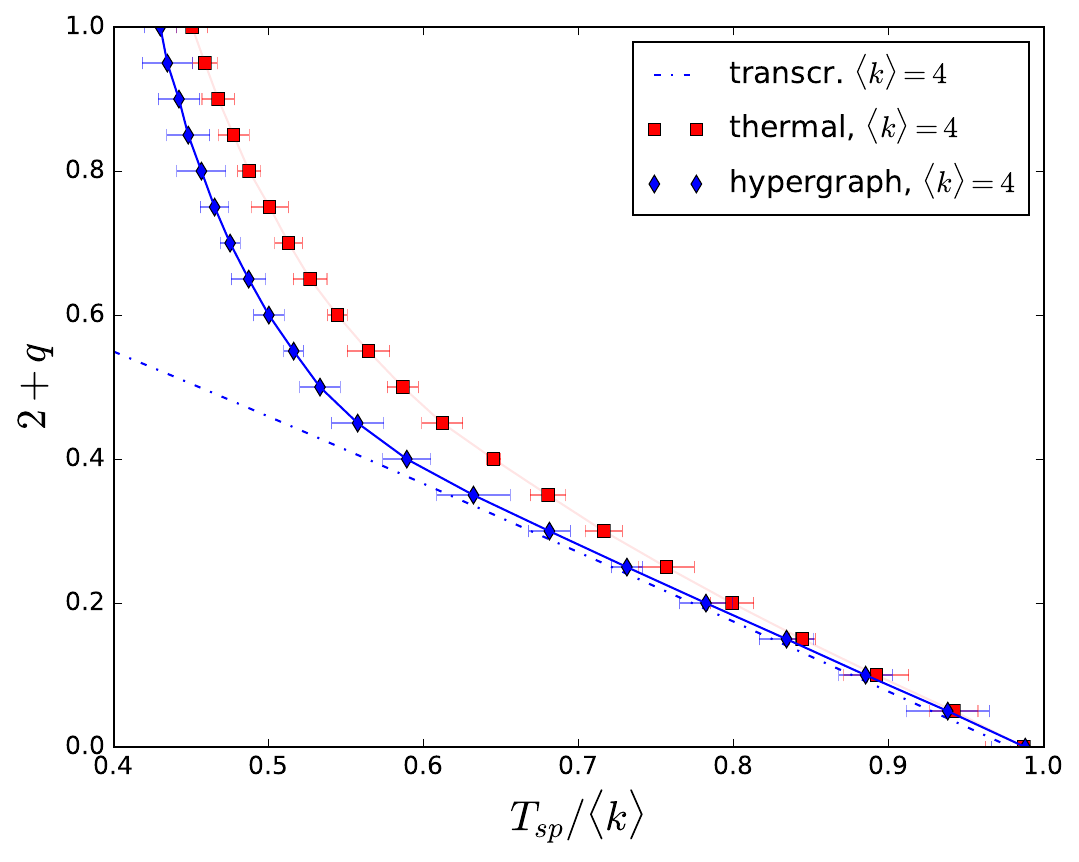}
    \includegraphics[width=0.245\linewidth]{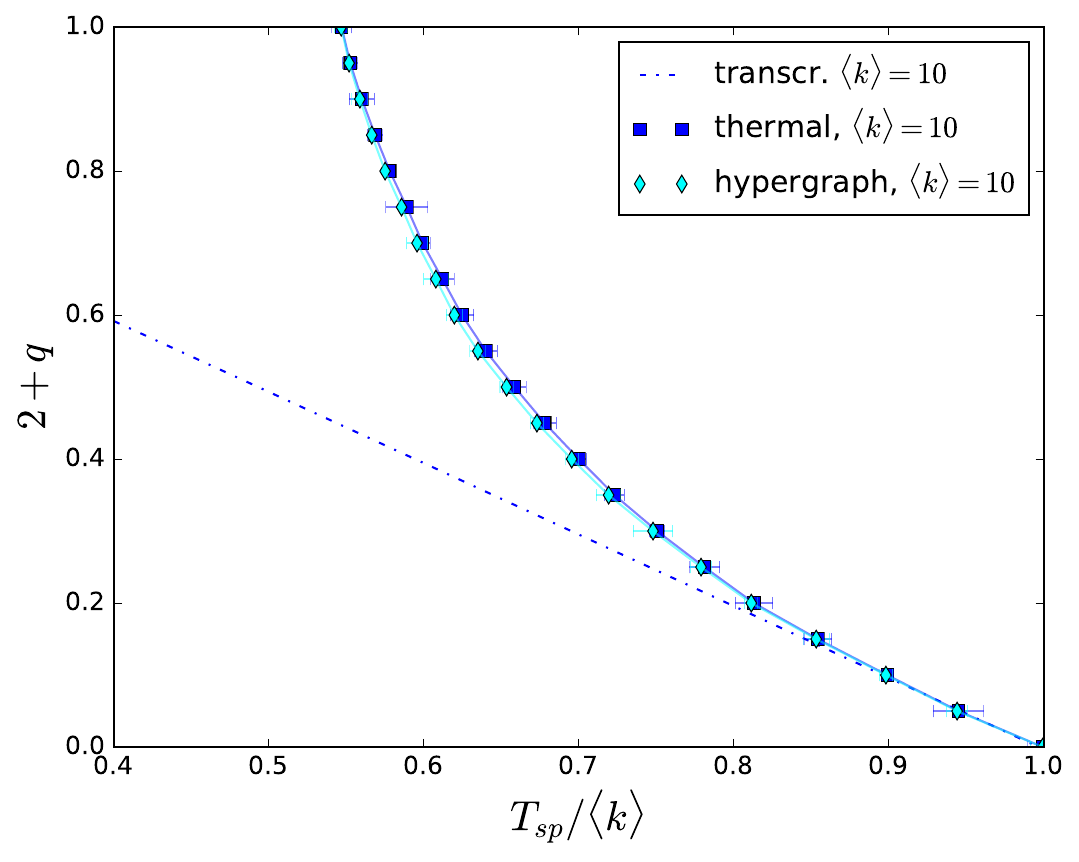}
    \includegraphics[width=0.245\linewidth]{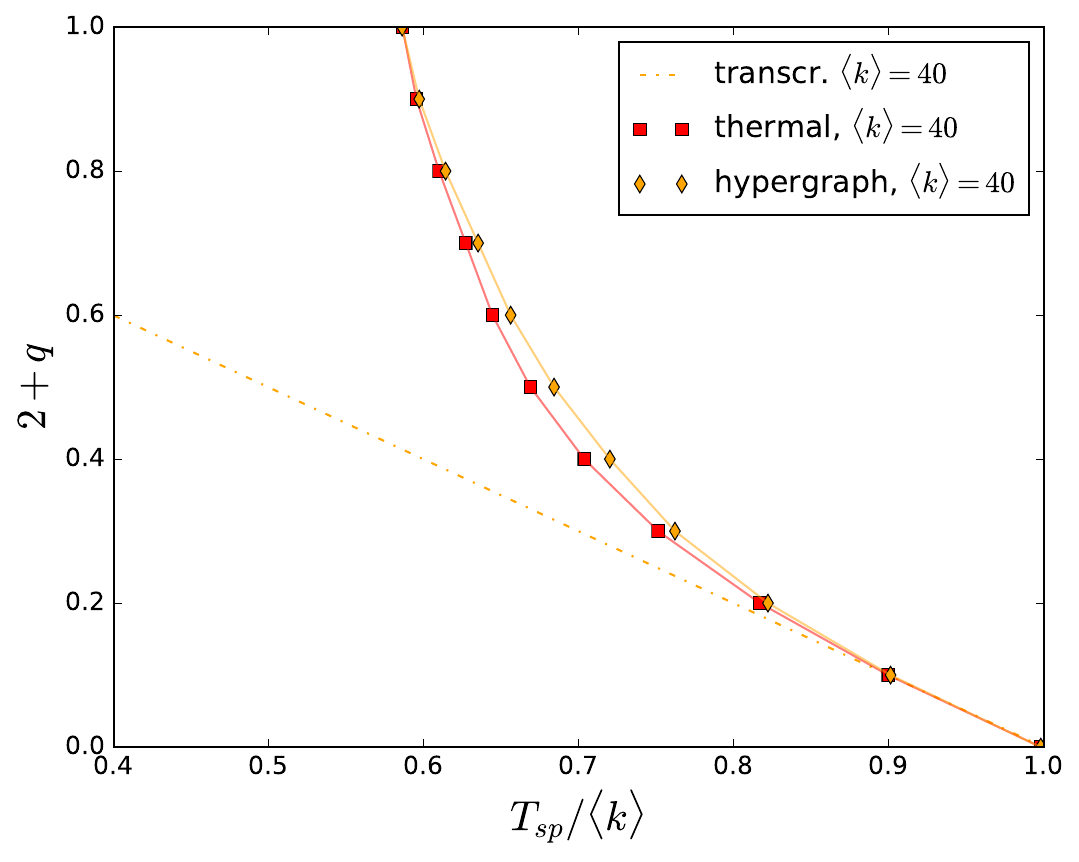}
    \includegraphics[width=0.245\linewidth]{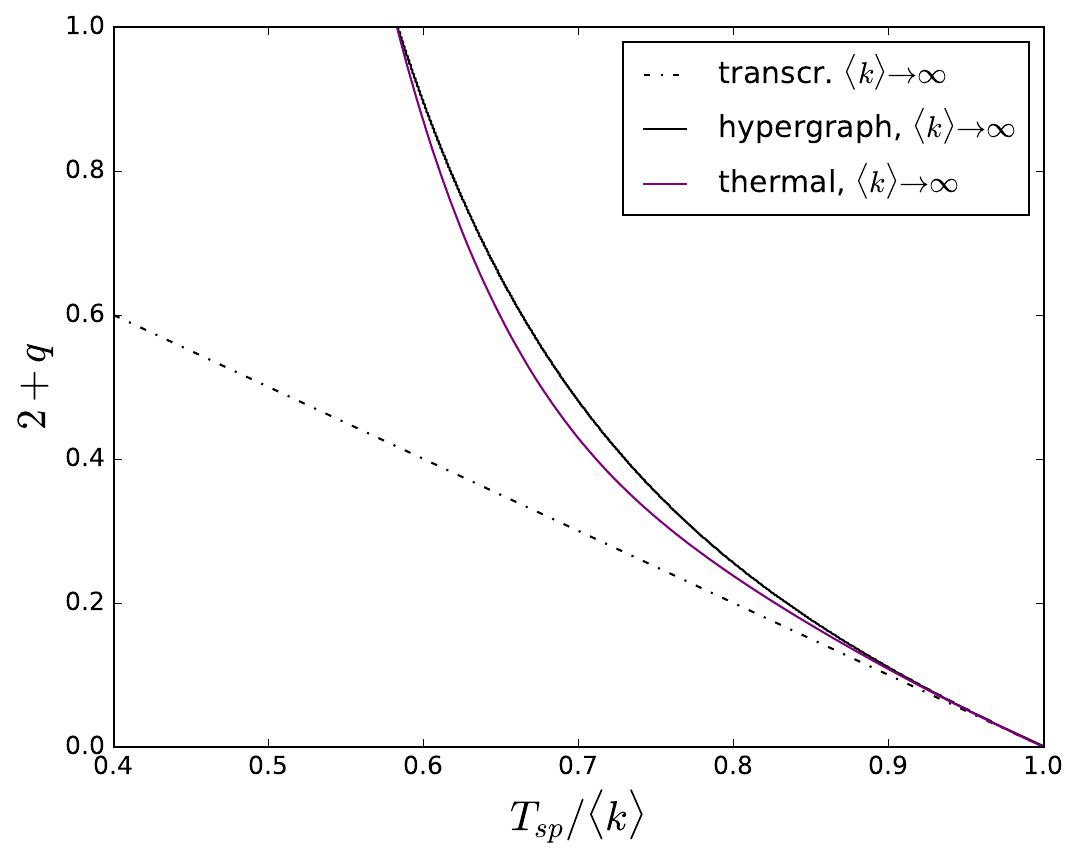}\\
    \includegraphics[width=0.245\linewidth]{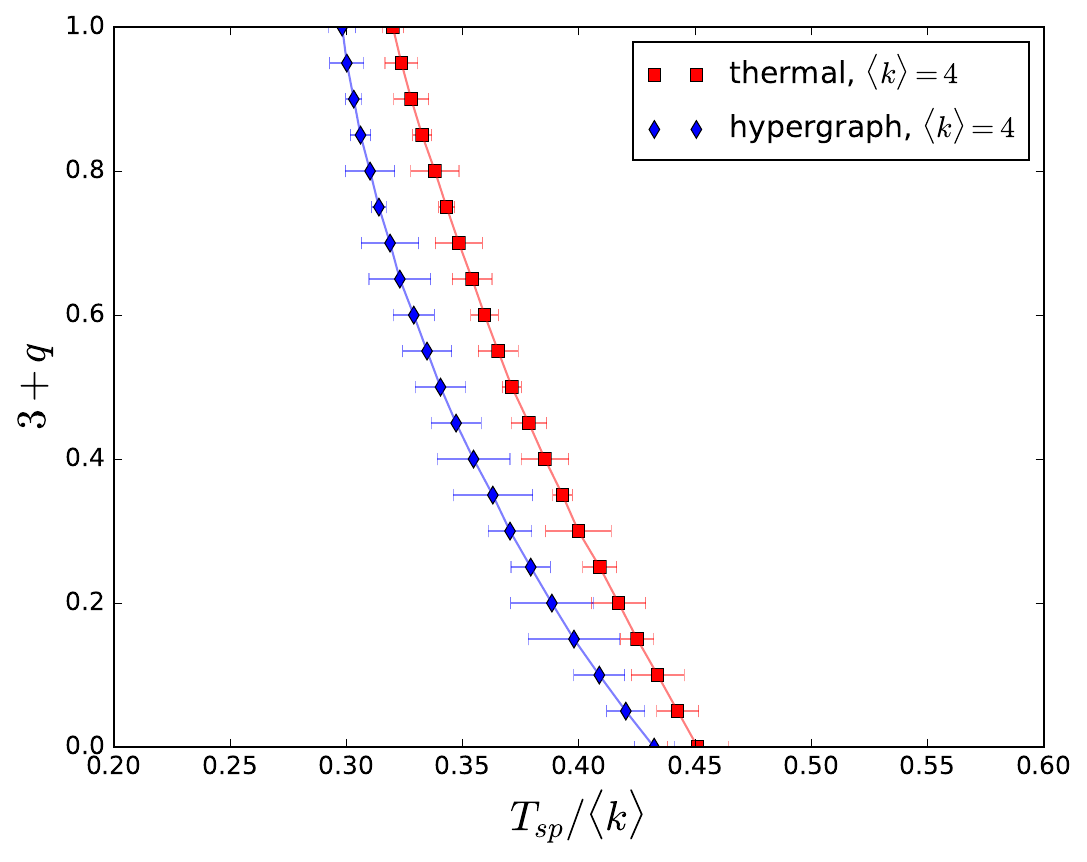}
    \includegraphics[width=0.245\linewidth]{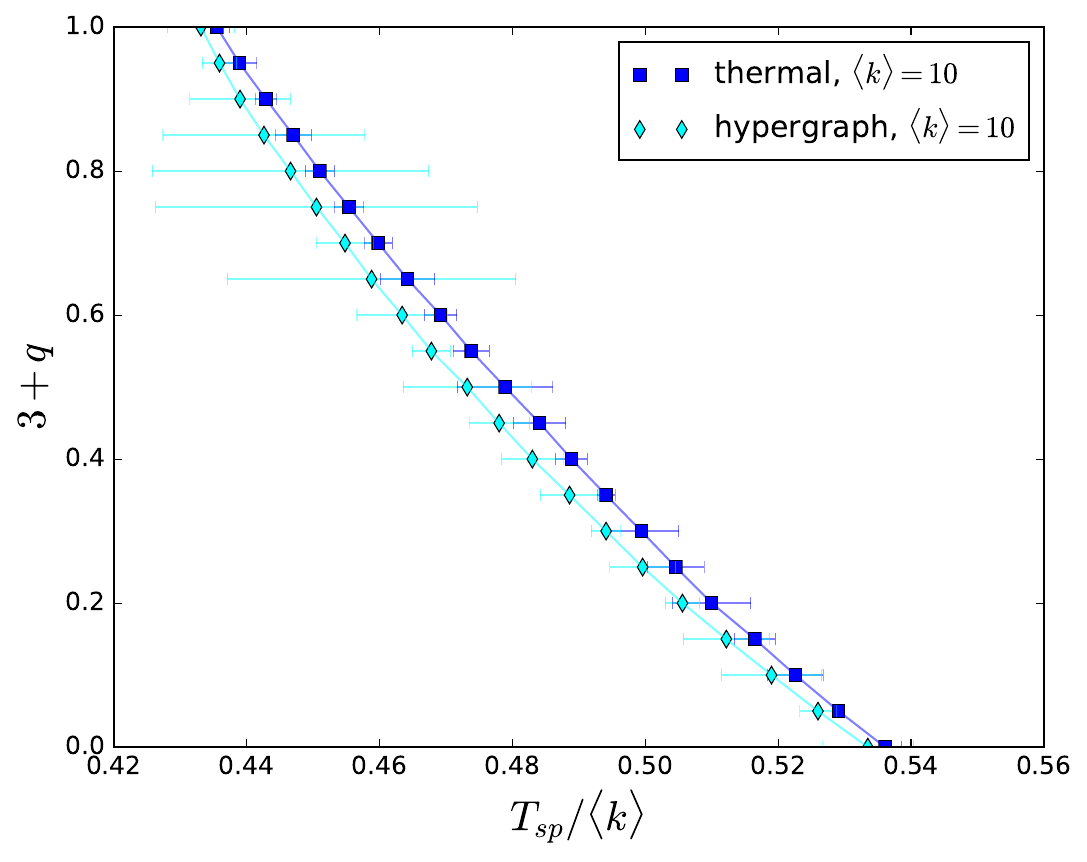}
    \includegraphics[width=0.245\linewidth]{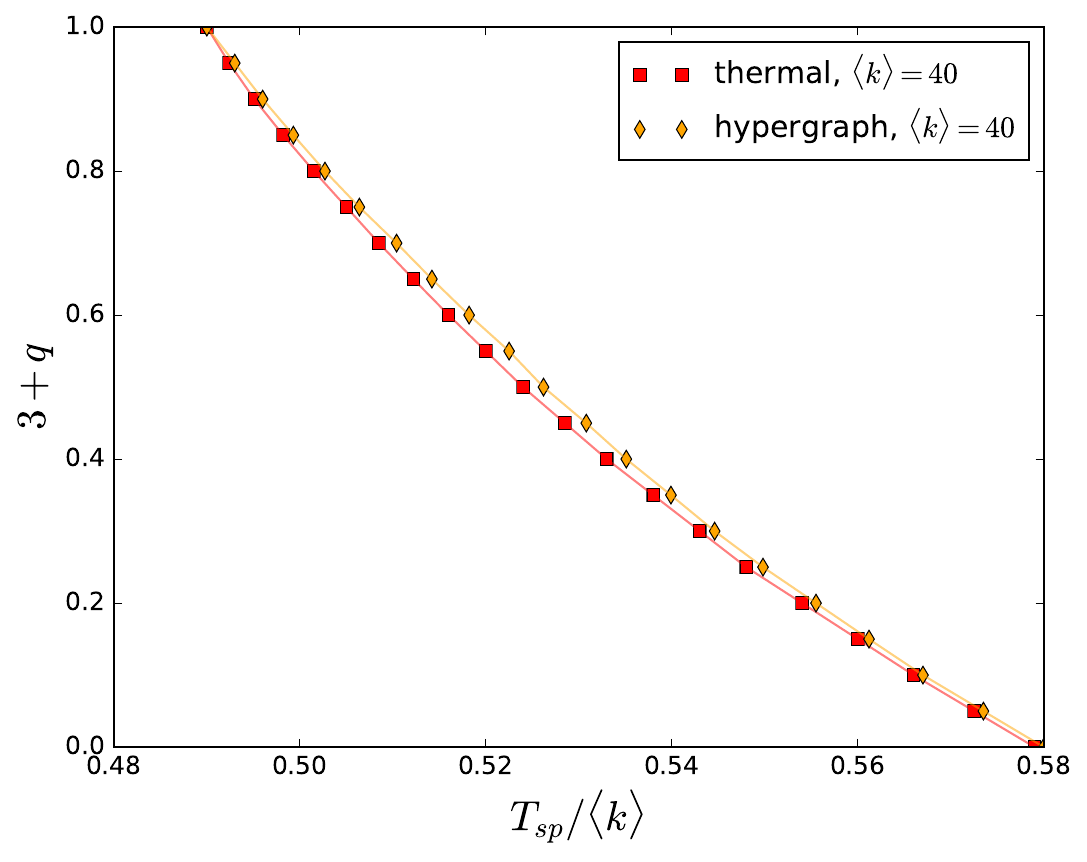}
    \includegraphics[width=0.245\linewidth]{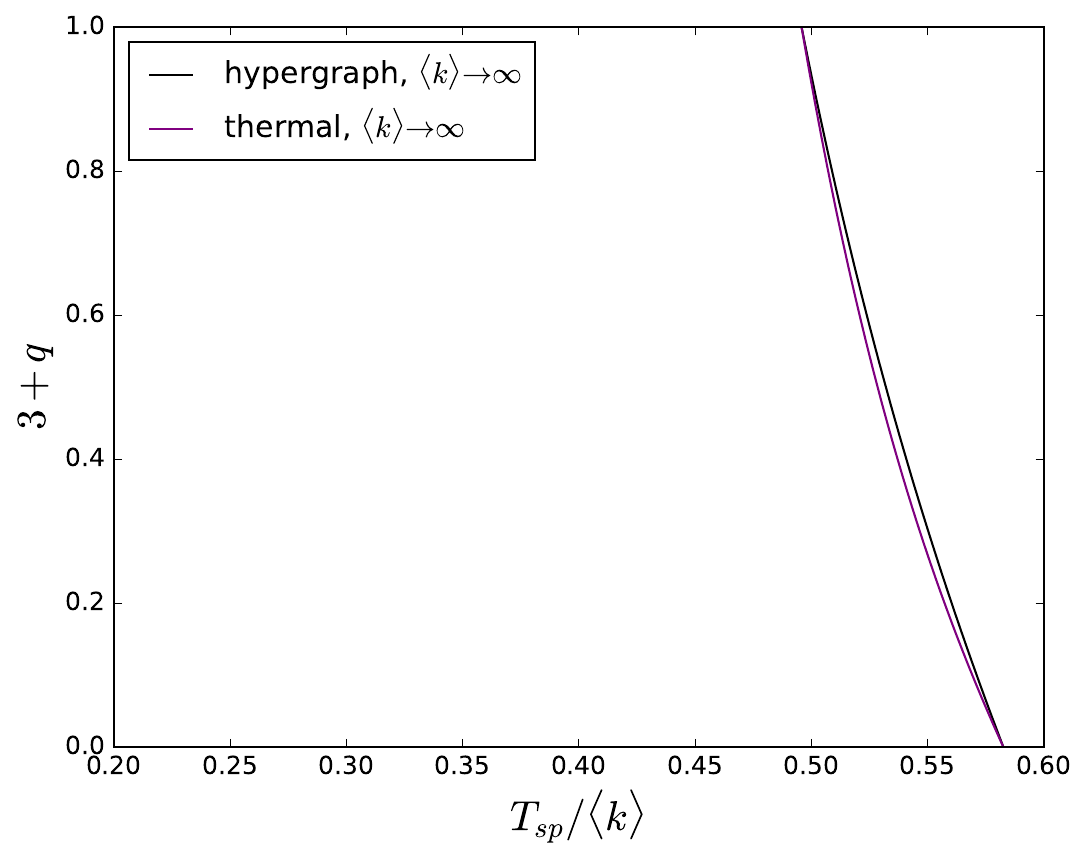}
            \caption{\small \textbf{Bulk-melting comparison of random $(K+q)$- vs. $(M,q)$- spin models.} (Color online) Ferromagnetic spinodal thresholds obtained via population dynamics (PD) for random ferromagnetic $(K+q)$-spin models and via interdependent population dynamics (iPD) for $M$ partially-coupled ferromagnetic spin networks. Hypergraph results were obtained via PDs with equilibration time $T_{eq}=10^4$ and a networked population of $N=10^4$ nodes. Interdependent network results were obtained via iPDs with equilibration $NOI=10^2$, equilibration time for each magnetization step $T_{eq}=10^3$ and networked populations of $N=10^4$ nodes. From the top left to the bottom right: $(2+q)$ vs.\ $(2,q)$ for $\langle k\rangle=4,10,40$ and for the fully MF limit $\langle k\rangle\to\infty$, and $(3+q)$ vs.\ $(3,q)$ for $\langle k\rangle=4,10,40$ and for the fully MF limit $\langle k\rangle\to\infty$. Notice the spinodals' inversion occurring at connectivities above $\langle k\rangle=10$ in both the $K=2$ ($M=2$) and $K=3$ ($M=3$) cases. }
        \label{fig:S3}\vspace*{-0.2cm}
\end{figure}

\indent
$\bullet$ \emph{\bf Bulk-melting comparison: $(K+q)$ vs.\ $(M,q)$ interdependent spin models}. 
In the main text we have showed that, for $M$ fully-coupled randomly interdependent ER networks of Ising spins, the interdependent population dynamics (iPDs) algorithm (see Methods) yields bulk-melting spinodals (Fig.~\ref{fig:2}\textbf{a},\textbf{b}) in agreement with those for $K$-spin models (Fig.~\ref{fig:S1}) with $K=M+1$. 
We have further observed that the mapping breaks down when adiabatically cooling the system from its paramagnetic phase---we did not find striking evidences of a spin-glass phase at low temperatures due to the very high, $\gamma=\bar{k}^M$, density of hyper-edges generated by the interdependent links. 
In partial interdependent spin networks, the mapping is broken also for the bulk-melting spinodal thresholds, as shown for all the studied cases (i.e.\ $\bar{k}=4,10,40$ and $\bar{k}\to+\infty$, for $M=2$ and $M=3$ layers) summarized in Fig.~\ref{fig:S3}. Differences between the bulk-melting spinodals are reported in inset to Fig.~\ref{fig:3}\textbf{c} in the main text.
Notice that, surprisingly, while at low $\bar{k}$ the hypergraph's bulk-melting spinodals are always below the fully-interdependent ones, at larger values the opposite occurs with a threshold inversion occurring around the average connectivity $\bar{k}=10$. 
In fact, the analytical curves obtained in the fully mean-field (MF) limit, i.e.\ for $\bar{k}\to\infty$ (Eqs.~\eqref{eq:M8}, \eqref{eq:M9} in the Methods) highlight the origin behind this discrepancy: in the hypergraph mixture the spins can be simultaneously part of a $K$-tuples and $(K+1)$-tuples, while in partially interdependent systems this never occurs since interdependent spins see only incoming higher-order interactions and independent spins always see only incoming hyperedges of lower order. 
Differently put, in interdependent spin networks, one might identify a community of independent spins and another one of interdependent spins, while the two are completely shuffled in the $(K+q)$-spin mixture. 

\section{Vanishing tricritical point in the $(2+q)$-case}\label{sec:S3}
Let us recall that, in the random $(2+q)$-spin case, the cavities' first moment is given by
\begin{equation}\label{eq:S21}
\langle h\rangle=\frac{1}{2\beta}\bigg(\bar{k}_2(1-q)\int\dd hP(h)\mathpzc{g}_2(\beta; h)+\bar{k}_3q\iint \dd h\dd h'P(h)P(h')\mathpzc{g}_3(\beta; h, h')\bigg). 
\end{equation}
\noindent 
To locate the tricritical points, let us perturb the cavity fields distribution around the RS-paramagnetic solution, i.e.\ $P(h)=\delta(h)+\mu(h)$ with $\mu(h)$ a small distributional perturbation. 
Further recall that $\mathpzc{g}_2(\beta; h)|_{h=0}=0$ and, likewise, $\mathpzc{g}_{3}(\beta; h, h')|_{h=0}=\mathpzc{g}_3(\beta; h,h')|_{h'=0}=\mathpzc{g}_3(\beta; h, h')|_{h,h'=0}=0$. 
In light of the above, Eq.~\eqref{eq:S21} can be expanded as follows:
\begin{displaymath}
\begin{aligned}
\langle h\rangle_\mu\equiv\int\dd h\mu(h)h=\bar{k}_2(1-q)&\,\Big(\langle h\rangle_\mu\mathrm{tanh}\beta+\mathpzc{f}_1(\beta)\langle h^3\rangle_\mu +\mathcal{O}(\langle h^5\rangle_\mu)\Big)+\\
&\,+\bar{k}_3q\Big(\beta\mathrm{tanh}\beta\langle h\rangle^2_\mu+\mathpzc{f}_2(\beta)\langle h\rangle_\mu\langle h^3\rangle_\mu+\mathpzc{f}_3(\beta)\langle h^3\rangle_\mu^2+\mathcal{O}(\langle h^4\rangle_\mu^2)\Big),
\end{aligned}
\end{displaymath}
\noindent 
where $\mathpzc{f}_1(\beta)\equiv -\frac{1}{3}\beta^2\mathrm{tanh}\beta(1-\mathrm{tanh}^2\beta)$, $\mathpzc{f}_2(\beta)\equiv-\frac{2}{3}\beta^3\mathrm{tanh}\beta$ and $\mathpzc{f}_3(\beta)\equiv -\frac{1}{9}\beta^5(1-2\mathrm{cosh}2\beta)\mathrm{sech}^2\beta\mathrm{tanh}\beta$.\\ 
\indent 
Inspecting linear terms yields the critical surface determined by Eq.~\eqref{eq:S20}. 
We are therefore interested in second order terms below $\beta_{tc}=\mathrm{atanh}(1/c_{tc})$, with $c_{tc}\equiv\bar{k}_2(1-q_{tc})$ which make the OP to undergo a finite jump. 
Following a reasoning analogous to the one in Ref.~\cite{monasson1998tricritical}, let us write $\mu(h)=\mathpzc{x}\eta(h)$ with $\mathpzc{x}\ll1$ and let us notice that $\mathcal{M}_{sp}$ is finite if and only if $\langle h\rangle_\mu=\mathpzc{x}\langle h\rangle_\eta\equiv\mathpzc{x}\int h\eta(h) \dd h$ is finite. 
Noticing that this occurs if $\eta(h)\to+\infty$ as $\mathpzc{x}\to0^+$, we can find the parameters at which the perturbation $\eta$ is singular. 
Tracking the same infinitesimal orders, we thus find: 
\begin{equation}\label{eq:S22}
\mathcal{O}(\mathpzc{x}^2)\quad:\qquad 
0=-c\Big(1-\frac{1}{c^2}\Big)\langle h\rangle_\eta+\frac{1}{c}\bar{k}_3\mathrm{atanh}(1/c) \langle h\rangle_\eta^2.
\end{equation}
\noindent 
Neglecting the trivial solution $\langle h\rangle_\eta=0$, Eq.~\eqref{eq:S22} yields 
$$
\langle h\rangle_\eta=\frac{c^2-1}{q\bar{k}_3\mathrm{atanh}(1/c)}\bigg|_{c\equiv \bar{k}_2(1-q)},
$$
\noindent 
so that $\langle h\rangle_\eta\to+\infty$ iif $q\to0^+$, implying that the system has a vanishing tricritical point $q_c\to0^+$ for any $\bar{k}_2,\bar{k}_3$. 
In the inset of Fig.~\ref{fig:3}\textbf{d} we corroborate this analytical result via iPDs and PDs in, respectively, the randomly interdependent $(2,q)$-spin model with local and global interactions and in the random $(2+q)$-spin model. 
The result agrees with the fully-MF behaviour of both hypergraph and interdependent models, where also $q_c^{MF}\to0^+$ (see  $\S$\ref{sec:S4}). 

\section{Ground states and interdependent percolation}\label{sec:S4}
A central result of our work is the mapping of interdependent percolation~\cite{bianconi2018multilayer} in $K$ partially interacting random graphs---to which we shall refer hereafter as to random $(K,q)$-\textsc{dep}---onto the equations governing the zero-temperature ground state (GSs) of random $(K+q)$-spin models with ferromagnetic interactions. 
The significance of this mapping lies on the long-celebrated discovery that random $(K+q)$-spin GSs are in one-to-one correspondence~\cite{kirkpatrick1983optimization,monasson1992relation,monasson1999determining,mezard2002analytic,krzakala2016statistical} with random $(K+q)$-\textsc{xor-sat}, a paradigmatic class of random constraint satisfaction problems (CSPs). 
This connection has enabled us to determine the free-energy density (i.e.\ Eq.~\eqref{eq:6} and Eq.~\eqref{eq:M6}) of interdependent percolation on randomly coupled ER graphs by considering the configurational entropy of the $T=0$ GSs of their computational complexity analogue, thus allowing to find the boundaries in the phase space of the structural metastability regime (Fig.~\ref{fig:4}\textbf{a},\textbf{b}) for this class of interacting systems. 
In this section we review the main calculations underlying the formal isomorphism between random $(K,q)$-\textsc{dep} and random $(K+q)$-\textsc{xor-sat} problems.\\

\indent 
$\bullet$ {$K$-\textsc{xor-sat}/$M$-\textsc{dep}}. In the $T=0$ limit (i.e.\ $\beta\to+\infty$), the RS-solution to Eqs.~\eqref{eq:S14} can be found analytically by noticing that the cavity fields take only integer values, i.e.\ their distribution has the form $P(h)=\sum_{\ell\in\mathbb{N}_0}r_{\ell}\delta(h-\ell)$. 
Moreover, as $\beta\to\infty$, we find that the integral kernel $\mathpzc{T}_K$ in Eq.~\eqref{eq:S14} is a piecewise function given by 
\begin{equation}\label{eq:S23}
\mathpzc{T}_K\big(h_1,\dots,h_{K-1}\big)\equiv\lim_{\beta\to+\infty}\frac{1}{2\beta}
\ln\frac{\alpha_{K,+}(\beta;h_{1},\dots,h_{K-1})}{\alpha_{K,-}(\beta;h_{1},\dots,h_{K-1})}=
\mathrm{min}\big\{1,h_1,\dots,h_{K-1}\big\}\mathrm{Sg}\left(\prod_{\ell=1}^{K-1} h_\ell\right),
\end{equation}
\noindent 
where we used the fact that $\mathrm{cosh}\beta h\to\frac{1}{2}e^{\beta|h|}$ as $\beta\to\infty$. 
The limiting behaviour in Eq.~\eqref{eq:S23} can be obtained by induction starting from the solutions obtained for $K=2,3,4$. 
In particular, one finds: 
\begin{displaymath}
\begin{aligned}
\mathpzc{T}_2\big(h\big)&\,\equiv\lim_{\beta\to+\infty}\frac{1}{2\beta}
\ln\frac{\mathrm{cosh}\beta (h+1)}{\mathrm{cosh}\beta (h-1)}=\lim_{\beta\to\infty}\frac{1}{2\beta}\ln\frac{e^{\beta|h+1|}}{e^{\beta|h-1|}}=\frac{1}{2}\big(|h+1|-|h-1|\big)=\mathrm{Sg}\big(h\big)\mathrm{min}\{1,|h|\},\\
\mathpzc{T}_3\big(h',h\big)&\,\equiv\lim_{\beta\to+\infty}\frac{1}{2\beta}
\ln\frac{e^{\beta|h'+h|}+e^{\beta|h'-h|-2\beta}}{e^{\beta|h'-h|}+e^{\beta|h'+h|-2\beta}}=\mathrm{Sg}\big(hh'\big)\mathrm{min}\{1,|h|,|h'|\},\\
\end{aligned}
\end{displaymath}
\noindent 
and similarly for $K=4$. 
Inserting the limiting kernel, Eq.~\eqref{eq:S23}, into Eq.~\eqref{eq:S14} one finds that the configuration of cavity fields corresponding to the $T=0$ paramagnetic phase is characterized by the condition 
\begin{equation}\label{eq:S24}
\sum_{\ell=1}^d\mathrm{Sg}\bigg(\prod_{\mu=1}^{K-1}h_{\mu,\ell}\bigg)=0.
\end{equation}
\noindent 
Recalling that the sum runs over the number, $d=|\partial i|$, of hyperedges incident on a randomly chosen node $i$, and that $\{h_{\mu,\ell}\}_{\mu=1,\dots,K-1}$ is the set of cavity fields acting on that node from each of its hyperedges, one can conclude that the node's magnetic state is paramagnetic (i.e.\ malfunctioning), $h=0$, if {\em at least one} of the $K-1$ neighbouring cavity fields vanishes. 
Notice that the latter can be analogized to the functional constraint imposed by interdependent couplings in interdependent percolation, setting a local logical \textsc{and} gate between the cavity fields' states.\\
\indent 
The combinatoric factor counting the number of configurations of cavity fields satisfying Eq.~\eqref{eq:S24} can be found probabilistically by considering, e.g.\ the contributions to the paramagnetic factor $r_0$ in Eq.~\eqref{eq:S14}.
Being the model purely random, one can conclude that $1-(1-r_0)^{K-1}$ represents the probability that {\em at least one} cavity field around a randomly chosen node is zero. 
Assuming hence that all cavities are statistically independent (i.e.\ the underlying structure is locally tree-like), it follows that $(1-(1-r_0)^{K-1})^d$ is (w.h.p.) the average number of cavity configurations 

\begin{wrapfigure}[33]{l}{0.42\textwidth}
	\centering 
		\includegraphics[width=0.41\textwidth]{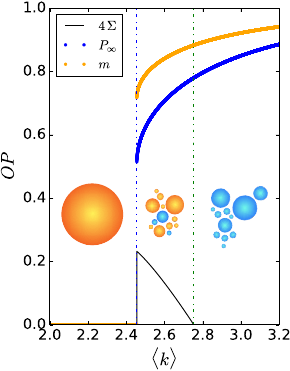}
		\caption{\small \textbf{Complexity in random $3$-\textsc{xor-sat}/$2$-\textsc{dep}.} (Color online) Observable order parameters (OPs) depicting the hyperbackbone, $m$, (see Eq.~\eqref{eq:S25}) and to the fraction of hard fields, $P_\infty$, in the 1RSB solution at $m=1$ (see Eq.~\eqref{eq:S28}) as functions of the average connectivity $\langle k\rangle$. Notice that $P_\infty$ is the observable OP of interdependent percolation in random $2$-\textsc{dep}, i.e.\ the mutual giant component (\textsc{mgc}). The black line shows the complexity function, Eq.~\eqref{eq:S27} with $K=2$, magnified by an arbitrary factor.}
	\label{fig:S4}
\end{wrapfigure}
\noindent 
leading to $h=0$ in Eq.~\eqref{eq:S14} and soo the contributions to $r_0$ satisfy the self-consistent equation 
$$
r_0=e^{-\langle k\rangle}\sum_{d\in\mathbb{N}_0}\frac{\langle k\rangle^d}{d!}\Big(1-(1-r_0)^{K-1}\Big)^d=e^{-\langle k\rangle(1-r_0)^{K-1}}. 
$$
\noindent 
Since $h=0$ corresponds to a paramagnetic state, the complement to $r_0$, i.e.\ $m\equiv\sum_{\ell\geq1}r_\ell\delta(h-\ell)=1-r_0$ must describe the fraction of ferromagnetic nodes. 
The zero-temperature $K$-spin average magnetization is therefore
\begin{equation}\label{eq:S25}
m=1-\mathrm{exp}\Big\{-\langle k\rangle m^{K-1}\Big\},\qquad \forall\,K\geq2.
\end{equation}
\indent 
As highlighted in the main text, the largest root of Eq.~\eqref{eq:S25} (see Fig.~\ref{fig:S4}) corresponds to the $2$-core of the underlying random hypergraph or, equivalently, to the fraction of frozen sites in random $K$-\textsc{xor-sat}~\cite{mezard2003two}. 
The sudden emergence of the $2$-core at $\langle k\rangle_{sp}$ (e.g.\ $\langle k\rangle_{sp}=2.4554(1)$ for $K=3$) corresponds to the spontaneous breaking of ergodicity in the system's phase space, resulting in a clustering phenomenon: for average connectivities $\langle k\rangle<\langle k\rangle_{sp}$ all the GSs for a unique cluster (orange disk, Fig.~\ref{fig:S4}), while for $\langle k\rangle>\langle k\rangle_{sp}$ they split into an exponentially large (in $N$) number of clusters (orange disks, Fig.~\ref{fig:S4}), followed by the proliferation of metastable states (blue disks, Fig.~\ref{fig:S4}). 
That is, the spinodal threshold $\langle k\rangle_{sp}$ corresponds to the onset of the computational complexity in random $K$-\textsc{xor-sat}, since above this threshold classical greedy search algorithms (e.g.\ \textsc{walk-sat}) remain exponentially often trapped into metastable minima.
Two relevant observations are therefore in order.

\begin{wrapfigure}{r}{0.45\textwidth}
	\centering 
		\includegraphics[width=0.45\textwidth]{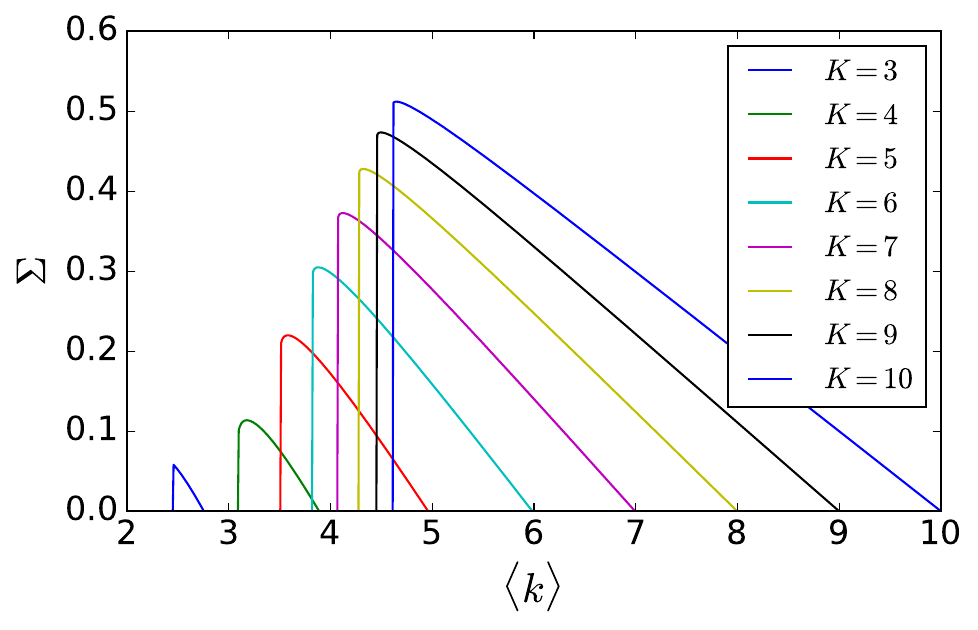}
		\caption{\small \textbf{GSs configurational entropy.} (Color online) The curves satisfy Eq.~\eqref{eq:S27} for increasing values of $K$. Notice that $\langle k\rangle_{cx,K}\sim K$ as $K$ grows.}
	\label{fig:S5}
\end{wrapfigure}

\textbf{1}) \underline{\textsc{sat-unsat}}. The onset of complexity at $\langle k\rangle_{sp,K}$ is followed by a second relevant phase transition in the space of $K$-\textsc{xor-sat} solutions, occurring at $\langle k\rangle_{cx,K}>\langle k\rangle_{sp,K}$, which locates the limit of satisfiability~\cite{mezard2009information}. 
This critical point corresponds, in fact, to the zero-temperature coexistence threshold of the random ferromagnetic $K$-spin model where a sharp first-order transition (for any $K\geq3$) occurs. 
In particular, it can be shown~\cite{ricci2001simplest} that this corresponds to the percolation of hyperloops in the underlying hypergraph, whose emergence percolate frustration in the system.  
Stated equivalently, $\langle k\rangle_{cx}$ is the location of the $T=0$ Gardner phase transition, while $\langle k\rangle_{sp}$ is the $T=0$ limit of the Kauzman line~\cite{semerjian2008freezing,krzakala2011melting1,krzakala2016statistical}.\\
\indent 
The location of $\langle k\rangle_{cx}$ can be obtained~\cite{franz2001exact, leone2001phase, mezard2003two} either via the $T=0$ GS free-energy $\mathcal{F}$ of the system or by computing the GSs configurational entropy $\Sigma$ (a.k.a.\ the system's {\em complexity}). 
The former can be found straightforwardly since the GS energy of the ferromagnetic model is zero, so that the GS free-energy density, $\mathcal{F}$, is simply given by 
\begin{equation}\label{eq:S26}
-\frac{\mathcal{F}_K[r_0;\langle k\rangle]}{\ln2}=r_0\big(1-\ln r_0\big)-\frac{\langle k\rangle}{K}\Big[1-(1-r_0)^K\Big]\
\end{equation}
\noindent 
where $r_0$ is the largest positive root of $r_0=\mathrm{exp}\{-\langle k\rangle(1-r_0)^{K-1}\}$. 
The complexity, $\Sigma$, corresponds to the logarithm of the number of clusters per spin~\cite{franz2001exact} and it can be calculated in the 1RSB (i.e.\ one-step replica symmetric breaking) formalism~\cite{monasson1995structural}. 
In random $K$-\textsc{xor-sat}, this can be done equivalently (thanks to the triviality of the paramagnetic phase of the model), by noticing that $\Sigma$ corresponds to the $T=0$ GS entropy of the ferromagnetic phase:
\begin{equation}\label{eq:S27}
\frac{\Sigma_K[m;\langle k\rangle]}{\ln2}=m\Big[1+\langle k\rangle(1-m)^{K-1}\Big]-\frac{\langle k\rangle}{K}\Big[1+\langle k\rangle(1-m)^{K}\Big]
\end{equation}
\noindent 
where $m$ is the largest root solving Eq.~\eqref{eq:S25}. 
In Fig.~\ref{fig:S4} we show the behaviour of $\Sigma_3$ (rescaled by an arbitrary factor to magnify its shape) in comparison to the thresholds identified by system's hyper-backbone. 
Notice that, at $\langle k\rangle_{sp,3}=2.4554(1)$, the complexity function jumps to a finite value, indicating the onset of computational hardness, whilst it vanishes at a certain connectivity threshold, $\langle k\rangle_{cx,3}$, locating instead the \textsc{sat-unsat} transition; for $K=3$, the latter is given by $\langle k\rangle_{cx,K}=2.7538(0)$.
Fig.~\ref{fig:S5} displays the various shapes and thresholds of the complexity $\Sigma_K$ for increasing values of $K$: notice that, as highlighted in the main text, while the spinodal thresholds $\langle k\rangle_{sp,K}\sim\ln K$ at large $K$, the \textsc{sat-unsat} thresholds grow linearly with the order $K$ of the problem, i.e.\ $\langle k\rangle_{cx,K}\sim K$ for $K\gg1$. 

\textbf{2}) \underline{\textsc{Order parameters}}. When comparing the above picture with interdependent percolation, one can readily notice that Eq.~\eqref{eq:S25} is precisely the self-consistent equation governing the behaviour of its branching order parameter (OP) in random $M$-\textsc{dep}, called $1-f$ in Refs.~\cite{buldyrev-nature2010, parshani-prl2010}. 
This corresponds to the fraction of edges reached by following the end of a randomly chosen node that lead to the {\em observable} OP in $M$-\textsc{dep}, i.e.\ the mutual giant component (\textsc{MGC}), $P_\infty$. 
For randomly coupled ER networks it is straightforward to show~\cite{bianconi2018multilayer} that this observable satisfies the self-consistent equation 
\begin{equation}\label{eq:S28}
P_\infty=\Big(1-e^{-\langle k\rangle P_\infty}\Big)^{M},\qquad M\geq2,
\end{equation}
\noindent
whose relation with the OP for the fraction of frozen sites, $m$, in random $K$-\textsc{xor-sat} is $m=P_\infty^M$ with $M\equiv K-1$. 
When studying within the RSB formalism to the clustering transition in random $K$-\textsc{xor-sat}~\cite{krzakala2007gibbs}, however, one can show that the underlying fundamental OP of the process is given by the fraction of hard cavity fields, $P_\infty$, whose behaviour can be described~\cite{zdeborova2007phase, semerjian2008freezing} via the 1SRB solution with Parisi parameter $m=1$, which coincides precisely with the \textsc{mgc} in random $M$-\textsc{dep}, i.e.\ Eq.~\eqref{eq:S28}. 
In other words, one can conclude that the {\em observable} OP in random $M$-\textsc{dep} coincides with the fraction of hard fields in random $K$-\textsc{xor-sat}, i.e.\ the 1RSB solution with $m=1$ on random trees. 
The very same equations characterize the freezing threshold in other paradigmatic CSPs, like $Q$-\textsc{col}oring on random graphs~\cite{zdeborova2007phase} and in the problem of tree-reconstruction~\cite{mezard2006reconstruction}.
In this respect, {\em interdependent percolation algorithms on randomly coupled ER graphs can be adopted as an unconventional whitening protocol to study the statistics of hard fields at the freezing transitions of different paradigmatic CSPs}.
Let us finally stress that, similar results are expected to be found for any local tree-like structure, which we leave here as an hypothesis i.e.\ random $K$-\textsc{xor-sat} on scale-free networks and on random regular graphs maps onto random $M$-\textsc{dep} between scale-free and random regular networks having the {\em same} structural parameters. 
Interdependent percolation on random mixed structures (e.g.\ scale-free graphs randomly coupled with ER graphs, etc.) would instead offer the solutions for novel classes of random \textsc{xor-sat} problems. \vspace*{+0.25cm}

\indent
$\bullet$ {$(K+q)$-\textsc{xor-sat}/$(M,q)$-\textsc{dep}}. 
Following a reasoning analogous to the one adopted for the random ferromagnetic $K$-spin case, we can obtain analytical results about the $(K+q)$-spin model zero-temperature GS by searching for distributions $P(h)=\sum_{\ell\in\mathbb{N}_0}r_\ell\delta(h-\ell)$ since the cavity fields take only integer values when $\beta\to\infty$. 
Inserting Eq.~\eqref{eq:S24} into the saddle-point equations in Eq.~\eqref{eq:S18}, we can then keep track of the contributions to $r_0$ by considering the possible combinations of cavity fields that make $h=0$. 
Since the fraction $q$ of $K$-tuples and the fraction of $(K+1)$-tuples are statistically independent, the cavity field $h$ acting on a randomly chosen node vanishes if at least $K-1$ of its $d$ incident $K$-tuples has a zero cavity field {\em and} at least one of its $\ell$ incident $K$-tuples has a zero cavity field. 
These conditions can be simply written as:
\begin{displaymath}
\begin{aligned}
r_0=e^{\gamma(K+q)}&\,\sum_{d\in\mathbb{N}_0}\frac{\bar{k}_K^d}{d!}(1-q)^d\Big(1-\big(1-r_0\big)^{K-2}\Big)^d
\sum_{\ell\in\mathbb{N}_0}\frac{\bar{k}_{K+1}^\ell}{\ell!}q^\ell\Big(1-\big(1-r_0\big)^{K-1}\Big)^\ell\\
&\,=\bigg[e^{-\bar{k}_K(1-q)}\sum_{d\in\mathbb{N}_0}\frac{\big(\bar{k}_K(1-q)\big)^d}{d!}\Big(1-(1-r_0)^{K-2}\Big)^d\bigg]
\bigg[e^{-\bar{k}_{K+1}q}\sum_{\ell\in\mathbb{N}_0}\frac{\big(\bar{k}_{K+1}q\big)^\ell}{\ell!}\Big(1-(1-r_0)^{K-1}\Big)^\ell\bigg]\\
&\,=\mathrm{exp}\bigg\{-\bar{k}_K(1-q)(1-r_0)^{K-2}-\bar{k}_{K+1}q(1-r_0)^{K-1}\bigg\}.
\end{aligned}
\end{displaymath}
\noindent 
This equation generalizes Eq.~(12) in Ref.~\cite{leone2001phase} to the $(K+q)$-spin case. 
Similarly to the above, noticing that $m=1-r_0$ is the $T=0$ average magnetization of the system, we find the self-consistent equation
\begin{equation}\label{eq:S29}
m=1-\mathrm{exp}\Big\{-(1-q)\bar{k}_Km^{K-2}-q\bar{k}_{K+1}m^{K-1}\Big\},\qquad K\geq3.
\end{equation}
\noindent 
When $\langle k\rangle=\bar{k}_K=\bar{k}_{K+1}$, Eq.~\eqref{eq:S29} is precisely the one characterizing the branching OP of interdependent percolation~\cite{parshani-prl2010} in $M$ randomly and partially-coupled ER graphs (i.e.\ random $(M,q)$-\textsc{dep}). 
For a review, we refer the interested reader to Ref.~\cite{gao-naturephysics2012} or to the monograph in Ref.~\cite{bianconi2018multilayer}. 
Similarly to the $K$-\textsc{xor-sat} case, Eq.~\eqref{eq:S29} characterizes the onset of computational complexity $\langle k\rangle_{sp,K,q}$ in the random $(K+q)$-\textsc{xor-sat} problem. 
The corresponding \textsc{sat-unsat} threshold, $\langle k\rangle_{cx,K,q}$, can be computed by analyzing the landscape of the $T=0$ GS's free-energy 
\begin{displaymath}
-\frac{\mathcal{F}_{K,q}[r_0; \langle k\rangle_{K},\langle k\rangle_{K+1}]}{\ln2}=r_0(1-\ln r_0)-\frac{\langle k\rangle}{K}q\big[1-(1-r_0)^{K}\big]-\frac{\langle k\rangle}{K-1}(1-q)\big[1-(1-r_0)^{K-1}\big].
\end{displaymath}
\noindent 
In the main text, we have performed this analysis by searching the values of $\langle k\rangle$ that make the \textsc{sat} (i.e.\ the paramagnetic) and the \textsc{unsat} (i.e.\ the ferromagnetic) phases {\em iso-energetic} (see Fig.~\ref{fig:4}\textbf{a}, inset).\vspace*{+0.25cm}

\indent
$\bullet$ {\textsc{$(M,q)$-dep}-spin GSs}. In the main text, we showed that our random $M$-\textsc{dep}-spin model maps its bulk-melting ferromagnetic spinodals (see Fig.~\ref{fig:S1}) onto those of random $K$-spin models, with $K=M+1$. 
We showed that this exact mapping breaks down and it becomes only qualitative when considering partial interdependent couplings (Fig.~\ref{fig:S3}) and we have characterized the thresholds' separation (main text, Fig.~\ref{fig:2}\textbf{c} inset) between the two models. \\
\indent 
We then expect that $(M,q)$-\textsc{dep}-spin models and $(K+q)$-spin models host, generally, different zero-temperature GSs configurations for intermediate values of $q\in(0,1)$, i.e.\ {\em different percolation processes underly the stability of their higher-order structures}. 
To verify this statement and compute the percolation equations characterizing the $T=0$ GS of the random $(M,q)$-\textsc{dep}-spin model, we can rely on the $s\to1$ limit of its corresponding Potts model~\cite{wu1982potts,dorogovtsev2004potts}. 
Similarly to the procedure proposed in the main text for Ising spins, let us set dependency links between networks of Potts-spins $\sigma_i\in\{1,2,\dots,s\}$ via their spin-flip probabilities. 
In particular, assuming that spin $i$ in network $\mu$ depends on the Potts-magnetic state of node $i'$ in network $\mu'$, we can write an {\em adaptive} Glauber rate~\cite{mendes1991dynamics} 
\begin{equation}\label{eq:S30}
\mathpzc{W}_{i,s_\mu}^{\mu\leftarrow \mu'}\big(\sigma_i\to\alpha_i\big|\mathpzc{x}_{\,\sigma_{i'}}^{\mu'}\big)=
\frac{e^{-\mathcal{J}_{i\leftarrow j}^\mu\textcolor{blue}{\mathpzc{x}_{\,\sigma_{i'}}^{\mu'}}\mathpzc{x}_{\,\alpha_i}^\mu}}{\sum_{\alpha_i=1,\dots,s_\mu}e^{-\mathcal{J}_{i\leftarrow j}^\mu\textcolor{blue}{\mathpzc{x}_{\,\sigma_{i'}}^{\mu'}}\mathpzc{x}_{\,\alpha_i}^\mu}},\qquad 
\mathpzc{x}_{\,\alpha_i}^\mu\equiv\frac{1}{k_i}\sum_{j=1}^N A_{ij}^{\mu}\delta_{\alpha_i\sigma_j},
\end{equation}
\noindent 
where $\mathpzc{x}_{\,\alpha}^\mu$ is the fraction of spins in the $\alpha$ state around spin $i$ in the $\mu$-th network. 
To ease the interpretation of Eq.~\eqref{eq:S30}, we have highlighted in blue the terms setting interdependent couplings. 
Notice that $\sum_s\mathpzc{x}_{\,\sigma}^\mu=1$, i.e.\ the state of each node is constrained to an $(s-1)$-dimensional manifold in the space of configurations. \\
\indent
The fractions $\mathpzc{x}_{\,\alpha}$ play the role of local OP components and they enlarge our interdependent spin framework by adding the possibility of coupling networks whose local functionality is defined via any of the $s$ Potts' spins modes! 
In isolated systems, one finds that $s-1$ decoupled spin ``transverse'' modes exponentially decay to equilibrium while only one ``longitudinal'' mode (i.e.\ the proper OP) exhibits critical slowing down~\cite{mendes1991dynamics}. 
{\em Interdependent Potts networks, on the other hand, can couple transverse modes in one layer with the longitudinal one in another layer, adding therefore room to novel critical features depending on the mode-to-mode cross-layer coupling considered}.
The general case can be studied in a deterministic way by analyzing the system of coupled Langevin equations
\begin{equation}\label{eq:S31}
\begin{aligned}
\dot{\mathpzc{x}}_{\,\sigma_i}^\mu(t)&\,=\sum_{\alpha_i=1}^{s}\big(1-s_\mu\delta_{\alpha_i\sigma_i}\big)\mathpzc{x}_{\,\alpha_i}^\mu\mathpzc{W}_{i,s_\mu}^{\mu\leftarrow\mu'}\big(\sigma_i\to\alpha_i\big|\mathpzc{x}_{\,\sigma_{i'}}^{\mu'}\big), \qquad i=1,2,\dots,N_\mu,\\
\dot{\mathpzc{x}}_{\,\sigma_{i'}}^{\mu'}(t)&\,=\sum_{\alpha_{i'}=1}^s\big(1-s_{\mu'}\delta_{\alpha_{i'}\sigma_{i'}}\big)\mathpzc{x}_{\,\alpha_{i'}}^{\mu'}\mathpzc{W}_{i',s_{\mu'}}^{\mu'\leftarrow\mu}\big(\sigma_{i'}\to\alpha_{i'}\big|\mathpzc{x}_{\,\sigma_i}^\mu\big),\qquad i'=1,2,\dots,N_{\mu'},
\end{aligned}
\end{equation}
\noindent 
where the local longitudinal mode $\mathpzc{X}_{i}$ and the local transverse modes $\mathpzc{Y}_{\alpha_i}$ can be written via $\mathpzc{x}_{\alpha_i}$ as~\cite{mendes1991dynamics}
$$
\mathpzc{x}_{\alpha_i}=\frac{1}{s}\Big(1+\mathpzc{X}_{i}\big(s\delta_{1\alpha_i}-1\big)\Big)+\mathpzc{Y}_{\alpha_i}.
$$
\indent 
We will study the general case, Eq.~\eqref{eq:S31}, elsewhere. 
Here, we focus on longitudinal-to-longitudinal mode interdependent couplings between Potts-spin networks, in which case the Potts magnetic state of node $i$ is characterized by the local OP $\mathpzc{X}_{\alpha_i}$. 
In the particular case of ER graphs, we can rely on the annealed network approximation, which enables to replace the adjacency matrix of each network, $A_{ij}^\mu$, with the corresponding wiring probability matrix $\mathpzc{p}_{ij}:=\langle A_{ij}\rangle_\mathfrak{G}\simeq \langle k\rangle/N$. 
From the structural perspective, this is equivalent to replacing the network of interactions with a complete graph whose links have weights $\mathpzc{p}_{ij}$; a byproduct of this approximation is that the local OPs, $\mathcal{O}_i\equiv\frac{1}{k_i}\sum_jA_{ij}\mathpzc{f}_j$, and the global OP, $\mathcal{O}\equiv\sum_{i,j}A_{ij}\mathpzc{f}_j/\sum_{i,j}A_{ij}$, coincide. 
In light of the above observations, the magnetization evolution of a system of $2$ randomly (longitudinally) interdependent ER networks of Potts-spins is given by the coupled coarse-grained Langevin equations
\begin{gather}
-\dot{\mathpzc{X}}_1(t)=\frac{1+(s_1-1)e^{-\beta J_1s_1\langle k\rangle_1\mathpzc{X}_1\textcolor{blue}{\mathpzc{X}_2}}}{e^{-\beta J_1s_1\langle k\rangle_1\mathpzc{X}_1\textcolor{blue}{\mathpzc{X}_2}}+s_1-1}\mathpzc{X}_1+\frac{e^{-\beta J_1s_1\langle k\rangle_1\mathpzc{X}_1\textcolor{blue}{\mathpzc{X}_2}}-1}{e^{-\beta J_1s_1\langle k\rangle_1\mathpzc{X}_1\textcolor{blue}{\mathpzc{X}_2}}+s_1-1},\label{eq:S32}\\
-\dot{\mathpzc{X}}_2(t)=\frac{1+(s_2-1)e^{-\beta J_2s_2\langle k\rangle_2\mathpzc{X}_2\textcolor{blue}{\mathpzc{X}_1}}}{e^{-\beta J_2s_2\langle k\rangle_2\mathpzc{X}_2\textcolor{blue}{\mathpzc{X}_1}}+s_2-1}\mathpzc{X}_2+\frac{e^{-\beta J_2s_2\langle k\rangle_2\mathpzc{X}_2\textcolor{blue}{\mathpzc{X}_1}}-1}{e^{-\beta J_2s_2\langle k\rangle_2\mathpzc{X}_2\textcolor{blue}{\mathpzc{X}_1}}+s_2-1}.\label{eq:S32}
\end{gather}
\noindent
In the particular case of symmetric networks of Potts spins, i.e.\ $\langle k\rangle\equiv\langle k\rangle_1=\langle k\rangle_2$ and $s\equiv s_1=s_2$, the above equations collapse into the single differential equation
$$
-\dot{\mathpzc{X}}(t)=\frac{1+(s-1)e^{-s\beta \langle k\rangle\mathpzc{X}^2}}{e^{-s\beta \langle k\rangle\mathpzc{X}^2}+s-1}\mathpzc{X}+\frac{e^{-s\beta \langle k\rangle\mathpzc{X}^2}-1}{e^{-s\beta \langle k\rangle\mathpzc{X}^2}+s-1},
$$
\noindent 
where $\mathpzc{X}\equiv\mathpzc{X}_1=\mathpzc{X}_2$ and, for simplicity, $J_1=J_2=1$. 
At equilibrium, we hence find 
\begin{equation}\label{eq:S32}
\mathpzc{X}^{(s)}(\beta)=\frac{1-e^{s\beta \langle k\rangle\mathpzc{X}^2}}{e^{s\beta \langle k\rangle\mathpzc{X}^2}+s-1},\qquad s=1,2,\dots
\end{equation}
\indent 
If $s=2$, Eq.~\eqref{eq:S32} describes the fully mean-field limit of a $2$-\textsc{dep}-spin model on randomly coupled ER graphs with average degree $\langle k\rangle$, i.e. $\mathcal{M}=\mathrm{tanh}(\beta\langle k\rangle \mathcal{M}^2)$ with $\mathcal{M}\equiv\mathpzc{X}^{(2)}$. 
The limit $s\to1$ of Eq.~\eqref{eq:S32} at $\beta\equiv1$ yields, instead, the exact solution for the GS of the random ferromagnetic $3$-spin model, i.e.\ the relative size of the hyperbackbone of a random hypergraph made of triads given in Eq.~\eqref{eq:S25} with $K=3$. 
Incidentally, in fact, the $s\to1$ limit of the fully MF equations for the Potts model on ER graphs leads to the exact equations characterizing the underlying percolation problem at finite connectivities. 
This coincidence enables to determine the percolation equations characterizing the $T=0$ GS limit of our random $(M,q)$-\textsc{dep}-spin model on ER networks. 
Following a reasoning analogous to the above for the case of e.g.\ $M=2$ symmetric ER networks of Potts spins coupled by a fraction $q\in[0,1]$ of interdependent links, one can readily find
\begin{equation}\label{eq:S35}
-\dot{\mathpzc{X}}^{(s)}(t)=q\mathcal{R}^{(s)}(\mathpzc{X}^2)+(1-q)\mathcal{R}^{(s)}(\mathpzc{X}),\qquad
\mathcal{R}^{(s)}(\mathpzc{x})\equiv
\bigg(\frac{1+(s-1)e^{-s\beta \langle k\rangle\mathpzc{x}}}{e^{-s\beta \langle k\rangle\mathpzc{x}}+s-1}\mathpzc{x}+\frac{e^{-s\beta \langle k\rangle\mathpzc{x}}-1}{e^{-s\beta \langle k\rangle\mathpzc{x}}+s-1}\bigg),
\end{equation}
\noindent
For $s=2$ we find the fully MF equations describing the magnetization evolution of $M=2$ partially interdependent ER networks with same average degree $\langle k\rangle$ (see Eq.~\eqref{eq:M8} in the Methods), whilst for $s\to1$ and $\beta\equiv1$ we obtain 
\begin{equation}\label{eq:S33}
P_\infty=(1-q)\Big(1-e^{-\langle k\rangle P_\infty}\Big)+q\Big(1-e^{-\langle k\rangle P_\infty^2}\Big).
\end{equation}
\noindent 
Eq.~\eqref{eq:S33} has to be compared with Eq.~\eqref{eq:S29} with $K=3$ and $\langle k\rangle\equiv \bar{k}_2=\bar{k}_3$, which characterizes the GS of the random ferromagnetic $(2+q)$-spin model or, equivalently, the exact solution for interdependent percolation on $M=2$
\begin{wrapfigure}[21]{l}{0.42\textwidth}
	\centering 
		\includegraphics[width=0.42\textwidth]{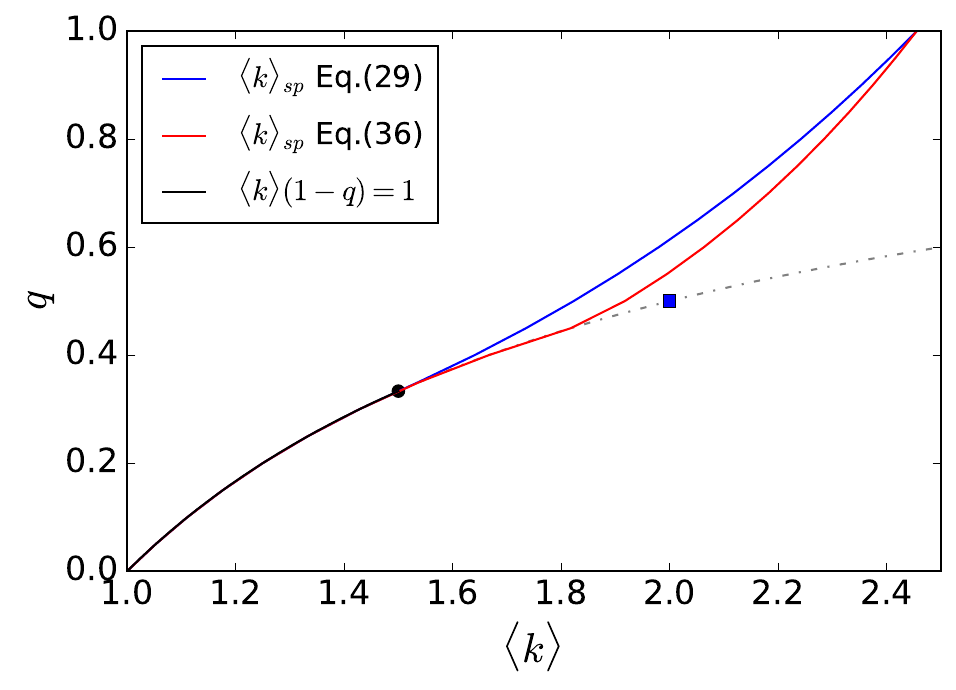}\vspace*{-0.25cm}
		\caption{\small \textbf{$(M,q)$-dep-spin $T=0$ GS.} (Color online) Structural thresholds characterizing interdependent percolation, Eq.~\eqref{eq:S29} with $K=3$ and the $T=0$ GS of our random $(M,q)$-\textsc{dep}-spin model on $M=2$ ER graphs with the same average degree $\langle k\rangle$. Notice that the discrepancy of the tricritical point $(1/2,2)$ for the $T=0$ GS of $(2,q)$-\textsc{dep}-spin with what depicted by the red curve is due to numerical precision.}
	\label{fig:S6}
\end{wrapfigure}
\noindent 
symmetric and partially coupled ER networks. 
The extreme limits $q=1$ and $q=0$ coincide with the corresponding limiting behaviours of interdependent percolation. 
For intermediate values of $q\in(0,1)$, however, the two percolation processes differ, hinting at the fact that the two underlying higher-order structures have different topological properties. 
In Fig.~\ref{fig:S6} we show the numerical thresholds characterizing the structural spinodals of Eq.~\eqref{eq:S33} with respect to those of Eq.~\eqref{eq:S29} with $K=3$. 
Notice, in particular, that whilst interdependent percolation has a tricritical point at $(1/3, 3/2)$ in the $(q,\langle k\rangle)$ phase plane (blue symbol, Fig.~\ref{fig:S6}), the $T=0$ GS of our random $(2,q)$-\textsc{dep}-spin model features a tricritical point at $(1/2,2)$ (red symbol, Fig.~\ref{fig:S6}).

\end{document}